%% file: APRreview_rev2.tex
\documentclass[%
 aip,
 amsmath,amssymb,
 reprint,%
]{revtex4-2}

\usepackage{graphicx}
\graphicspath{{figures/}{.}}
\usepackage{dcolumn}
\usepackage{bm}

\usepackage[utf8]{inputenc}
\usepackage[T1]{fontenc}
\usepackage{mathptmx}
\usepackage{etoolbox}

\usepackage{hyperref}
\renewcommand{\doi}[1]{\textsc{doi}: \href{http://dx.doi.org/#1}{\nolinkurl{#1}}}

\usepackage{bbold} 
\newcommand{\given}{\vert}
\newcommand{\be}{\begin{equation}}
\newcommand{\ee}{\end{equation}}

\makeatletter
\def\@email#1#2{%
 \endgroup
 \patchcmd{\titleblock@produce}
  {\frontmatter@RRAPformat}
  {\frontmatter@RRAPformat{\produce@RRAP{*#1\href{mailto:#2}{#2}}}\frontmatter@RRAPformat}
  {}{}
}%
\makeatother

\begin{document}

\preprint{AIP/123-QED}

\title[Metastable dynamics of neural circuits and networks]{Metastable dynamics of neural circuits and networks}

\author{B. A. W. Brinkman}
\altaffiliation{These authors contributed equally as first authors.}
\affiliation{Department of Neurobiology and Behavior \& Program in Neuroscience, Stony Brook University, Stony Brook, NY 11794, USA}
\affiliation{Center for Neural Circuit Dynamics, Stony Brook University, Stony Brook, NY 11794, USA}

\author{H. Yan}%
\altaffiliation{These authors contributed equally as first authors.}
\affiliation{State Key Laboratory of Electroanalytical Chemistry, Changchun Institute of Applied Chemistry, Chinese Academy of Sciences, Changchun, Jilin, 130022, P.R. China}

\author{A. Maffei}
\affiliation{Department of Neurobiology and Behavior \& Program in Neuroscience, Stony Brook University, Stony Brook, NY 11794, USA}
\affiliation{Center for Neural Circuit Dynamics, Stony Brook University, Stony Brook, NY 11794, USA}

\author{I. M. Park}
\affiliation{Department of Neurobiology and Behavior \& Program in Neuroscience, Stony Brook University, Stony Brook, NY 11794, USA}
\affiliation{Center for Neural Circuit Dynamics, Stony Brook University, Stony Brook, NY 11794, USA}

\author{A. Fontanini}
\affiliation{Department of Neurobiology and Behavior \& Program in Neuroscience, Stony Brook University, Stony Brook, NY 11794, USA}
\affiliation{Center for Neural Circuit Dynamics, Stony Brook University, Stony Brook, NY 11794, USA}

\author{J. Wang}
\altaffiliation[Authors to whom correspondence should be addressed: ]{jin.wang.1@stonybrook.edu (J.W.) or giancarlo.lacamera@stonybrook.edu (G.L.C.)}
\affiliation{Center for Neural Circuit Dynamics, Stony Brook University, Stony Brook, NY 11794, USA}
\affiliation{Department of Physics and Astronomy and Department of Chemistry, Stony Brook University, Stony Brook, NY 11794, USA}

\author{G. La Camera}
\altaffiliation[Authors to whom correspondence should be addressed: ]{jin.wang.1@stonybrook.edu (J.W.) or giancarlo.lacamera@stonybrook.edu (G.L.C.)}
\affiliation{Department of Neurobiology and Behavior \& Program in Neuroscience, Stony Brook University, Stony Brook, NY 11794, USA}
\affiliation{Center for Neural Circuit Dynamics, Stony Brook University, Stony Brook, NY 11794, USA}

\date{\today}

\begin{abstract}
Cortical neurons emit seemingly erratic trains of action potentials, or `spikes', and neural network dynamics emerge from the coordinated spiking activity within neural circuits. These rich dynamics manifest themselves in a variety of patterns which emerge spontaneously or in response to incoming activity produced by sensory inputs. In this review, we focus on neural dynamics that is best understood as a sequence of repeated activations of a number of discrete hidden states. These transiently occupied states are termed `metastable' and have been linked to important sensory and cognitive functions. In the rodent gustatory cortex, for instance, metastable dynamics have been associated with stimulus coding, with states of expectation, and with decision making. In frontal, parietal and motor areas of macaques, metastable activity has been related to behavioral performance, choice behavior, task difficulty, and attention. In this article, we review the experimental evidence for neural metastable dynamics together with theoretical approaches to the study of metastable activity in neural circuits. These approaches include: (i) a theoretical framework based on non-equilibrium statistical physics for network dynamics; (ii) statistical approaches to extract information about metastable states from a variety of neural signals, and (iii) recent neural network approaches, informed by experimental results, to model the emergence of metastable dynamics. By discussing these topics, we aim to provide a cohesive view of how transitions between different states of activity may provide the neural underpinnings for essential functions such as perception, memory, expectation or decision making, and more generally, how the study of metastable neural activity may advance our understanding of neural circuit function in health and disease.
\end{abstract}

\maketitle
\tableofcontents 

\section{Introduction} \label{sec:intro}
\begin{figure*}
\centering
\includegraphics[width=1\textwidth]{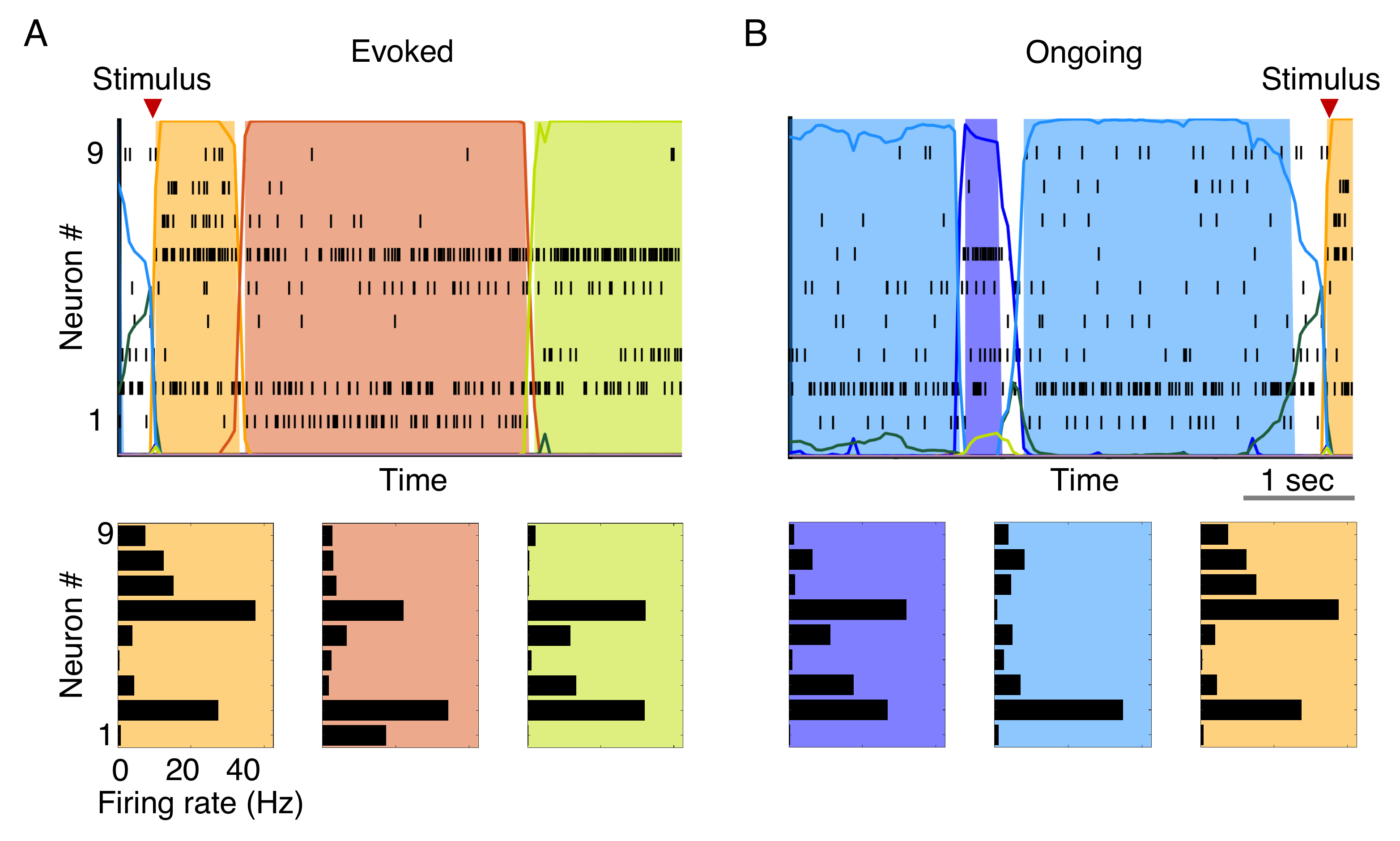}
\caption[]{\small Example of metastable neural dynamics. A: Top panel: segmentation of neural activity from 9 simultaneously recorded neurons in the rat gustatory cortex. Each line is a spike train, i.e., a sequence of spike times from one of the 9 neurons. Recordings were taken as the animal waited and then received a tastant in its mouth at random times (`Stimulus'). Colored areas correspond to hidden states of the neural activity, each color representing a different state. A bin of data was assigned to a state if the probability of being in that state, given the data, was higher than $0.8$ (colored lines). Bottom panel: the hidden states can be represented as vectors of firing rates across the 9 neurons. B: Same as A for `ongoing' neural activity, i.e., for neural activity in the `idle time' between two stimuli. See Sec.~\ref{sec:HMM} for details.}
\label{fig:hmm1}
\end{figure*}

Metastability of neural dynamics is receiving growing recognition for its role in cortical computations \cite{Rabinovich2008-kg,Durstewitz:2008we,Miller2010-xy,La-Camera:2019zm,Cao:2020pd}. Aspects of sensory processing, attention, expectation and decision making are increasingly found to be explained in terms of neural activity transitioning through sequences of metastable states, and by the temporal modulation of sequences dynamics. In a prototypical situation, metastable states are patterns of firing rates across simultaneously recorded neurons which linger for 300~ms -- 3~sec prior to transitioning to a new pattern. An example is shown in Fig.~\ref{fig:hmm1}, where the electrophysiogical activity of 9 neurons from the gustatory cortex of behaving rats is shown together with its segmentation in a sequence of metastable states. These hidden state patterns have been detected in cortical and hippocampal areas of  monkeys and rodents engaged in a variety of tasks, as well as during periods of spontaneous, ongoing activity \cite{La-Camera:2019zm}. Recently, evidence of metastable states preferentially associated with different task conditions has also been found in humans performing a working memory task \cite{Taghia2018-gp}; and metastable states in the monkey dorsal premotor cortex have been used to decode the intention to plan a movement in brain-machine prosthetic devices \cite{Kemere:2008uq}.

What advantages might metastable dynamics provide to a physical or biological system -- such as the brain -- that processes information and performs complex tasks? To understand the function of these dynamics it may be useful to begin describing when they occur. Metastable states can be induced by external stimuli but can also be generated spontaneously, in the absence of external stimulation (see e.g. the activity prior to `Stimulus' in Fig.~\ref{fig:hmm1}B). In the presence of stimulation, new states occur and coexist with the internally generated ones. For instance, in the rat gustatory insular cortex (GC), some of these metastable states occur more frequently in the presence of a particular taste stimulus. These have been dubbed `coding states', as they convey information about the stimulus \cite{Mazzucato:2019wi}. Metastable states have been found to code for more abstract concepts such as the relative distance of two target stimuli based on stimulus features \cite{Benozzo:2021en}.  Besides the meaning of coding states, it is their organization in sequences that promises the largest benefit in terms of coding. In one example, when states coding for different stimulus features \cite{Abeles1995-zb,Seidemann1996-bf,Jones:2007ff} or different decisions \cite{Bollimunta:2012rt,Rich2016-iq} occur in the same sequence, they allow the possibility to code for all options relevant to a particular task, even while the subject is being presented with a subset of them. This presence of multiple switching states could therefore represent the neural substrate of keeping a menu of options in mind for the purpose of making decisions. 

Metastable dynamics also presents advantages from the point of coding for temporal events. Hidden states are not precisely locked to external triggers even when induced by external stimuli \cite{Abeles1995-zb,Seidemann1996-bf,Jones:2007ff,Ponce-Alvarez2012-zi}. States related to internal deliberations have variable onset times which can be taken as a proxy for the timing of deliberations, allowing one to pinpoint the timing of the decision. This timing is flexible and can be modulated globally by stretching or shrinking the metastable sequences in which they occur. For instance, in GC, coding states for specific tastants tend to be within the first 0.5~s following stimulus presentations, but shift towards earlier onset times in trials when a stimulus is expected -- providing a potential neural substrate of expectation \cite{Mazzucato:2019wi}. On the other hand, monkeys performing a distance-discrimination task tend to make errors when state sequences stretch out in time, i.e., when the metastable dynamics slows down \cite{Benozzo:2021en}. 

These and related findings -- discussed in more detail in Sec.~\ref{sec:evidence} -- suggest an important role for the temporal modulations of sequences of metastable states rather than, or in addition to, the identity of states coding for specific features at specific points in time. Little is known, however, about the mechanistic origin of these metastable states. This problem has been addressed with computational modeling, starting from the work of \cite{Miller2010-xy} that has clarified the benefits of metastable activity for categorical decision making. This and subsequent related models are based on biologically plausible spiking network models that allow to predict the results of specific experiments (reviewed here in Sec.~\ref{sec:spiking_net}). In particular, the metastable activity observed in electrophysiological experiments can be explained by spiking network models with a clustered architecture \cite{Deco2012-wg,Litwin-Kumar:2012ty,Mazzucato:2015jk,Cao:2016bf,Setareh:2017km,Rostami:2020wx}. A clustered network consists of groups of excitatory and inhibitory neurons that are preferentially connected to one another inside each group. When the mean strength of the synaptic weights inside clusters exceeds a critical point, a mean field analysis shows the existence of a large number of activity configurations characterized by the number of active clusters \cite{Mazzucato:2015jk}. In networks of finite size these configurations become metastable, as shown in numerical simulations. This model has so far explained a wealth of data, mostly obtained in the GC of rodents, including the temporal modulation of transition rates due to expectation \cite{Mazzucato:2019wi} as well as the reduced dimensionality of the neural activity evoked by a stimulus compared to ongoing activity \cite{Mazzucato2016-hl}. 

In this article we give a detailed and up-to-date description of metastability in cortical circuits together with current modeling efforts. We start from a definition of metastability in physics and neuroscience and with a clarification of the kind of metastability that is the main focus of this review (Sec.~\ref{sec:definition}): the one characterized by repeatable metastable transitions, rather than metastability en route to a ground state configuration. We exemplify this notion in a classical spin system in Sec.~\ref{sec:spinmodels}. We then review evidence of metastable dynamics in neural circuits and describe how such metastable dynamics can explain important features of sensory and cognitive processes (Sec.~\ref{sec:evidence}). We then present statistical models of metastable dynamical systems and methods for their analysis, with an emphasis on hidden state models (Sec.~\ref{sec:statmodels}). This section is followed by a section on theoretical models of metastable dynamics (Sec.~\ref{sec:theorymodels}), proceeding from cortical networks of spiking neurons to more formal models interpretable as coarse-grained descriptions of population activity. Mean field reductions of these models are essential for understanding their behavior and typically result in firing rate models of spiking networks. We also present a path integral formalism for studying metastability in non-equilibrium systems lacking detailed balance, an approach known as the landscape and flux theory of neural networks \cite{Yan:2013rs,YH25}. The last section will focus on the problem of learning and plasticity, specifically, how metastable circuits can be formed via experience-dependent plasticity and can sustain themselves in the face of ongoing metastable activity (Sec.~\ref{sec:plasticity}). We will review the available evidence for neural clusters and present a concrete example of the existing models focusing on this problem, as well as theoretical investigations of the consequences of learning in models of memory, decision making and fear expression. Finally, in the `Summary and conclusions' section (Sec.~\ref{sec:summary}), we summarize the main points reviewed in this article and appraise the potential role of metastable dynamics in neural coding and cortical computation in comparison to earlier views.

\section{Definitions of metastable dynamics} \label{sec:definition}

In physical systems, metastability typically refers to the long-lived occupation of a state with higher energy than the lowest energy state \cite{penrose1971,capocaccia1974}. For simple biological and chemical systems, such as the case of isomerization, this definition also applies. The long time spent in the metastable state is due to the presence of effective energy barriers that prevent the system from easily making transitions to lower energy states. Thermal agitation or external perturbation can induce the system to escape the metastable state. In systems with many local energy minima, metastable dynamics may ensue as transitions among states with lower energy after some amount of lingering in each metastable state, eventually reaching the lowest energy state (potentially after an asymptotically long time) \cite{BinderRevModPhys1986,BOUCHAUDBook}. It is possible, however, that there are many minima of comparable energy, and noise fluctuations may be able to knock the system between these different configurations repeatedly. More generally, any stochastic process in which many configurations are comparable in probability can feature such metastable transitions between such configurations---it is this aspect of metastability that is the primary focus of this review and whose implications for neuroscience we will expound on. This extends metastability to complex biological or chemical systems in which the energetics of a process may not be known or well-defined but the dynamics can be modeled using stochastic processes \cite{dinner1998,brazhkin2006}. In this more general context of stochastic dynamical systems, metastable transitions exist due to the existence of stable fixed points in the deterministic dynamics, which are then perturbed by noise fluctuations, with large enough fluctuations allowing the system to escape the basin of attraction of one fixed point and be drawn towards another \cite{serdukova2016}.

Metastable dynamics also occur in deterministic systems which are not characterized by a notion of energy. One example is the Volterra-Lotka system in high dimensions, which has been studied in the context of brain dynamics by Rabinovich, Abarbanel, Laurent and collaborators \cite{Rabinovich:2001ex,Rabinovich2006-lw,Laurent:2001dn}. In this deterministic system, the trajectories proceed along saddle points, i.e., points that attract the flow of the dynamical system along some directions, spending a transient time near the saddle point before being repelled along an unstable direction. Systems characterized by a large number of these unstable equilibria will tend to follow erratic trajectories. 

Finally, a third class of dynamical systems that exhibit metastable dynamics straddles the line between the previous two examples: large but finite deterministic systems with quenched disorder often behave effectively like stochastic systems \cite{parisibook}. A particular example that we will discuss in detail in this review is a spiking neural network with random connectivity but organized in clusters of strongly connected neurons. Such a network can linger in multiple different metastable patterns of firing rates across its neurons. Only a handful of such patterns are observed, which are explored in a way that resembles the dynamics of a finite Markov chain. Note that the last two examples are fully deterministic dynamical systems, and yet they produce metastable dynamics with seemingly random transition times. 

In an effort to clarify the notion of metastable dynamics that is the main focus of this review, in Sec.~\ref{sec:spinmodels} we discuss some elementary examples of this phenomenon in elementary physical models, namely the Ising model and variations of it -- again, with a focus on repeatable metastable transitions, not just metastability en route to a ground state configuration. Following this introduction, in Sec.~\ref{sec:evidence} we review the evidence for metastability in neural circuitry, followed by Sec.~\ref{sec:models} in which we review methods for statistical analysis (Sec.~\ref{sec:statmodels}) and modeling of this data and, in general, metastable dynamics in the brain (Sec.~\ref{sec:theorymodels}). Readers familiar with metastability in physical systems may skip ahead to these sections.

\section{Metastable dynamics in classical spin systems} 
\label{sec:spinmodels}
Metastability has long been a topic of interest in the physics of disordered systems \cite{BinderRevModPhys1986,BOUCHAUDBook,BovierBook2009}, chemical reaction networks \cite{WangJChemPhys2017}, and population biology \cite{AssafJPhysA2017}. Many applications in physics focus on metastable transitions from high energy (low probability) states to the ground state configuration. In disordered systems these transitions can take much longer than a typical experiment or even a typical graduate student Ph.D. \cite{BinderRevModPhys1986,BOUCHAUDBook}. However, the more interesting phenomena from the point of view of neuroscience is repeatable transitions between configurations of similar probability, caused by some source of external or internal fluctuations. Finite-size spin models like the Ising model display such repeatable transitions between states of opposite magnetization, and we begin by briefly reviewing results on metastability in the Ising model, highlighting features of the metastable statistics that are observed more generally. This section is useful for physics readers unfamiliar with reversible metastable transitions and neuroscience readers unfamiliar with spin models of neural activity.

\subsection{A prototypical example: spin models} 
Spin models, while originally developed to understand magnetization, have also enjoyed extensive use as models in neuroscience \cite{amari72a,h82,YH2,amitBook89}. In the simplest cases, a magnetic material can be modeled as being composed of many magnetic domains, each of which has a local magnetic moment, referred to as `spins'. In many applications of interest the spin of a domain points either `up' or `down.' We can therefore assign to each domain a binary variable $s$ such that $s = +1$ if the domain's spin is pointing up and $s = -1$ if the spin is pointing down. In neuroscience applications these binary variables may be interpreted as representing `active' or `inactive' neurons, respectively\footnote{The choice of $s= \pm 1$ is not fundamental, and in neuroscience the values of $0$ (inactive) and $1$ (active) are often conventionally used instead. The two choices are related by a linear change of variables. We use the physics convention as the symmetry of the model is clearer in this representation.} \cite{Schneidman:2006fg,MackeNeurIPS2011}.

The magnetic properties of a material are determined by the overall configuration and alignment of spins, $\mathbf{s} = \left\{-1,1\right\}^N$ (a vector of $N$ binary elements), where $N$ is the number of magnetic domains (or neurons). The configurations are determined by a competition between the magnetic interactions between spins and thermal fluctuations. The simplest models quantify the total configuration energy of the spins as
\begin{equation} \label{eqn:spinHamiltonian}
E(\mathbf{s}) = -\frac{1}{2} \sum_{i,j=1}^N  J_{ij} s_i s_j - \sum_{i=1}^N H_i s_i,
\end{equation}
where $J_{ij}$ is the interaction strength between spins $i$ and $j$ and $H_i$ is the magnetic field felt by spin $i$, which can vary from spin to spin due to impurities in the material \cite{BinderRevModPhys1986,BOUCHAUDBook,GoldenfeldBook1992,NATTERMANNbook}. This form of model has also been used extensively in neuroscience, in which $J_{ij}$ represents pairwise synaptic interactions between neurons and $H_i$ mimics the effects of external driving currents. One of the earliest uses was the celebrated Amari-Hopfield network model of associative memory \cite{amari72a,h82}, and in recent years the model (\ref{eqn:spinHamiltonian}) has also emerged in data-driven applications as the maximum-entropy model that exactly matches the empirically observed mean firing rates and pairwise covariances of a neural population \cite{Schneidman:2006fg,MackeNeurIPS2011}. We will elaborate on these connections in Sec.~\ref{sec:fluxtheory}.

Configurations of spins requiring the least amount of energy to be maintained are the most stable; accordingly, strong positive bonds $J_{ij} > 0$ favor the alignment of spins $i$ and $j$ in the same direction, while strong negative bonds $J_{ij} < 0$ favor opposite alignments. Similarly, spins will tend to align with strong fields $H_i$. For a fixed set of spin-spin interactions $J_{ij}$ and magnetic fields $H_i$ this function quantifies the  `energy landscape' of the magnet. Energetically, the most favorable configuration of spins is that with the lowest energy, the global minimum of the landscape. However, strong enough thermal fluctuations can provide enough energy to flip spins into energetically unfavorable configurations. Precisely, if the magnet is held at a fixed temperature $T$, then the configuration of spins will equilibrate to a distribution of the form
\begin{equation} \label{eqn:eqspinmodel}
P(\mathbf{s}) = \frac{1}{Z} e^{-E(\mathbf{s})/k_B T },
\end{equation}
where $k_B$ is Boltzmann's constant and the normalization $Z$ is the `partition function', defined by
\begin{equation}
Z = \sum_{\mathbf{s}} e^{-E(\mathbf{s})/k_B T },
\label{eqn:partition}
\end{equation}
so that the probability $P(\mathbf{s})$ is properly normalized. The logarithm of the partition function determines the free energy of the system, $F \equiv -k_B T \ln Z$, both of which play a central role in equilibrium statistical mechanics. Specifically, all statistical information about the model (means, covariances, etc.) can be obtained by differentiating the free energy with respect to the parameters $J$ or $H$. This feature is due to the fact that the exponential form of the distribution means that the external fields $H_i$ or $J_{ij}$ can play the role of source terms in the definition of the moment generating function, imbuing the partition function with related properties. Similarly, the logarithm of the moment generating function is the cumulant generating function, derivatives of which produce cumulants (covariances, etc.) of the distribution, a property inherited by the free energy.

The form of Eq.~(\ref{eqn:eqspinmodel}) reveals that configurations with comparable energy will have comparable probabilities. For finite $N < \infty$ the system is ergodic, meaning that $P(\mathbf{s})$ can be interpreted equivalently as the frequency of different spin configurations across an infinite ensemble of magnets with the same temperature and energy landscape \emph{or} the steady state distribution of a single spin system at long times, such that the frequency of configurations of this single system at different snapshots in time will be distributed according to Eq.~(\ref{eqn:eqspinmodel}), or a combination of these two interpretations. Consequently, long time averages of the value of any spin will be equal to the average value of that spin across an ensemble of identically prepared magnets. 

\begin{figure*}
\centering
\includegraphics[width=1\textwidth]{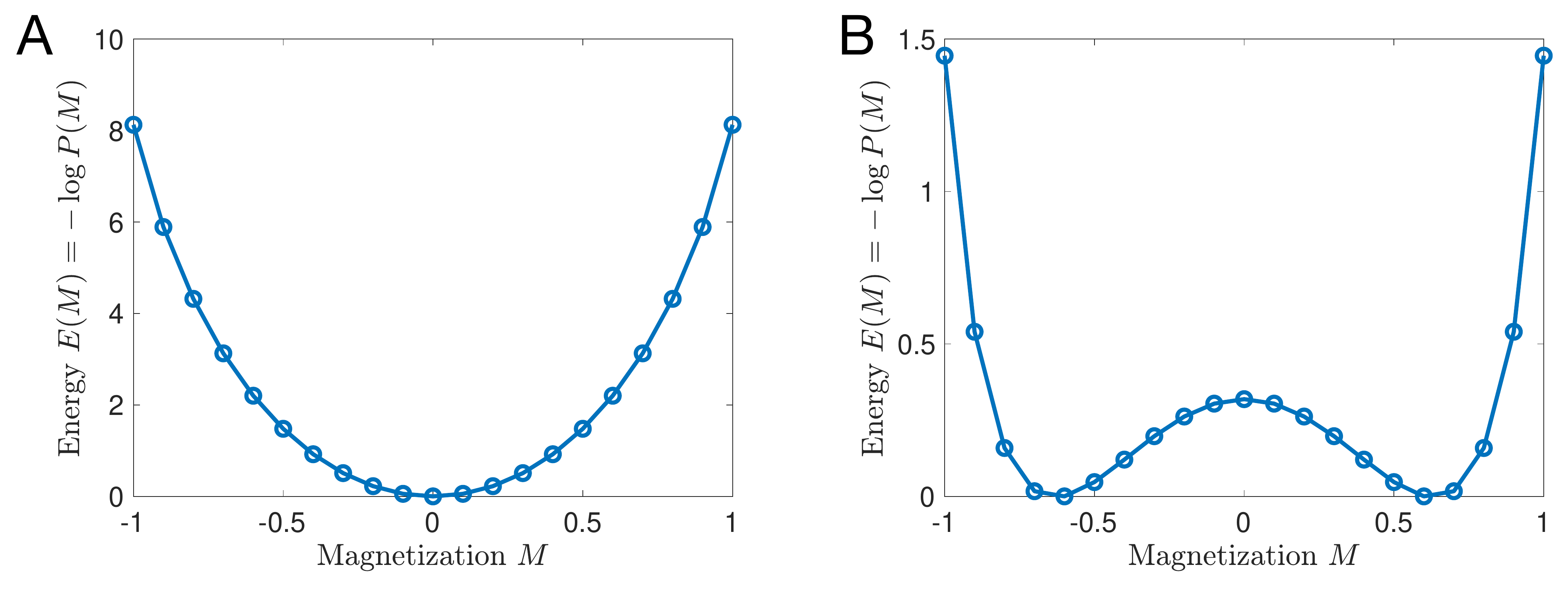}
\caption[]{\small
    Energy landscape in an Ising model of $N = 20$ spins, for $T > T_c$ (A) and $T < T_c$ (B). At high temperatures (panel A) there is a single minimum at $M = 0$ and the mean magnetization is $\langle M \rangle = 0$. At low temperatures (panel B) there are two equally probable magnetizations at $M = \pm M_{\rm sp}$. Thermal fluctuations will cause the magnet to reverse orientation occasionally, such that over time the average magnetization is $\langle M \rangle = 0$ for any finite $N$.}
\label{fig:EMmodel}
\end{figure*}

The most probable configurations of the spin system are those with \emph{local} energy minima, as depicted in Fig.~\ref{fig:EMmodel}. In a stochastic process with Eq.~(\ref{eqn:eqspinmodel}) as its steady state distribution the system will spend extended periods of time near each of these locally probable/energetically favorable configurations, until thermal fluctuations cause the system to escape and transition to a different state---i.e., the existence of local energy minima/probability maxima gives rise to metastability. This said, the metastability of spin systems is generally impossible to observe due to the impracticality of measuring the microscopic configuration of every spin in a magnet. Instead, one would typically measure `macroscopic' properties. In the case of ferromagnetic materials, the primary quantity of interest is the overall magnetization, the population average of the spins:
\begin{equation}
M = \frac{1}{N} \sum_{i=1}^N s_i,
\label{eqn:magnetizationdefn}
\end{equation}
where $s_i$ is the state of the $i^{\rm th}$ spin. If a majority of spins are up  then $M > 0$, and $M < 0$ if a majority of the spins are down. In neuroscience applications, $(M+1)/2$ would represent the fraction of active neurons (using the $s_i = \pm 1$ convention). There are typically combinatorially many microscopic configurations of spins that yield any given value of $M$, the exceptions being values near the extremes of $M = \pm 1$, for which there are only a relatively small number of configurations. In general, metastability occurs even at the level of the total magnetization. We may formally derive the distribution of magnetizations from the distribution of spin configurations,
\begin{equation}
P(M) = \sum_{\mathbf{s}} \mathbb{1} \left(M = \frac{1}{N}\sum_{i=1}^N s_i \right) P(\mathbf{s}),
\label{eqn:Mprob}
\end{equation}
where the indicator function $\mathbb{1}\left(M = \frac{1}{N}\sum_{i=1}^N s_i \right)$ ensures that only configurations of spins with the specified magnetization $M$ contribute in the sum.
One can define an energy landscape for the magnetization by $E(M) = -\log P(M)$; see Fig.~\ref{fig:EMmodel}.
Local minima of this function will correspond to metastable states in the macroscopic magnetization. In the next section, we specialize to the case of the Ising model to illustrate metastable transitions in the magnetization.

\subsection{Metastable transitions in the Ising model} 
For concreteness, we consider the Ising model Eq.~(\ref{eqn:spinHamiltonian}) with nearest neighbors ferromagnetic interactions, in which only adjacent spins (`nearest neighbors') interact. Specifically, $J_{ij} = J>0$ when spins $i$ and $j$ are adjacent and $J_{ij} = 0$ otherwise. A non-zero magnet field $H_i = H$ would bias the magnetization towards ${\rm sign}(H)$; however, in the absence of an external field every configuration of spins $\mathbf{s}$ has an energetically equivalent---and hence equally probable---configuration obtained by reversing the direction of each spin; i.e., $P(\mathbf{s}) = P(-\mathbf{s})$. It follows that $P(M)=P(-M)$. At sufficiently large temperatures this property has no impact on the most likely configuration of the system: the distribution $P(M)$ is unimodal (Fig.~\ref{fig:EMmodel}A), peaked at $M = 0$; i.e., the most energetically favorable configurations of spins are those with an equal number of spins pointing up and down. However, in spatial dimensions $d \geq 2$ there is a critical temperature $T_c$,  below which $P(M)$ becomes bimodal, with peaks at $M = \pm M_{\rm sp}$, the `spontaneous magnetization' (Fig.~\ref{fig:EMmodel}B). These two peaks represent two equally probable (energetically favorable) metastable states, and predicts that an Ising magnet should occasionally reverse its overall magnetization, flipping from $M = + M_{\rm sp}$ to $-M_{\rm sp}$ or vice versa. Readers familiar with the ferromagnetic transition may find this paradoxical: the conventional wisdom is that as the temperature is lowered below a critical temperature $T_c$ the mean magnetization should change from $0$ to a non-zero value, either $M_{\rm sp}$ or $-M_{\rm sp}$; however, if the Ising magnet is constantly switching magnetization, then the time-averaged magnetization should be $\frac{1}{2}M_{\rm sp} + \frac{1}{2}(-M_{\rm sp}) = 0$, even for temperatures below $T_c$!

The resolution of this apparent paradox is related to the thermodynamic limit $N \rightarrow \infty$, in which the `spontaneous symmetry breaking' is accompanied by an ergodicity breaking: the dynamics of the spins become trapped in either the $M > 0$ or $M < 0$ phase space, with an infinite energy barrier between them.  We can see how this barrier develops for finite $N$ by investigating the stochastic process of switching from one metastable state to another, and estimating the rate of these transitions. For the Ising magnet this can be implemented using, for example, the Metropolis-Hastings algorithm for simulating stochastic spin flips \cite{BinderRevModPhys1986}. Mathematically these stochastic dynamics can be studied using a master equation formalism \cite{Kampen:2007cr,SmadbeckPNAS2013}, which allows one to calculate the probability that thermal fluctuations will flip a sufficiently large cluster of spins to reverse the sign of the magnetization of an Ising magnet with magnetization near $M = \pm M_{\rm sp}$.  In a $d$-dimensional hypercubic lattice of $N$ spins, the rate $r$ at which the magnetization flips scales as \cite{GoldenfeldBook1992}
\begin{align}
r \sim \exp\left(-{c}N^{(d-1)/d}/k_BT\right),
\label{eqn:Isinglifetime}
\end{align}
where the constant $c$ depends on the coupling $J$ and other parameters (which determine $M_{\rm sp}$ along with the temperature) and $N^{(d-1)/d}$ is the surface area of the cluster size necessary to reverse the sign of the magnetization. 

The form of Eq.~(\ref{eqn:Isinglifetime}) is typical for metastable transition rates in many models, not just spin models; readers may recognize that Eq.~(\ref{eqn:Isinglifetime}) is of the same form as the Arrhenius law in chemical reactions \cite{GoldenfeldBook1992}, and in Sec.~\ref{sec:WCmodel} we give another example. The key feature of the transition rates $r$ is the \emph{exponential} dependence on the system size $N$ and inverse temperature $1/T$. As a result, for large systems or small temperatures metastable transitions will be rare, but over a long enough observation period the Ising system would spend equal amounts of time in each metastable state and hence the time-average of the magnetization will be zero. However, in the thermodynamic limit $N \rightarrow \infty$ the rate $r$ of magnetization flips vanishes, and the Ising magnet will be indefinitely `trapped' in one of its two metastable states. This is often interpreted as the result of an infinitely large energetic barrier that thermal fluctuations would need to overcome in order to flip the magnetization. As a result of this barrier, ergodicity and the spin-reversal symmetry are broken in the thermodynamic limit, and the time average of the magnetization will be equal to $+M_{\rm sp}$ or $-M_{\rm sp}$, depending on which state was chosen by the initial conditions. Accordingly, in the thermodynamic limit $N \rightarrow \infty$ the total magnetization is an `order parameter' for the ferromagnetic-paramagnetic transition: when $M = \pm M_{\rm sp} \neq 0$ the model is in a ferromagnetic phase, and when $M = 0$ the model is in a paramagnetic phase.

The exponential dependence on the number of elements (spins, neurons, etc.) of metastable transition rates like Eq.~(\ref{eqn:Isinglifetime}) illustrates one of the key mysteries of metastability in the brain: in neural populations of thousands or tens-of-thousands of neurons, why do we observe frequent and repeatable metastable transitions over experimentally accessible timescales? One possibility is that spontaneous transitions are rare, but external signals cause transitions (which could be modeled, e.g., in the Ising model by using the external field $H$ to force the system into the the desired state). Another possibility is that there are so many possible metastable states that the total transition rate out of any given state is not negligible. Available experimental evidence, which we review next, suggests that spontaneous transitions do occur in neural circuitry, and provides more clues to the role that metastability might play in neural computation. While spin models capture several key features of metastable dynamics, more detailed dynamical systems and statistical modeling are necessary to describe such data, which we review in Sec.~\ref{sec:models}.

\section{Metastable dynamics in neural circuits} \label{sec:evidence}
The analysis of neural activity from several cortical areas indicates the existence of discrete transitions between different collective neural states. In early pioneering work, Abeles, Tishby and collaborators found that activity in the prefrontal cortex of monkeys performing a delayed object localization task could be described as a sequence of metastable states, where each state was a collection of firing rates across simultaneously recorded neurons \cite{Gat1993-hs,Abeles1995-zb,Seidemann1996-bf,Gat1997-kp}. The authors analyzed the spike counts of simultaneously recorded neurons and demonstrated that a Hidden Markov model analysis---to be described in Sec.~\ref{sec:HMM}---could segment the neural time series data into separate epochs representing distinct (hidden) states. These hidden states appear as unstable attractors of the neural dynamics \cite{Miller2016-kx,Cao:2020pd}, in the sense that these patterns linger for a random time (from hundreds of ms to seconds) before quickly giving way to different patterns. Among the most significant results of these studies were the demonstration that (i) the hidden states identified in response to a given stimulus tend to recur during most of the later recorded activity, \emph{even in the absence of stimuli}, and (ii) pairwise correlations among simultaneously recorded neurons depend on the current hidden state, and not just on neural connectivity. These studies were among the first to shift the focus from stationary to dynamic patterns of neural activity as a means to represent relevant information, and have inspired more recent research that has uncovered multiple potential roles of metastability for sensory and cognitive processes. 

\subsection{Hidden states coding for sensory, motor and cognitive variables} \label{sec:cognitive}

Since these original works, metastable dynamics has been reported in rat gustatory cortex (GC) \cite{Jones:2007ff}, in monkey somatosensory, motor, and premotor cortex \cite{Kemere:2008uq,Ponce-Alvarez2012-zi,Mazurek:2018br}, in monkey's area V4 \cite{Engel2016-bb}, in monkey's orbitofrontal \cite{Rich2016-iq}, parietal \cite{Bollimunta:2012rt} and dorsolateral prefrontal cortex \cite{Benozzo:2021en}, in the hippocampus of rats \cite{Maboudi2018-ox}, in the zebrafish \cite{Marques:2020qe} and in multiple human brain areas \cite{Taghia2018-gp}.

In addition to sensory information, metastable activity has been implicated in changes of behavioral and cognitive states during tasks requiring attention \cite{Engel2016-bb}, expectation \cite{Mazzucato:2019wi}, decisions \cite{Ponce-Alvarez2012-zi,Rich2016-iq,Sadacca2016-yo,Benozzo:2021en}, spatial navigation \cite{Maboudi2018-ox,Marques:2020qe} and working memory \cite{Ponce-Alvarez2012-zi,Taghia2018-gp}. 

\begin{figure*}
\center
\includegraphics[width=1\textwidth]{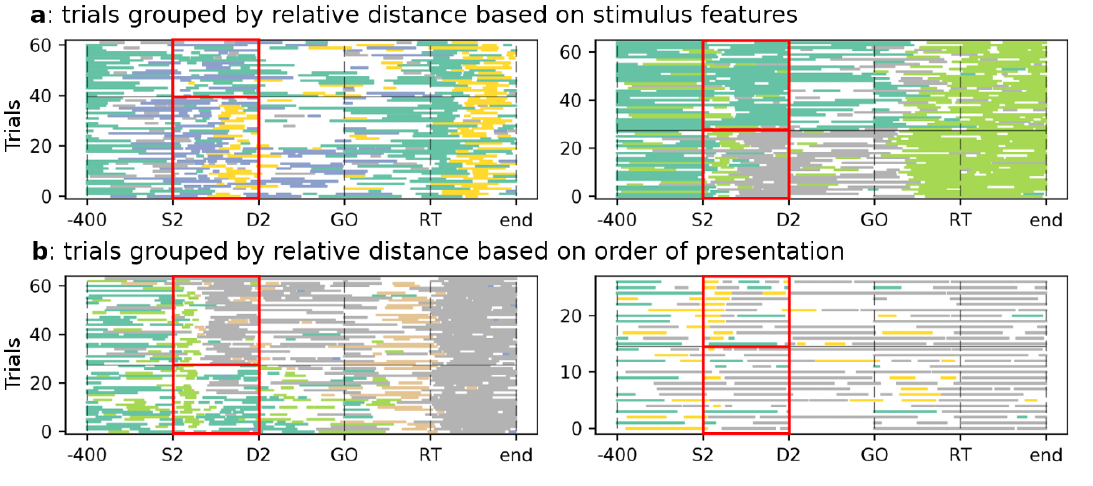}
\caption[]{\small Sequences of hidden states for the monkey experiment reported in \cite{Benozzo:2021en}. Each line is a trial, each colored segment is a hidden state. White segments correspond to epochs in which no hidden state could be assigned with the necessary confidence. (a) This panel shows hidden states that are coding states for relative distance based on stimulus features, occurring during the presentation of the second stimulus (`S2', red box). By definition, these hidden states were statistically more often present depending on whether the further stimulus from the center was a blue circle (bottom trials) or a red square (top trials). Two example sessions are shown; coding states are the dark green and yellow states in the left panel, and the dark green and gray states in the right panel. (b) Coding states for relative distance based on order of presentation during the second stimulus (2 example sessions shown). Coding states are the dark green, orange, and gray states in the left panel and the yellow state in the right panel. In this case, the coding states were more often present if the further stimulus appeared first (bottom trials) or last (top trials). In both panels, trials were grouped according to the coded variable and highlighted by the red box. The same colors in different panels do not imply the same state. Reproduced from D. Benozzo, G. La Camera, and A. Genovesio, Cell Rep 35, 108934 (2021). Licensed under a Creative Commons Attribution (CC BY-NC-ND 4.0) license. \cite{Benozzo:2021en}}
\label{fig:benozzo}
\end{figure*}

Most of these studies were electrophysiological studies where the hidden states were vectors of firing rates across neurons. Some of these states seem to convey more information than other states on the identity of specific stimuli, and have been dubbed `coding states'. Coding states in the rat GC have been found to code for specific taste stimuli \cite{Mazzucato:2019wi}. More recently, states coding for more abstract stimulus features have been found in a distance-discrimination task in which monkeys had to report which of two stimuli were farther from a central location on a computer screen \cite{Benozzo:2021en}. In this case, it was found that some states reflect the relative distance of the two stimuli based on their features (e.g., whether they were a blue circle or a red square, see Fig.~\ref{fig:benozzo}a). Similarly, other hidden states where found to code for relative distance based on which stimulus had been presented first to the monkey (Fig.~\ref{fig:benozzo}b). Note the trial-to-trial variability in the onset and duration of the hidden states, a hallmark of internally generated activity in neural circuits (more on this later). 

Hidden states have also been linked to acts of decisions and other internal deliberations. \cite{Bollimunta:2012rt} have described hidden states related to perceptual decisions in monkeys' parietal cortex. These authors found that, despite single neurons' firing rates tend to increase gradually as the subjects sample stimulus evidence to perform perceptual decisions, sharp transitions are occasionally observed among discrete states coding for specific decisions. This phenomenon has been interpreted as reflecting `changes of mind', a rather elusive internal process whose neural substrate is notoriously difficult to characterize. Abrupt transitions in neural states associated to changes of mind have also been reported in the medial prefrontal cortex of rats performing rule-based decisions \cite{Durstewitz2010-vh}.

In more recent work \cite{liamB20}, three types of hidden states were found in the GC of mice performing a discrimination task based on the identity of 4 tastants serving as decision cues \cite{Vincis:2020zv}. Two of the 4 tastants cued a `go left' action while the other two cued a `go right' action. Separate hidden states were found to code for the `quality' of tastants (bitter vs. sweet), for the cue value of the tastants (`go left' vs `go right'), and for the actual action taken (`left' vs `right'). Notably, the sequence of onset times of these coding states follows the demands of the task in an orderly fashion. 

More examples can be added to the list above. Specific coding states in the visual area V4 of monkeys (called `ON' states) were found to coexist with improved selective attention \cite{Engel2016-bb}. In the orbitofrontal cortex of monkeys, metastable states were found coding for the reward value of competing options in a choice task \cite{Rich2016-iq}. Specifically, the option chosen by the monkeys was the one associated with the hidden state present for a larger portion of time (i.e., with the larger occupancy rate) during deliberation. Interestingly, slower decisions tended to occur when the occupancy rates of the states were similar, regardless of the actual difficulty of the decision (as measured by whether or not two options had similar reward value). This suggests a link between dynamic aspects of metastable activity and the substrate of internal deliberations, which we review further in Sec.~\ref{sec:modulations} below.

Relevance of the occupancy rates of metastable states has also been found in humans engaged in a working memory task \cite{Taghia2018-gp}. Measurements of BOLD signals with functional magnetic resonance imaging (fMRI) have long uncovered rich ongoing dynamics spanning the entire brain \cite{karahanoglu:2017}. The main goal of fMRI studies is often the establishment of the neural substrate of functional connectivity, the pattern of correlations of neural activity in anatomically separated brain regions \cite{Rogers_2007,Heuvel:2010}. In the \cite{Taghia2018-gp} study, after HMM analysis was performed on the BOLD signal of various brain areas, different hidden states were preferentially associated to different task conditions, with occupancy rate in each state predicting better performance in the corresponding task. Remarkably, changes in patterns of functional connectivity across brain areas co-occurred more reliably with state transitions than with external triggers.

By linking hidden states with patterns of functional connectivity dependent on the particular task being performed, the \cite{Taghia2018-gp} study supports the notion that neural circuits may rapidly adapt to better support current task demands, an influential idea in systems neuroscience known as `cognitive control' \cite{Miller_2001}. If metastable activity can unfold along different sequences depending on task demands, metastability may provide a means to switch among relevant dynamical patterns according to specific features of a task. A related example comes form the rat hippocampus, where neural representation of spatial maps have long been known to aid navigation \cite{Knierim:1995iq,Moser:2008uq,Hartley:2014bg}. Intriguingly, hidden metastable states in rat hippocampus have been found to represent the position in a linear track and in an open field during a navigation task \cite{Maboudi2018-ox}. Importantly, these metastable states were recorded while the animal was idling (rather than while navigating the track or field), and could be used to reconstruct a map of place fields evoked during locomotion.

Analogous results are being found also outside the mammalian brain. Hidden brain states related to locomotion and hunting were recently found in zebrafish \cite{Marques:2020qe}. Zebrafish spontaneously alternate between two internal states during foraging for live prey, a state of `exploration' (locomotion-promoting) and a state of `exploitation' (hunting-promoting). These states were found with an HMM analysis and had exponentially-distributed duration. Clusters of neurons, especially in the ventrolateral habenula and dorsal raphe nuclei,  seemed to activate at the state transition from exploration to exploitation. Hidden behavioral states corresponding to different decision strategies have also been found in mice engaged in decision tasks \cite{Ashwood:2021wc} and can be modulated by the motivational level of novelty seeking \cite{Ahmadlou:2021bf}.

Finally, we mention that hidden metastable states may be the substrate of multistable perception \cite{mrr07,Cao:2016bf} as well as motor planning and execution \cite{Kemere:2008uq,Mattia:2013uy}. A recent study has found that multistable perception requires a discretely stochastic format of perceptual representations, which in turn could be supported by the metastable activity of cortical networks \cite{Cao:2020pd}.

\subsection{Temporal modulation of metastable dynamics} \label{sec:modulations}

An important feature of hidden states is their ability to link behavior with neural activity on a trial by trial basis, as no average across trials (a practice common in neuroscience) is needed. Analysis such as that of Fig.~\ref{fig:benozzo} reveals that while states are indeed triggered by behavioral events, their onset and offset times are variable and not precisely pinned to external triggers. It is tempting to speculate that the occurrence of these states allows one to pinpoint the time at which a sensory perception or an internal decision is being made in each trial \cite{Bollimunta:2012rt,Ponce-Alvarez2012-zi,Sadacca2016-yo}. If this were the case, the timing of state transitions, and not just the nature of the coding states, may be related to aspects of the decision or the perceptual process. Recent studies suggest that this is the case. In Ref.~\cite{Ponce-Alvarez2012-zi}, the authors found a novel correlate of trial difficulty in monkeys performing a delayed vibrotactile discrimination task: when trials involved a more difficult discrimination, the transition to a new hidden state took longer, on average, than when discriminations was easier. This occurred after the onset of the second stimulus (when deliberation takes place), and was found in neural ensembles of the motor and premotor cortex but not in the somatosensory cortex, showing that this phenomenon is peculiar to neural circuits involved in execution and planning, rather than discrimination. A similar result was also obtained in the dorsal prefrontal cortex of monkeys performing the distance discrimination task of Ref.~\cite{Benozzo:2021en} described in the previous section (see Fig.~\ref{fig:benozzo}). In this task, the more the two stimuli were similarly distant from the central spot, the longer the mean transition time to the next state at the time of deliberation \cite{Benozzo:2021en}. Notably, no correlate of trial difficulty was evident in any single neuron's activity, pointing to the importance of the ensemble nature of hidden states. It is tempting to link these results to the finding that, in the rat GC, state transitions after a taste stimulus are affected by learning and extinction \cite{Moran_2014}. 

A link between trial difficulty and the onset of decision-coding states has also been reported during perceptual decisions \cite{Bollimunta:2012rt} and naturalistic consumption decisions \cite{Sadacca2016-yo}. The former study found that rapid switches in coding states prior to the behavioral decision were more frequent in trials with more difficult discriminations. In Ref.~\cite{Sadacca2016-yo}, the authors studied the decision of whether to eject or consume a tastant based on its palatability, and found a correlation between the timing of the decision and the fast onset of a `dominant' state presumably coding for that decision.

In addition to modulations in specific state transitions, more global modulations of the metastable dynamics have also been observed. These global modulations have been found to underly states of expectation \cite{Mazzucato:2019wi} and behavioral performance \cite{Benozzo:2021en}. Mazzucato {\em et al} \cite{Mazzucato:2019wi} analyzed multi single unit recordings in rat GC and found that that metastable sequences sped up in trials when the rats expected a stimulus to be delivered, as opposed to trials in which a stimulus was delivered at an unexpected time. In the prefrontal study mentioned above, Benozzo {\em et al} \cite{Benozzo:2021en} found that longer state durations between the onset of the second stimulus and the GO signal (when the decision is due) are observed during incorrect trials (in both studies, the number of different hidden states does not change between conditions, only their mean state duration does). Importantly, longer state durations could predict error trials regardless of their difficulty (as measured by the relative distance of the two stimuli with respect to the central spot), and therefore reflect the internal deliberation rather than difficulty of the task, in a manner similar to what was found by \cite{Rich2016-iq} in their choice task. 

In their expectation study, Mazzucato {\em et al}~\cite{Mazzucato:2019wi} have shown that the speed-up of metastable dynamics can be understood as the consequence of lowering energy barriers between the local minima of a landscape of configurations of a clustered network of spiking neurons (more on the landscape of a cortical network in Sec.~\ref{sec:fluxtheory}). In turn, this causes taste-coding states to occur early during the trial, presumably reflecting a state of expectation. This has provided a mechanistic model for the neural substrate of expectation, a crucial mental process whose quantitative explanation has always been elusive. 

\subsection{Metastable dynamics and ongoing activity} \label{sec:ongoing}

As reviewed in the previous sections, metastable sequences have been found in taste-evoked patterns of activity and related to the coding of specific taste stimuli in rat GC cortex. Metastable sequences, however, have also been found during long inter-trial intervals when the animal is not experiencing taste stimuli or engaged in any task \cite{Mazzucato:2015jk}. The rich, structured neural activity found in the absence of an overt external stimulation is known as `spontaneous' or `ongoing' activity \cite{asga96,tkga99,kbtga03,Raichle:2006dq}. Ongoing activity has long been suspected to have a role in memory consolidation and synaptic pruning, especially during sleep \cite{Huber:2004gw}. In rat auditory and somatosensory cortex, transient $50$-$100$ ms packets of spiking activity have been suggested to serve as a repertoire of available `symbols' with which to build the representation of sensory stimuli \cite{Luczak:2009wr}. These ongoing patterns of activity have also been interpreted as the sporadic opening of a `gate' allowing auditory cortex to broadcast a representation of external sounds to other brain regions \cite{Luczak:2013gg}. More generally, ongoing activity may contain an internal model of the environment and serve as `context' for interpreting incoming input and/or prepare forthcoming decisions \cite{kbtga03,arieli-ws04,Berkes:2011ul}. 

Since ongoing activity shares some common features and similar transient states with activity evoked by external inputs \cite{Abeles1995-zb,kbtga03,Berkes:2011ul}, studies attempting to quantify the subtle interaction between ongoing and evoked activity have emerged. The simultaneous hidden-state analysis of both ongoing and evoked activity could provide a quantitative account of this interaction. For example, \cite{Mazzucato2016-hl} found that the dimensionality of neural activity -- a measure of the number of independent degrees of freedom sufficient to characterize it  \cite{Rigotti:2013zp,Mazzucato2016-hl,Williamson:2016rc,Gao:2017} -- is larger during ongoing activity and it is quenched by the arrival of an external stimulus. This result has been captured by the same spiking network model put forward to explain expectation, suggesting that stimulus-driven reduction of dimensionality could be an inherent feature of metastable dynamics (but see \cite{Abbott2011-kh} for a model with continuous trajectories). The quantitative characterization of the interplay between ongoing and evoked activity has just begun and much more needs to be done, but by its own nature such interaction is likely to be rooted in the dynamics of latent brain states (whether continuous or discrete states).

\subsection{Other types of observed neural dynamics} \label{sec:other_observed}

In a series of studies, Laurent, Rabinovich, Abarbanel and colleagues have investigated the metastable nature of neural activity in the olfactory system of the locust \cite{Rabinovich:2001ex,Rabinovich2006-lw,Laurent:2001dn}. In the antennal lobe of these insects, spatiotemporal patterns of spike trains follow heteroclinic trajectories. The dimension of the space occupied by these trajectories is large enough to be able to separate the representations of different odors in separate trajectories. Although reminiscent of neural trajectories observed in the primary motor cortex of monkeys (see Sec.~\ref{sec:statmodels}), the heteroclinic trajectories of the antennal lobe have a metastable character that can be explained by a high dimensional Volterra-Lotka system.
This is a deterministic system wherein the trajectories proceed along saddle points rather than (locally) stable equilibria, transiently hovering around a saddle before moving towards the next along an unstable direction. Deciding whether neural data can be best described by this type of dynamical system, a system with continuous trajectories (Sec.~\ref{sec:trajectorymodeling}), or a system with metastable discrete states is a challenging and subtle task.

Continuous trajectories can be approximately described by discrete states and vice versa, therefore, the practical choice between discrete and continuous modeling depends on the spatiotemporal structure and signal-to-noise ratio of the neural data (Sec.~\ref{sec:statmodels}). 
Continuous trajectories have been effective descriptions in motor \cite{Churchland2012-ul}, cognitive \cite{Sohn2019-ht}, and sensory cortices \cite{Chowdhury2020-ud} -- leading to the concept of neural manifolds~\cite{Jazayeri2017-vm}.
However, continuous trajectories and dynamics do not preclude metastability, since continuous dynamical system features such as hyperbolic fixed points and multistable limit cycles can exist ~\cite{Zhao2016d,Jordan2019a}.
We further discuss methods that can analyze continuous trajectories in Section~\ref{sec:statmodels}.

Other types of observed neural dynamics include: `avalanches' of neural activity \cite{Beggs:2003zc,Ponce-Alvarez:2018dn}; states of slow oscillations \cite{Sanchez-Vives:2017ng}; UP and DOWN states \cite{sm00,Jercog:2017lq,Setareh:2017km}; traveling waves \cite{Muller:2018pe} and various combinations of these phenomena \cite{Luczak:2007wp}. These dynamics have been observed in experiments in behaving humans and animals probed with different methods spanning multi-single units recordings, multi-electrode arrays, local field potentials, voltage sensitive dies, calcium imaging, electroencephalography, electrocorticography, and fMRI. They have all given impetus to shifting the focus of research from stationary to dynamic patterns of neural activity and can all manifest metastable dynamics. In this review we focus mainly on metastable dynamics uncovered by electrophysiological recordings of multiple single units (for a broader view, see e.g. Ref.~\cite{metastable_frontiers_book:2018}). This technique can resolve neural spiking activity at sub-millisecond precision, allowing the determination of fast transitions among discrete metastable states. While the advent of neuropixel technology \cite{Jun:2017zd} promises the possibility to record from hundreds or thousands of neurons simultaneously, most of the results reviewed in this paper come from recordings of the range of ten to a few dozen neurons.

\section{Modeling metastable dynamics in neural circuits and networks}
\label{sec:models}
In this section we provide a survey of recent statistical and dynamical systems models used infer and replicate metastable dynamics observed in cortical data. The statistical models reviewed here are applicable to many kinds of data; we focus on models which infer low dimensional representations of high dimensional data like simultaneously recorded neural spike trains. We then review models built in the tradition of computational neuroscience and based on populations of spiking neurons. These models can reproduce many features of the metastable dynamics observed in electrophysiological recordings and allow for predictions that are most closely testable in experiment, given the biological detail of these models. These spiking models are complex and difficult to analyze, but mean field techniques and the path-integral-based landscape flux theory, which are also reviewed here, allow for tractable progress in understanding what properties of cortical networks may be necessary or sufficient to generate metastable dynamics. 

\subsection{Statistical inference and state space analysis of neural data}
\label{sec:statmodels}
If the underlying population dynamics are metastable, how can we detect evidence for this metastability in neural recordings?
Furthermore, can we infer the unobserved---or `latent'---dynamical systems underlying these recordings?
In this section, we describe data-driven approaches that use a state space formulation to describe the time evolution of the \emph{collective} neural population state.
The idea of `state space' in neuroscience is borrowed from the signal processing literature, and is related to the notion of phase space in physics.
The central assumption of these methods is the existence of a concise Markovian description, typically in the form of a low-dimensional continuous or discrete state space with a small number of states. In other words, the neural state description $\bm{x}_t$ at time $t$ is sufficient to describe the future neural population activity. In probabilistic form, we can express the Markovian assumption as $P(\bm{y}^\ast_{> t} |\bm{x}_t,\bm{x}_{< t}) = P(\bm{y}^\ast_{> t} |\bm{x}_t)$, where $\bm{y}^\ast_{> t}$ denotes all future neural activity of interest and $\bm{x}_{< t}$ is the past history of the process $\bm{x}_t$. Thus, the goal of state space analysis approaches is to infer the time evolution of the neural state $\bm{x}_t$  corresponding to the duration of the neural recording (trajectory modeling) or to infer the dynamical structure of the state space in the form of time evolution operator $P(\bm{x}_{t+1}|\bm{x}_t)$ (dynamical systems modeling).

\subsubsection{Trajectory modeling}
\label{sec:trajectorymodeling}
The trajectory of the neural state evolving over time will linger for extended periods before escaping from a metastable state. Therefore, extracting the neural trajectory from recordings can provide evidence for metastable dynamics. There are two main challenges of statistical nature in performing such inference. First, with current technologies, only partial neural observations are possible, meaning that only a small number of neurons or neural signals can be measured relative to the full population. It may thus not be possible to fully reconstruct the state space, and it is beneficial to have as many simultaneous recording dimensions as possible. Second, neural recordings are noisy reflections of the neural population states, such that two measured neural recordings corresponding to identical underlying neural trajectories are not identical. Sources of variability in neural activity include spiking noise, irrelevant neural activity that is not of interest, as well as measurement noise. Traditional Takens' style delay-embedding methods for recovering (chaotic) attractors popular in dynamical systems analysis~\cite{Broomhead1986-oe} can be difficult to apply in the presence of noise and metastable states. Therefore, additional assumptions must be made to reduce the noise to recover a concise, denoised trajectory. We discuss popular approaches to this statistical inference problem.

\subsubsection{Virtual ensemble}
One common approach to deal with both statistical issues is to average over repeated trials. With a strong assumption that neural trajectories are repeated indistinguishably through controlled experimental manipulations, the average neural response will have reduced noise. Furthermore, one can combine trial-averaged neural recordings that are not simultaneously recorded together to form a virtual ensemble. This approach is widely used, for example, in olfaction~\cite{Mazor2005-pg,Bathellier2008-er}, motor~\cite{Shenoy2011-dx}, contextual decision-making~\cite{Mante2013-em,Akam2021-ms}, and timing~\cite{Wang2018-wm}.

Even for heterogeneous trials, there are regression models that allow the extraction of low-dimensional components (see below)~\cite{Aoi2018-bo}, although it is unclear how to interpret the resulting family of deterministic (average) trajectories. It is important to note that the variability in each channel of neural signal is treated as independent (no covariability), and the trial-to-trial deviations from the average trajectory are ignored in this analysis. The metastable activity seen in the neural trajectory may still reflect the stereotypical nonlinear dynamical features as long as the assumptions hold. This strategy cannot be used for spontaneous activity because in the absence of trial structure there is no meaningful way to align the data.

\subsubsection{Principal Component Analysis and related dimensionality reduction methods} \label{sec:PCA}
Currently, the most popular method for continuous trajectory modeling is principal component analysis (PCA). PCA is used as a dimensionality reduction and denoising tool where a high-dimensional time series of neural recordings is explained as a linear combination of a much smaller number of latent processes. The principal components (PCs) that only contribute to a small amount of the total variance are dropped, resulting in a lower dimensional neural trajectory spanned by the PCs. This process requires that the number of observed neural recordings be sufficiently larger than the state space. PCA implicitly assumes the observations have independent additive noise with same variance across channels. Therefore, when applied to spike trains, it is typical to use large time bins and averages across trials when possible, using the virtual ensemble method.

Gaussian Process Factor analysis (GPFA) is a related method which weakens the equal noise variance assumption and assumes continuous changes in time~\cite{Yu2009-qp}. The Gaussian process prior $P(\{\bm{x}_t\}_{t\in T})$ explicitly puts higher probability on latent trajectories $\bm{x}_t$ in the time window $t \in T$ that have specific temporal smoothness. The temporal smoothness (hyper-)parameters are inferred from the data. This provides GPFA the power to prioritize inferring slowly changing factors automatically such that it smoothly interpolates the time series. When the system is in a metastable or stable state, the neural trajectory evolves slowly, consistent with the prior assumption of GPFA. However, when the metastable states are short-lived and transitions are fast, GPFA may not provide additional benefits or even be counterproductive. Moreover, GPFA, like PCA, assumes additive Gaussian observation noise, which is not suitable for spike train analysis. Extensions of GPFA to Poisson observations~\cite{Nam2015-vn,Zhao2016a} and more general counting distributions~\cite{Keeley2020-vn} have been developed for these cases.

The aforementioned PCA and related methods look for a linear subspace in the population neural activity. However, the relation between the state space and the observations may be highly nonlinear, rendering linear methods less useful for identifying metastable states. Nonlinear dimensionality methods such as MDS, t-SNE, UMAP~\cite{Becht2018-rr}, manifold learning tools such as Isomap, LLE (e.g., see~\cite{Chaudhuri2019-kb}), and probabilistic modeling tools (GPLVM~\cite{Wu2017-az}) are used to recover neural trajectories in these cases.

\subsubsection{Kalman filtering and smoothing}
The temporal smoothness in continuous trajectory inference can be achieved with structured smoothing methods. The state space model due to Kalman is a linear dynamical system with additive white gaussian noise and a linear observation model:
\begin{align}\label{eq:lds}
   \bm{x}_{t+1} &= 
    	A\bm{x}_t + \bm{\eta}_t, \quad &\text{(linear dynamics)}\\
   \bm{y}_t &= 
    	C\bm{x}_t + \bm{\nu}_t, \quad &\text{(linear observation)}
\end{align}
again with Gaussian noise~\cite{Wiener1965-xv,Kalman1966-xk,Haykin2002-qv}.
Typically the linear dynamics matrix $A$ is a scaled identity matrix such that the trajectory $\bm{x}_t$ retains temporal smoothness.
The optimal inference algorithm for causal inference (given data up to current time) is the celebrated Kalman filtering algorithm, and the inference given the entire time series is the Kalman smoothing algorithm.
These methods provide fast estimates of the smooth neural trajectory and are widely used in neuroscience~\cite{Yang2021-fj}.
To obtain the best parameters of the linear state space model~\eqref{eq:lds}, expectation-maximization~\cite{demp77} or spectral subspace identification methods can be used~\cite{Katayama2005-jt}.
PCA and FA can be written as special cases of inference within this linear dynamical system framework.

\subsubsection{Clustering based approaches}
When the transitions between the metastable states are short, the neural trajectory may spend most of the time at metastable states.
In this regime where the metastable states dominate the dynamics, it is beneficial to directly model the metastable states as discrete entities rather than continuous trajectories.
Clustering algorithms such as $k$-means~\cite{Duda2006-wd} can be used to detect the metastable states. In Ref.~\cite{Hudson2014-na}, the authors used PCA combined with k-means on spectral feature vectors, and found metastable states (with dwell times in the order of minutes) corresponding to the stochastic transition from anesthesia to wakefulness.
Further state velocity analysis in the PCA space supported that highly occupied clusters (states) were stable~\cite{Hudson2014-na}.
In clustering approaches, once feature vectors are formed, their temporal order is ignored.
As we discuss in Secs.~\ref{sec:dyn-syst-model}--\ref{sec:HMM}, the Hidden Markov model (HMM) extends simple clustering with state dependent probabilistic transitions.

\subsubsection{Dynamical system modeling} \label{sec:dyn-syst-model}
The methods and models assumed in the previous sections ignore the time evolution of metastable states.
This is evident from the fact that even the models that can generate data would not generate anything resembling metastable dynamics because they lack non-trivial structure in $P(\bm{x}_{t+1} \given \bm{x}_t)$.
This dynamical law is assumed to be consistently applied to the neural state for all time, forming the basis of higher frequency of repeated spatiotemporal patterns. The linear dynamical system assumed in the Kalman filter and variants can only have 1 isolated fixed point, hence metastability cannot be expressed.
This does not mean they are not useful tools to analyze neural trajectories, but it means that they are not appropriate tools for modeling the metastable dynamics as a dynamical system. Statistically inferring the nonlinear probabilistic state transition $P(\bm{x}_{t+1} \given\bm{x}_t)$ or implicitly assuming its existence is at the core of dynamical system based modeling.
In the following subsections, we discuss continuous and discrete forms of state representations.

\begin{figure*}
\includegraphics[width=\textwidth]{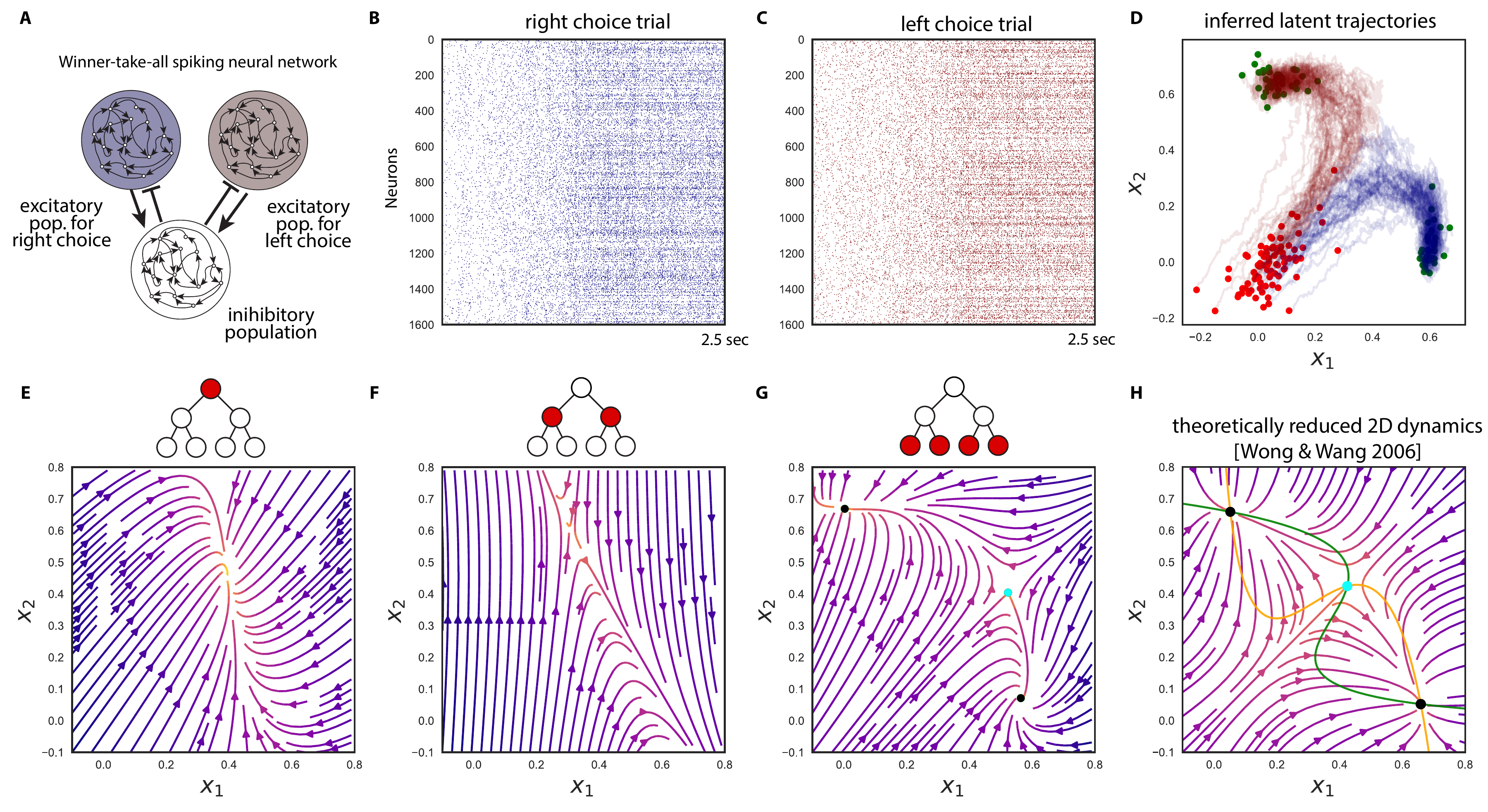}
\caption[]{
    Inferring metastable dynamics from spike train observations only can recover the theoretical phase portrait.
    The spike trains were generated from a winner-take-all decision making model implemented with spiking neural network~\cite{Wang2002-he}.
    The tree-structured recurrent switching linear dynamical system (TrSLDS) model is fit to subsampled spike trains~\cite{Nassar2018a,Nassar2018b}.
    Inference was performed using augmented Gibbs-sampling.
    (\textbf{A}) Overview of the connectivity structure of the spiking neural network.
    (\textbf{B,C}) raster plots of excitatory neurons for 2 random trials.
    (\textbf{D}) the latent trajectories converge to either one of the two of sinks at the end of trial (green).
     Each trajectory is colored by their final choice.
    (\textbf{E-G}) Dynamics inferred by each level of the tree structure provide a multi-scale view. The most detailed view in (G) exhibits one saddle (cyan) and two stable fixed points (black).
    \textbf{(H)} Theoretically reduced 2-dimensional phase portrait of the spiking neural network dynamics given the full specification and no data~\cite{Wong2006-ey}.
    The green and yellow curves are nullclines.
    Note the similarity between (G) and (H). Reproduced with permission from J. Nassar, S. Linderman, Y. Zhao, M. Bugallo, and I. M. Park, 52nd Asilomar Conference on Signals, Systems and Computers (2018). Copyright 2018 Institute of Electrical and Electronics Engineers. \cite{Nassar2018a}
}
\label{fig:TRSLDS}
\end{figure*}

\subsubsection{Latent nonlinear continuous dynamical systems modeling} \label{sec:latent-cont}
If the trajectories are modeled as continuous, the corresponding model for dynamics $P(\bm{x}_{t+1} \given\bm{x}_t)$ is assumed to originate from an autonomous ordinary differential equation (ODE) of the form $\dot{\bm{x}} =\bm{f}(\bm{x})$, or a stochastic differential equation (SDE) with the presence of state noise.
As discussed previously, metastability can originate in various ways including from multiple isolated saddle points, stable fixed points, slow regions, and continuous attractors.
The function $\bm{f}(\bm{x})$, also referred to as `flow field', captures the velocity of the neural state $\bm{x}_t$'s time evolution governed by the dynamical system for continuous time $t$.
Therefore, recovering $\bm{f}(\bm{x})$ is key to understanding the nature of metastability and the topological relation between metastable states.
An arbitrary form of $\bm{f}$ may seem theoretically attractive, however, in practice allowing infinite flexibility is an ill-posed problem, not to mention doomed to overfit the data.
Therefore, various methods have been proposed that assume an \textit{a priori} structure for $\bm{f}$.
All practical methods in this class assume Lipschitz continuity in $\bm{f}(\bm{x})$, as this guarantees that the neural trajectories which are solutions to the ODE do not cross themselves in finite time and are uniquely specified by any neural state $\bm{x}(t)$. One can evaluate the degree to which an inferred neural trajectory is tangled with itself to support the dynamical systems view of neural signals \cite{Russo:2018ni}. When it comes to parameterizing $\bm{f}(\bm{x})$, there are two camps, the low-dimensional camp where the complexity of $\bm{f}$ is high but the dimensionality of $\bm{x}$ is small, and the high-dimensional camp where $\bm{f}$ is only weakly nonlinear but the latent state $\bm{x}$ is of high-dimension. The former approach focuses on interpretability of the state space, while the latter pivots on the success of recurrent neural networks as a black-box predictor in machine learning. We will discuss both approaches here.

Latent nonlinear continuous dynamical systems methods fall in the general Bayesian state space modeling framework where the generative model is given by a dynamics model (written in discrete time for convenience),
$\bm{x}_{t+1} \sim P(\bm{f}(\bm{x}_t), \bm{\theta})$,
and an observation model,
$\bm{y}_t \sim P(\bm{g}(\bm{x}_t), \bm{\phi})$,
where $\bm{\theta}$ and $\bm{\phi}$ parametrize their corresponding distributions~\cite{Haykin1998-up}.
When one is interested in causal information, i.e. inference only using the data from the past to the current time point, this inference is referred to as (Bayesian) filtering, which can be implemented by a recursive update of the posterior over $\bm{x}$ and other parameters one step at a time:
\begin{subequations}\label{eq:latent:BayesianUpdate:filtering}
        \begin{align}
                P( \bm{x}_t, \bm{\Theta} \mid \bm{y}_{\le t} ) &=
                \underbrace{P( \bm{y}_t \mid \bm{x}_t, \bm{\Theta} )}_{\substack{\text{likelihood}}} \,
                \underbrace{P( \bm{x}_t, \bm{\Theta} \mid \bm{y}_{<t} )}_{\substack{\text{prior at time $t$}}}
		\,
                /
                \underbrace{P(\bm{y}_t \mid \bm{y}_{<t})}_{\substack{\text{marginal likelihood}}}
                \label{eq:latent:BayesianUpdate:bayes}
                \\
                P( \bm{x}_t, \bm{\Theta} \mid \bm{y}_{<t} )    &=
                \int
                \underbrace{P( \bm{x}_t \mid \bm{x}_{t-1}, \bm{\Theta})}_{\substack{\text{state dynamics}}} \,
                \underbrace{P( \bm{x}_{t-1}, \bm{\Theta} \mid \bm{y}_{<t} )}_{\substack{\text{previous posterior}}}\,\mathrm{d}\bm{x}_{t-1}
                \label{eq:latent:BayesianUpdate:marginal}
        \end{align}
\end{subequations}
where $\bm{\Theta} = \{ \bm{\theta}, \bm{\phi}, \bm{f}, \bm{g}\}$ is a collection of parameters. If one is interested in inferring based on all recorded neural data, it is referred to as (Bayesian) smoothing, and the corresponding forms are
\begin{widetext}
\begin{equation}\label{eq:latent:BayesianUpdate:smoothing}
P( \{\bm{x}_t\}_t, \bm{\Theta} \mid \{\bm{y}_t\}_t ) =
    \underbrace{P( \{\bm{y}_t\}_t \mid \{\bm{x}_t\}_t, \bm{\Theta} )}_{\substack{\text{likelihood}}} \,
    \underbrace{P( \{\bm{x}_t\}_t, \bm{\Theta})}_{\substack{\text{prior}}}
    \,
     /
    \underbrace{P( \{\bm{y}_t\}_t )}_{\substack{\text{marginal likelihood}}},
\end{equation}
\end{widetext}
where $\{\bm{x}_t\}_t$ is a shorthand for $\{\bm{x}_t\}_{t\in T}$.

Unfortunately, the analytical form of \eqref{eq:latent:BayesianUpdate:filtering} or \eqref{eq:latent:BayesianUpdate:smoothing} is typically not tractable, especially for flexible nonlinear dynamics models.
Therefore, algorithms either opt for Monte Carlo sampling~\cite{Nassar2018b}, variational inference~\cite{Archer2015a,Pandarinath2018-hg,Duncker2019-jq,Zhao2017c}, or hybrid~\cite{Zhao2019a} approaches.

In the low-dimensional models, the expressive power of the specific parameterization of $\bm{f}$ must be high enough to capture metastable dynamics.
Radial basis function networks, Gaussian processes with square-exponential kernels, linear-nonlinear forms with hyperbolic tangent function, switching linear dynamical systems, and gated recurrent units were investigated as flexible methods of parameterizing $\bm{f}$ and shown to have sufficient expressive power in the low-dimensional regime~\cite{Zhao2016d,Duncker2019-jq,Jordan2019a,Nassar2018b,Zhao2019a}.
Due to the high flexibility of the functional form, it is important to put sufficient emphasis on simpler, more robustly generalizing functions.
To control for the complexity of the function, various regularization methods such as penalty, simple initialization combined with early stopping strategy~\cite{Genkin2019-ff}, restricting the effective number of parameters, or simply imposing a prior distribution over functions are commonly used.
In~\cite{Zhao2019a}, the authors used a sparse Gaussian processes framework to represent (a belief distribution over) $\bm{f}$.
Given a partial observation from a simulated spiking neural network with input-dependent saddle and stable fixed points, this method was able to recover the general 2D phase space through approximate Bayesian filtering.
The spiking neural network~\cite{Wong2006-ey} implemented an integrator and decision-making process, and the metastability at the onset of each trial as well as the metastable (saddle) point that defined the mid-point between the two choices represented by two stable states were recovered.
Another approach to modeling is to softly divide local regions of the state space and endow them with a linear dynamical system~\cite{Linderman2017-tb,Taghia2018-gp,Duncker2019-jq}.
In~\cite{Nassar2018b}, the authors showed that by imposing a hierarchical division of the state space the method can represent the dynamics at multiple spatial scales, which provides further interpretability of the dynamics (Fig.~\ref{fig:TRSLDS}).
Linear dynamics around a fixed point can be easily understood as metastable states, however, due to non-trivial emergence of fixed points at the boundary between regions, this approach requires further research for analyzing metastable dynamics.

In the high-dimensional models, dynamical modeling is achieved by recurrent neural networks where the expressive power/flexibility is adjusted by the dimensionality of the hidden states.
In~\cite{Pandarinath2018-hg}, a gated-recurrent unit RNN was used together with a variational inference scheme at the core for the `latent factor analysis via dynamical systems' (LFADS) method.
They showed that LFADS was able to outperform the PCA-like smoothing methods for inferring continuous neural trajectories and for predicting held-out neurons' single trial spike trains~\cite{Pandarinath2018-hg}.
LFADS aims for better smoothing of trajectories, and the nonlinear dynamics captured by the RNN is not analyzed. Further analysis of the recovered dynamics to extract fixed points of trained RNNs was explored in~\cite{Sussillo2012-qb}.

\subsubsection{Hidden Markov modeling} \label{sec:HMM}

The Hidden Markov model (HMM) \cite{rab89,ZucchiniBook09} is an unsupervised method for segmenting a time series into intervals corresponding to distinct, discrete states. It has been widely used for phoneme segmentation in speech recognition, DNA sequence analysis, behavioral analysis, and many other applications \cite{Durbin1998-qk,Dymarski2011-ej}. After early work in the nineties \cite{Gat1993-hs,Radons_1994,Abeles1995-zb}, HMM is now probably the most widely used data analysis tool to uncover sequences of metastable states in ensembles of spikes trains from simultaneously recorded neurons. 

An HMM is characterized by $M$ (hidden) states and a matrix $\Gamma_{ij}$ of transition probabilities from state $i$ to state $j$ -- a Markov chain. This means that the next state depends only on the current state (the Markov assumption). Once in state $i$, the system emits an observation $O_i$ according to some probability distribution $\rho_i(O_i)$ that depends only on the current state $i$. The observation is therefore a noisy manifestation of the hidden state. For the application to neural spiking data discussed here, each state is a vector of `true' firing rates across neurons, $\bm{\lambda^i}=(\lambda_1^i, ..., \lambda_N^i)^T$, where $\lambda_n^i$ is the firing rate of neuron $n$ in state $i$, $N$ is the total number of neurons, and $T$ denotes matrix transposition. While in state $i$, neuron $n$ emits a spike train according to a Poisson process with firing rate $\lambda_n^i$, though the model can be extended to include refractory periods and other history-dependent factors \cite{Escola2011-fi}. 

The model can be defined in discrete or continuous time and here we assume the latter, but in practice, time is discretized in bins to be able to fit the model to the data. In continuous time, $\Gamma$ is a matrix of transition rates rather than probabilities; the probability of making a transition from $i$ to $j$ in a small interval $dt$ is $\Gamma_{ij} dt$, and state durations in between transitions are exponentially distributed. The model is completely specified by its $M$ states through $[\bm{\lambda^1}, ..., \bm{\lambda^M}]$ (an $N \times M$ matrix) and by the transition matrix $\Gamma$ (an $M \times M$ matrix).\footnote{The probability over the states at time zero would also be required to characterize the model; here we assume that the chain is in a predefined state (e.g., state 1) at a time prior to the beginning of the period of interest. This reflects the assumption that the initial state occurs a long time in the past and it effectively allows to do away with specifying its distribution. See e.g. \cite{Mazzucato:2015jk}.}

The states of the underlying Markov chain are `hidden' in the sense that only noisy observations of the states are available from experiments. For example, under the Poisson firing hypothesis, a neuron in a given state with firing rate $\lambda$ is expected to emit a spike train with about $\lambda \Delta t \pm \sqrt{\lambda \Delta t}$ spikes in a time window of length $\Delta t$. Every time the activity returns to the same state, the experimenter will measure different spike counts compatible with the true (hidden) firing rates. The challenge is to infer the true firing rates and the state transition rates from fitting the model to the data. This is usually accomplished via maximum likelihood. Although direct (numerical) maximization of the likelihood is possible and sometimes recommended (see e.g. \cite{ZucchiniBook09}), an expectation-maximization algorithm is typically used (called the Baum-Welch algorithm in this context; see e.g. \cite{rab89} for a clear description of the algorithm). 

The fitting procedure is repeated for different numbers of states $M$, and the optimal number of states $M^\ast$ is selected via cross-validation, i.e., by testing the model on a test set not used for fitting \cite{Maboudi2018-ox,Recanatesi:2020xw}. The purpose of cross-validation is to minimize the generalization error, however it is often the case, when fitting HMM to spike data, that the likelihood on the test set keeps increasing with the number of states. This fact would lead to models with a large number of states that overfit the data. A number of alternative strategies have therefore been adopted to avoid overfitting. In some studies, the number of states was fixed to a predefined value based on prior knowledge on stimuli or conditions in a task \cite{Abeles1995-zb,Seidemann1996-bf,Jones:2007ff,Engel2016-bb}. Other authors have chosen the value of $M$ that minimizes the Bayesian Information Criterion (BIC), but have also set an upper limit for $M$ \cite{Ponce-Alvarez2012-zi}. BIC penalizes the log-likelihood (LL) by a measure of the number of parameters to be estimated relative to the available data: $BIC = - 2LL + [M(M-1)+MN]\ln T$, where $T$ is the number of observations (which equals the number of trials times the number of bins in each trial). More recently, BIC has been combined with a procedure to remove states during {\em decoding}. Decoding is the process of assigning one of the HMM states to each data bin (see Figures~\ref{fig:hmm1} and \ref{fig:benozzo} for examples). The most basic form of decoding assigns a bin of data $x$ to state $i$ if the posterior probability of $i$ given $x$, $P(i|x)$, is maximal among all the posteriors. Typically, however, a more restrictive condition is used, one that requires $P(i|x)>0.8$ or even larger \cite{Jones:2007ff,Ponce-Alvarez2012-zi,Mazzucato:2015jk,Mazzucato2016-hl,Mazzucato:2019wi,Benozzo:2021en}. When none of the posteriors reaches this criterion, the state is not assigned (white spaces in Figures~\ref{fig:hmm1} and \ref{fig:benozzo}). More recently, authors have further required that, during decoding, only those states with probability exceeding 80\% in at least 50 consecutive ms are retained for further inference. This procedure eliminates states that appear only very transiently and with low probability, and it reduces further the chance of overfitting \cite{Mazzucato:2015jk,Mazzucato2016-hl,Mazzucato:2019wi,Benozzo:2021en}.

By definition, a good HMM model should result in fast transitions among the decoded states, as this is in keeping with the assumption that the neural activity remains in a state for some time, before quickly transitioning to another state. In several works \cite{Abeles1995-zb,Jones:2007ff,Ponce-Alvarez2012-zi,Sadacca2016-yo} it has been found that the transitions are one order of magnitude faster than the state durations, and are as fast as can be expected if the neural data with the same characteristics (e.g., the same firing rates) were transitioning instantaneously from one state to the next. State transitions were also significantly faster than in randomly shuffled datasets or in surrogate datasets with gradual state transitions -- in fact, the inferred transition times are close to their theoretically observable lower bound.

Inference based on the identity of the hidden states, as well as the temporal modulation of their sequences, has uncovered a significant number of results which we have reviewed in Sec.~\ref{sec:evidence}. To ensure that these results are not a side-effect of the fitting algorithm, i.e., that the states and their properties are true properties of the data, a commonly used control procedure is to compare the results of the same HMM analysis on the original and shuffled datasets, and show that the results obtained on the original data are lost when the data are randomized \cite{Seidemann1996-bf,Jones:2007ff,Ponce-Alvarez2012-zi,Sadacca2016-yo,Maboudi2018-ox,Recanatesi:2020xw,Benozzo:2021en}.

The strength of HMM analysis is that it is a principled, unsupervised method for segmenting neural activity into a sequence of discrete metastable states. The model can uncover transitions in neural activity that are not just triggered by external events, such as a stimulus or a reward, but are instead spontaneously generated and may occur anytime, including when the subject is idling and not engaged in a task \cite{Maboudi2018-ox}. Generalizations of the basic HMM reviewed here are possible in several directions, and include combinations with generalized linear models to account for non-stationarity \cite{Escola2011-fi,Ashwood:2021wc}, hidden semi-Markov models to account for non-exponential distributions of state durations \cite{yu_semiMarkov2010}, and Bayesian non-parametric HMMs which do not require separate model selection \cite{Chen:2013lb,Linderman2016-ab,Taghia2018-gp}.

\subsection{Theoretical models of cortical networks} \label{sec:theorymodels}

In this section we review a few prominent examples of neural network models that are relevant to the study of metastability in cortical circuits. These models differ from the models reviewed in Sec.~\ref{sec:statmodels} in that the latter are rooted in statistical descriptions (often in conjunction with dynamical systems theory), whereas in this section we consider network models that are closer to the biology and attempt a more mechanistic description of cortical circuits. The description of these models varies according to style and scope, so that `theoretical models' of brain function range from biologically detailed models of neural activity to more abstract or formal models where some level of biological detail is sacrificed for better analytical tractability. The more formal models are sometimes constructed to achieve a specific goal (such as phenomenologically reproducing the animal behavior observed in certain tasks) and, often invoking some first principles, attempt to derive constraints on neural circuitry and/or algorithms for achieving the desired result. This is e.g. the case of the Amari-Hopfield model \cite{amari72a,h82}, where the goal of modeling memories as stable attractors of the neural dynamics leads to assuming symmetric synaptic weights (more on this later), while the goal of embedding specific desired patterns as memories dictates the specific analytical form of the synaptic weights (\cite{h82}; see e.g. \cite{Hertz:1991lr} for a comprehensive treatment). Other examples include `normative' models, i.e., models derived from the minimization of a cost function (such as metabolic cost, information loss, or punishment). At the other end of the spectrum, models are based on the detailed description of individual neurons and their synaptic interconnections and tend to incorporate knowledge from anatomical and physiological data. Even in this case, biological detail is to some extent sacrificed in exchange for theoretical tractability, as is the case for networks of integrate-and-fire neurons discussed in Sec.~\ref{sec:spiking_net}. This modeling approach has provided us with concrete examples of the diversity of dynamics in single neurons and small groups of neurons with different types of connections, as well as on the emergence of various degrees of coordinated activity in large neural networks \cite{Vogels:2005sd,Rabinovich2006-lw,YH11,YH14,gerstner2book2014,Sanchez-Vives:2017ng}. Biologically detailed models, however, are computationally expensive to simulate and difficult to analyze, and a mean field theory of these models, when attainable, is often used. This effectively amounts to reducing the system to a set of coupled relevant parameters, such as the firing rates of subpopulations of neurons, and it exemplifies the fact that one may start with a detailed model which is then reduced to a more formal one. More abstract models share a similar coarse-grained description as these reduced models, but without being derived from a specific microscopic model. One advantage of more abstract models is a more immediate and transparent way to introduce the phase portrait and to analyze it in search for local and global changes of the dynamics brought about by varying control parameters \cite{amitBook89,YH18,YH19}.

As we have reviewed in Sec.~\ref{sec:evidence}, the activity of cortical networks often unfolds as a sequence of metastable states. These metastable states are linked to the existence of configurations that may attract or repel the dynamics along different directions. When dynamics is highly dissipative, it typically converges to attractor states. These attractors are generally modeled as fixed points of an effective dynamical system, and may lend themselves to an interpretation as an energy landscape, as in the Ising model. As in the example of the finite size Ising model outlined in Sec.~\ref{sec:spinmodels}, metastable transitions emerge due to intrinsic or external noise perturbing the dynamical system enough that it escapes the basin of attraction of one fixed point and is attracted towards another \cite{Braun:2010uq,Miller2016-kx}. We review these phenomena in three examples of neural population models, in order of increasing abstraction: a spiking network model in which the elementary units are neurons coupled by pairwise synaptic connections (Sec.~\ref{sec:spiking_net}), a population activity model in which the elementary units may be interpreted as small clusters of neurons (Sec.~\ref{sec:WCmodel}), and an energy-landscape model in which the elementary units can be interpreted as continuous coarse-grained neural activity states (Sec.~\ref{sec:fluxtheory}). These different approaches can also be combined, as illustrated in Sec.~\ref{sec:combined}. We finally summarize a very general framework for the non-equilibrium thermodynamics of general neural networks in Sec.~\ref{sec:noneq-thermo}.

\subsubsection{Spiking network models} \label{sec:spiking_net}

The origin of metastable activity has been investigated with some success in networks of simplified spiking neurons known as `integrate-and-fire' neurons \cite{Miller2010-xy,Deco2012-wg,Litwin-Kumar:2012ty,Mazzucato:2015jk,Mazzucato2016-hl,Cao:2016bf,Setareh:2017km,Mazzucato:2019wi}. Integrate-and-fire (IF) models are simplified descriptions of neural activity that are significantly easier to simulate and analyze mathematically than more biophysically detailed models such as the Hodgkin-Huxley (HH) model of action potential propagation \cite{HODGKIN:1952pb}. Yet, IF models retain essential features of real neurons such as a continuous-time membrane potential and the ability to mimic the emission of an action potential, more commonly called `spike' in this context, upon suitable perturbation. Therefore, networks of IF neurons present an excellent trade off between biological plausibility and amenability to theoretical analysis \cite{cmla04,ab97,at91b,t93,av93,fm99,b00,bh99}.  

\begin{figure*}
\begin{center}
\includegraphics[width=1\textwidth,angle=0]{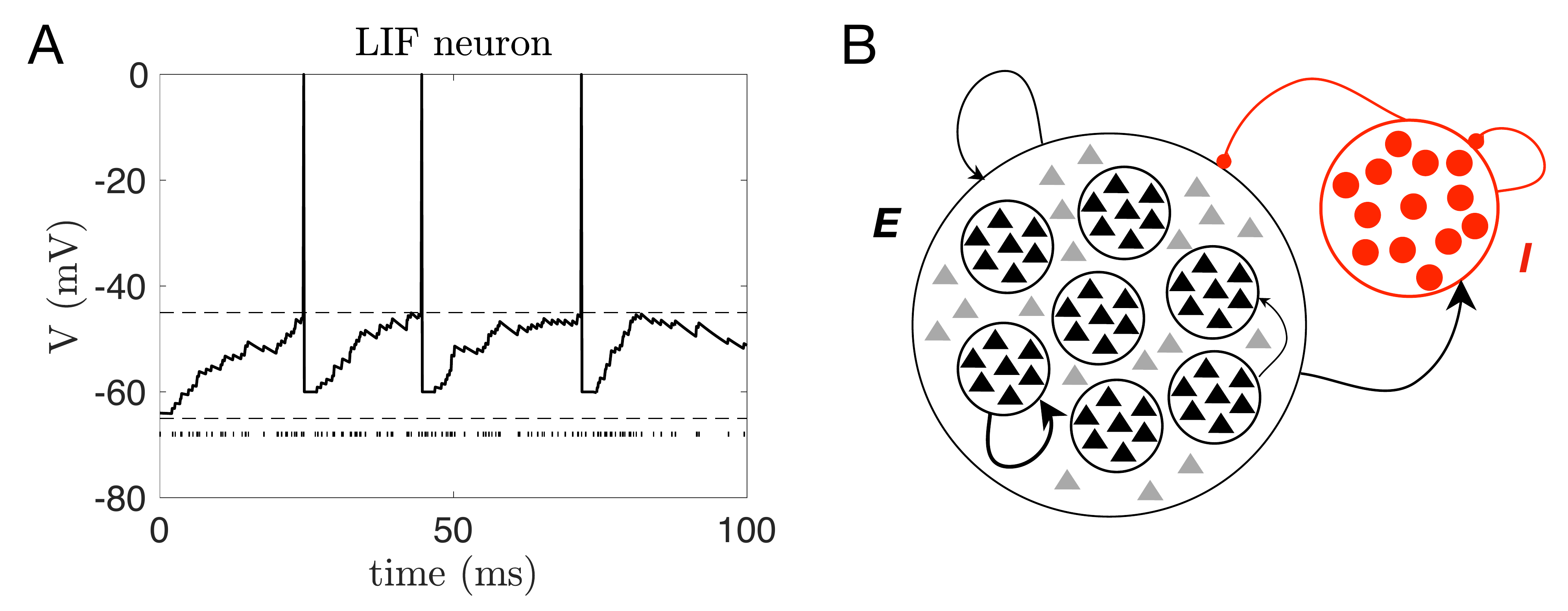}
\end{center}
\caption[]{\small A: simulation of a leaky integrate-and-fire (LIF) neuron in response to an excitatory Poisson spike train (sequence of tickmarks at the bottom). B: schematic diagram of a clustered spiking network. $E$=excitatory neurons, $I$=inhibitory neurons. See the text for details.}
\label{fig:spikingnet}
\end{figure*}

One of the simplest and most widely used IF models is the so-called leaky IF neuron (LIF), in which the membrane potential $V_i$ of each neuron $i \in \{1,\dots, N\}$ obeys a linear ordinary differential equation (ODE): 
\be \label{eq:LIF}
\tau \dot V_i = -(V_i-V_L) + I_{i,{\rm syn}} + I_{i,{\rm ext}},
\ee
where $\tau$ is the membrane time constant, $V_L$ is the resting potential, $I_{i,{\rm syn}}$ is the synaptic input current to neuron $i$, and $I_{i,{\rm ext}}$ is an external current which we take to be constant (both input currents are given here in units of voltage). This ODE is linear in the voltage and cannot generate an action potential, unlike `conductance-based' models such as HH. For this reason, spike emission is mimicked by appropriate boundary conditions: when $V_i$ reaches a threshold $V_{\rm spk}$ (from below), a spike is said to be emitted and the membrane potential is reset to a value $V_{\rm r} \sim V_{L}$ for a short interval $\tau_{\rm arp}\sim 2$-$5$~ms, after which the dynamics resumes according to Eq.~\ref{eq:LIF}. A simulation of this model neuron is shown in Fig.~\ref{fig:spikingnet}A.

The synaptic input current $I_{i,{\rm syn}}$ is the linear sum of inputs coming from the other neurons in the network connected to the postsynaptic neuron $i$. Synaptic inputs have finite (although rather short) rise and decay times, especially for current mediated by AMPA or GABA$_A$ receptors (some of the main mediators of excitatory and inhibitory synaptic inputs, respectively -- see e.g. \cite{Dayan:2001fk}); however, to simplify the analysis of the model, they are often modeled as sums of delta functions: 
\be \label{eqn:Isyneqn}
I_{i,{\rm syn}}=\sum_{j \in {\rm Exc}} J_{ij} \sum_k \delta(t-t^j_k) + \sum_{h \in {\rm Inh}}J_{ih} \sum_{l} \delta(t-t^{h}_{l}),
\ee
where we have separated the inputs coming from excitatory (Exc) and inhibitory (Inh) neurons (abbreviated in the following simply as $E$ and $I$ neurons). The time $t^j_k$ is the time of arrival of spike $\# k$ from presynaptic neuron $j$. According to this model, a presynaptic spike from $E$ neuron $j$ causes a positive jump $J_{ij}$ in the membrane potential $V_i$, whereas an input coming from $I$ neuron $h$ causes a negative jump $J_{ih}$ in $V_i$. The collection of all the $J_{ab}$ values is called the `synaptic matrix' as it contains the values of the synaptic strengths connecting any two neurons in the network.

In the example of Fig.~\ref{fig:spikingnet}A, the spike times obey a Poisson process with some given rate, which means that the inter-spike intervals (ISIs) are exponentially distributed and the spiking process is memoryless (see e.g. Vol.~2 of \cite{t88}); however, in a recurrent model network as well as in real cortical circuits, the ISI distribution is never exactly exponential, and it will depend on the collective behavior of the network. The more asynchronous the network activity, the more accurate the Poisson approximation (see e.g. \cite{sn94}). 

To mimic the heterogeneous connectivity of real cortical neurons, the neurons in the network are recurrently connected according to some random rule. Most frequently, any two neurons are connected with some given probability $c_{\alpha \beta}$ if belonging to populations $\alpha$ and $\beta$, respectively, where $\alpha, \beta \in \{E,I\}$ (also known as Erd\'os-Renyi connectivity). An example is shown in Fig.~\ref{fig:spikingnet}B, in which the excitatory neurons are in shown in black and grey and the inhibitory neurons are shown in red (see below for a description of the actual connectivity structure). The synaptic strengths $J_{ij}$ depend only on the identity of the pre- and postsynaptic neurons ($j$ and $i$, respectively), and may be chosen to be constant values or may be drawn from random distributions with specified mean and variance. Note that the only source of randomness in this model is in the connectivity of the network and potentially the distribution of synaptic strengths --- the dynamics is otherwise purely deterministic. Yet, due to the heterogeneity of synaptic connections and/or the finite size of the network, the neurons can display intrinsic variability in their spike times and/or their firing patterns \cite{vs98,Litwin-Kumar:2012ty}.

The synaptic input in Eq.~(\ref{eqn:Isyneqn}) is instantaneous, though a fixed delay in transmission can also be included. Network models of this kind can be generalized in many ways, from the use of different model neurons, to including biological features such as firing rate adaptation, short-term synaptic plasticity, or synaptic inputs with finite time constants and/or explicit voltage dependence (see, e.g., the book by \cite{gerstner2book2014} for a review). Although these networks may possess multiple configurations of constant firing rate activity (the fixed points of the dynamics) \cite{ab97,cmla04,Mazzucato:2015jk}, one does not typically observe metastable transitions between these states. Here we consider a minimal generalization that allows these networks to exhibit metastable dynamics. The original version of this model is the Amit-Brunel network \cite{ab97}, in which the excitatory population of neurons is partitioned into clusters as shown in Fig.~\ref{fig:spikingnet}B. In each cluster, the average synaptic strength of the weights $J_{ij}$ is potentiated to a value $J_+ J_{EE}$ with $J_+>1$, where $J_{EE}$ is the mean value in an analogous homogeneous network not partitioned in clusters; neurons in different clusters are instead weakly connected, with a mean synaptic strength $J_- J_{EE}<J_{EE}$. This type of network structuring is most often conceptualized as the consequence of training, specifically, as the consequence of being repetitively exposed to stimuli that each activate different subsets of neurons. We review in Sec.~\ref{sec:learning_clusters} how this could occur via experience-dependent synaptic plasticity.
  
\begin{figure*}
\begin{center}
\includegraphics[width=1\textwidth]{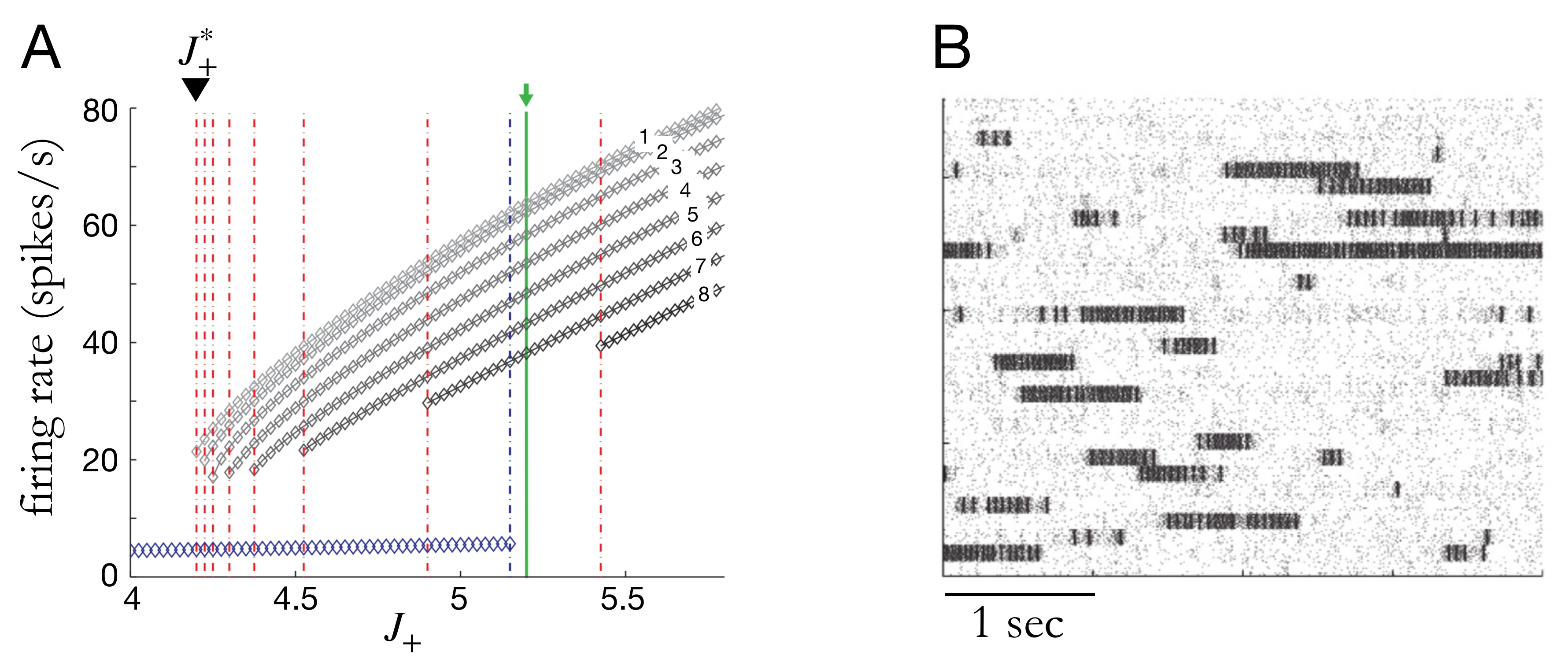}
\end{center}
\caption[]{\small A: Mean field analysis of the clustered spiking network of Fig.~\ref{fig:spikingnet}B with $Q=30$ clusters. $J_+^*$ is the first critical point for the mean synaptic weight inside each cluster; the blue diamonds represent activity configurations with no `active' clusters (i.e., the firing rate in each cluster remains at the level of `spontaneous activity'; see the text for details).  B: Rasterplot of the network in panel A, for a value of $J_+=5.2$ (green vertical line in panel A) illustrating metastable dynamics in this network. The network comprises $4,000$ $E$ neurons and $1,000$ $I$ neurons; the sequence of dots in each line is a spike train emitted by the neuron represented on that line. Transient activations of clusters of neurons is visible as darker bands (see the text for details). Reproduced from L. Mazzucato, A. Fontanini, and G. La Camera, J Neurosci 35, 8214--31 (2015). Copyright 2015 the authors. \cite{Mazzucato:2015jk}}
\label{fig:mfrasters}
\end{figure*}

The original motivation for this model was to obtain a realistic description of associative memory occurring on top of spontaneous brain activity, with the latter being a global attractor of the dynamics. In a later version in which neural clusters can overlap (modeling the fact that real neurons can code for more than one stimulus), this model is the closest biologically plausible analog so far of the influential Amari-Hopfield network (discussed below in Sec.~\ref{sec:fluxtheory}), and can store an extensive number of stimuli modeled as patterns of firing rates across neurons \cite{cmla04}.

This network can be analyzed with a mean-field approach to obtain a bifurcation diagram between different activity configurations of the network (Fig.~\ref{fig:mfrasters}A). The diagram shows that, when the mean synaptic strength inside each cluster exceeds a critical point, the network is multistable. It is convenient to use the mean synaptic potentiation factor $J_+$ as a measure of potentiation. When $J_+$ exceeds a critical point $J_+^\ast$, the network is bistable. In the lower branch of activity the network has a uniform low firing rate activity (`spontaneous' or `ongoing' activity, blue diamonds); configurations with one cluster's activity on the upper branch of the diagram are also possible, and there are $Q$ such configurations, one for each cluster of the network. This way Amit and Brunel described a realistic memory model capable of capturing experimental data from working memory experiments \cite{ab97}. 
As the mean intracluster strength $J_+$ increases further, the mean field analysis demonstrates the existence of a sequence of bifurcations, each of which gives rise to a larger number of additional activity configurations characterized by variable numbers of active clusters (see Fig.~\ref{fig:mfrasters}A). The firing rate in each active cluster depends on the current activity configuration of the network. Beyond a higher critical point, configurations with clusters at low firing rate (`inactive clusters', the blue diamonds in Fig.~\ref{fig:mfrasters}A) are not possible, and at least one cluster is always active. These configurations are stable in the infinite network as predicted by mean-field theory. However, because of random connectivity and recurrent inhibition, in finite networks these configurations become metastable, as shown in simulations (Fig.~\ref{fig:mfrasters}B).

The existence of this kind of metastable dynamics in a spiking network was first pointed out by \cite{Litwin-Kumar:2012ty} and \cite{Deco2012-wg}, who noticed that, because of metastable activity, the network produces slow fluctuations in the neural activity, much slower than the time scales of the single neurons---the origin of such timescales is a long-standing problem in theoretical neuroscience \cite{Murray:2014xd,Doiron:2016qd,Huang:2017hb}. They also showed that, unlike the case of a homogeneous excitatory population, a stimulus will suppress trial-to-trial fluctuations, another widespread phenomenon in cortical circuits \cite{Churchland:2010zp}. \cite{Mazzucato2016-hl} also found that, in this clustered network, the dimensionality of the neural activity (Sec.~\ref{sec:PCA}) is larger during ongoing metastable dynamics than when the network is externally stimulated.   

HMM analyses (Sec.~\ref{sec:HMM}) performed on simulated data of random subsets of neurons in the clustered network of Fig.~\ref{fig:mfrasters} has captured a wealth of features observed in experiments, particularly in the gustatory cortex of rodents, including the existence of hidden states coding for stimulus features and the speed-up of the metastable dynamics during states of expectation reviewed in Sec.~\ref{sec:evidence}. We stress that metastable dynamics occurs in this model despite it being completely deterministic, i.e., the fluctuations of the neural activity are endogenously generated through the quenched variability in the synaptic connections, as already mentioned in Sec.~\ref{sec:definition}.

\subsubsection{Master equation models of neural population activity and effective Markov chains}
\label{sec:WCmodel}
An influential phenomenological model of neural population dynamics is the Wilson-Cowan (WC) model and related variants \cite{wilson1972excitatory,wilson1973mathematical,ChowJNeurophys2020}. The WC model has been used to study coding in networks with continuous attractors (such as the bump attractor model \cite{BenYishaiPNAS1995,Wimmer:2014ie}), and pattern formation, including hallucinations \cite{ErmentroutBiolCyb1979,ButlerPNAS2012,Pearsonelife2016}. 

For discrete neuron-like units at spatial positions $\bm{x}_i$ the WC equations may be written
\begin{equation}
\frac{\partial u(\bm{x}_i,t)}{\partial t} = -\alpha u(\bm{x}_i,t) + f\left(\sum_j ~J(\bm{x}_i,\bm{x}_j) u(\bm{x}_j,t) + I(\bm{x}_i,t) \right).
\label{eqn:WCneuralfield}
\end{equation}
Here, $u(\bm{x}_i,t)$ can be thought of as the activity of a (coarse-grained) neuron located at position $\bm{x}_i$, $\alpha$ is the rate at which this activity decays, $f(\cdot)$ is a nonlinear transformation of the neuron's input, which consists of synaptic-like input from other neurons $(\sum_j ~J(\bm{x}_i,\bm{x}_j) u(\bm{x}_j,t)$ and possible external inputs $I(\bm{x}_i,t)$). Many investigations using the WC equations formally take the continuum space limit $\bm{x}_i \rightarrow \bm{x}$, replacing $\sum_j ~J(\bm{x}_i,\bm{x}_j) u(\bm{x}_j,t)$ with $\int_{D} d\bm{x}\prime~J(\bm{x},\bm{x}\prime) u(\bm{x}\prime,t)$ where $D$ is the spatial domain. Typically the space is assumed to be translation and rotation invariant, such that $J(\bm{x},\bm{x}\prime) = J(|\bm{x}-\bm{x}\prime|)$. Eq.~(\ref{eqn:WCneuralfield}) has also been modified by adding fields for separate types of neurons, typically excitatory and inhibitory cell types. 

While the WC model has been useful for investigating many different types of population activity, to study phenomena like the kind of metastability discussed in this review requires a stochastic version of these dynamics, either at the continuum neural field level or the level of discrete neuron-like units. Many investigations of noisy WC models add stochasticity \emph{ad hoc}, for example by adding a Langevin-type drive to the deterministic equations, similar to the cases to be discussed in Sec.~\ref{sec:fluxtheory}. Several studies have taken an alternate route of constructing stochastic models that yield the WC equations as the mean-field approximation in order to study the effects of stochasticity that better match the variability seen in real data. For example, one approach is to use a master equation formalism, similar to that used to describe the stochastic dynamics of the Ising model \cite{BinderRevModPhys1986}. Briefly, a stochastic master equation model is a system of differential equations for the probability $P(\bm{n},t)$ that a system is found in a particular state $\bm{n}$ at time $t$,
\begin{equation}
\frac{d P(\bm{n},t)}{dt} = \sum_{\bm{n}'} \Big\{T(\bm{n} \leftarrow \bm{n}')P(\bm{n}',t) - T(\bm{n}' \leftarrow \bm{n}) P(\bm{n},t) \Big\},
\label{eqn:mastereqn}
\end{equation} 
where the first term $T(\bm{n} \leftarrow \bm{n}')P(\bm{n}',t)$ describes the flow of probability into the configuration $\bm{n}$ and the second term describes the flow of probability out of configuration $\bm{n}$, such that the total probability is conserved in time, $\frac{d}{dt}\left( \sum_{\bm{n}}P(\bm{n},t)\right) = \frac{d}{dt}\left(1\right) = 0$.

\cite{BuicePRE2007} and \cite{BressloffPRE2010} have used this master equation formalism to model populations of neurons in which the state $\bm{n}$ represents the number of `active' neurons. This characterization does not derive from the individual spiking events, and this model is therefore best thought of as a coarse-grained phenomenological model of collective activity dynamics. In particular, \cite{BressloffPRE2010} investigated metastable transitions in these population models, and showed how a theoretical analysis of such models can be used to derive a reduced Markov chain representation of transitions between fixed point states, which correspond to steady-state solutions of the WC equations. We briefly review \cite{BressloffPRE2010}'s results for a single population of neurons described by Eq.~(\ref{eqn:mastereqn}); see \cite{BressloffPRE2010} for details on extensions of the analysis to excitatory-inhibitory populations. 

In the single population model \cite{BressloffPRE2010} considers, the spatial organization of the population is neglected, and $\bm{n}$ may be taken to be a scalar $n$ that simply counts the number of active neurons in the population. The probability per unit time that an inactive neuron becomes active is given by $T(n + 1 \leftarrow n) = N f(n/N)$, for some nonlinear activation function $f$ and a large parameter $N$ (which could be the number of neurons or the expected number of synaptic inputs), and the probability per unit time that any active neuron becomes inactive is $T(n-1 \leftarrow n) = \alpha n$, where $\alpha$ is the decay rate. By multiplying the master equation by $n$, summing over all possible values of $n$, and neglecting correlations ($\langle f(n/N)\rangle \approx f(\langle n \rangle/N)$ yields the mean field approximation of the stochastic dynamics,
\begin{equation}
\frac{du}{dt} = -\alpha u + f(u)
\label{eqn:BressloffMF}
\end{equation}
where $u = \langle n \rangle/N$. This is the zero-dimensional version of the WC equations (\ref{eqn:WCneuralfield}). If the nonlinearity is taken to be sigmoidal, $f(u) = f_0/(1+\exp(-\gamma(u-\theta))$, where $f_0$ is the amplitude of the transition rate per neuron, $\gamma$ is a gain factor, and $\theta$ is a soft threshold, then for certain parameter choices Eq.~(\ref{eqn:BressloffMF}) has two fixed points $u^\ast_\pm$ and one unstable fixed point $u^\ast_0$. The stable fixed point $u^\ast_-$ corresponds to a small fraction of active neurons, the unstable fixed point $u^\ast_0 > u^\ast_-$ corresponds to an intermediate number of active neurons, and the other stable fixed point $u^\ast_+ > u^\ast_0$ corresponds to a large fraction of active neurons. By using a WKB (Wentzel–Kramers–Brillouin) approximation \cite{kurchan2009equilibrium,AssafJPhysA2017} one can calculate the escape rates from $u^\ast_-$ to $u^\ast_+$ and $u^\ast_+$ to $u^\ast_-$; \cite{BressloffPRE2010} found them to be of the form
\begin{equation}
r_{\pm} \sim \exp\left(-N(\mathcal I(u^\ast_0)-\mathcal I(u^\ast_\pm)) \right),
\label{eqn:Bressloffescaperates}
\end{equation} 
where $\mathcal I(u) = \int^u dy~\ln (y)/f(y)$ is a large deviation function. As in Eq.~(\ref{eqn:Isinglifetime}), the transition rates $r_\pm$ are exponentially dependent on the large parameter $N$ (the number or neurons or number of input connections, depending on model interpretation), indicating that a large population will remain in either of the fixed point states for long periods of time. Note that, unlike the single rate given in Eq.~(\ref{eqn:Isinglifetime}) for the Ising model transition, in this network model the transition rates between the two metastable states need not be equal: $r_+ \neq r_-$, and the system spends different amounts of time in each state.

\begin{figure*}
\includegraphics[width=1\textwidth]{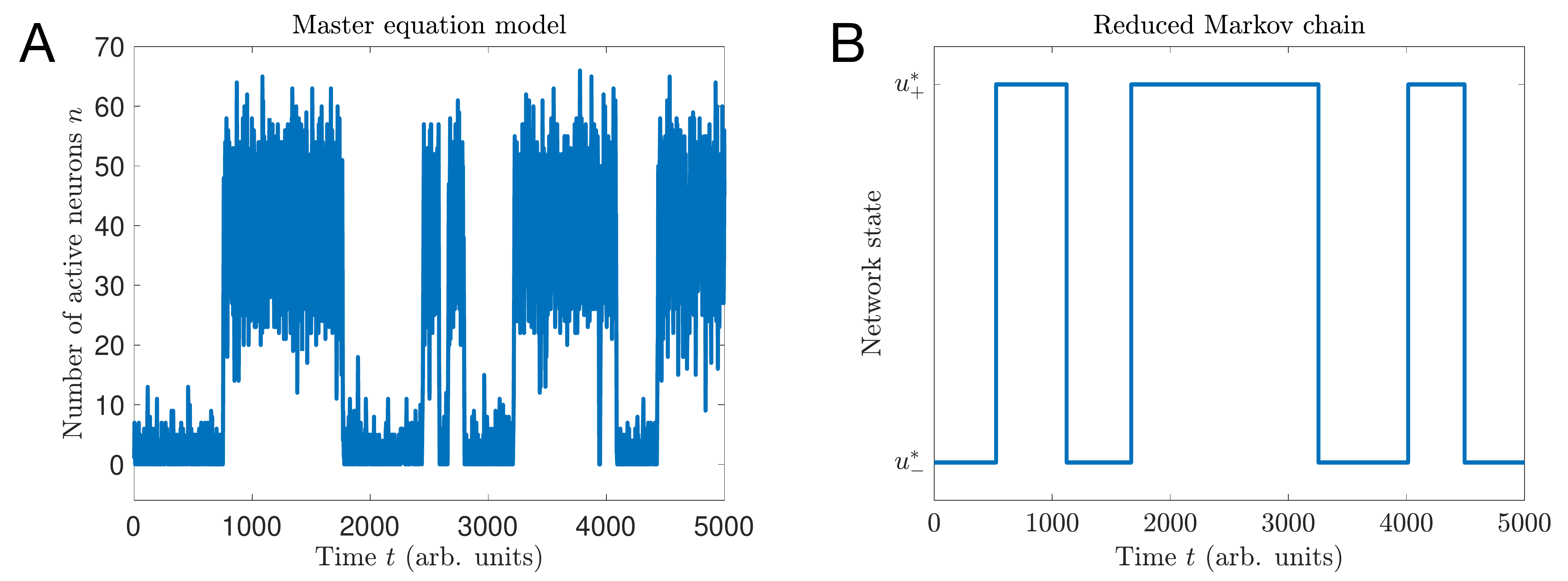}
\caption[]{
    Simulations of the stochastic dynamics of a single population of active and inactive neurons. A: Full simulation of the master equation model Eq.~(\ref{eqn:mastereqn}) with $f_0 = 2$, $\gamma = 1$, $\theta = 0.86$, and $N = 20$ (interpreted as the average number of synaptic inputs each neuron receives). B: Reduced Markov chain (Eq.~(\ref{eqn:BressloffMarkov})) using the estimated escape rates $r_\pm$ from each of the fixed point states $u^\ast_+$ to $u^\ast_-$ or vice versa. See text for details of the meaning of each parameter. This figure mimics the results shown in Fig.~6 of \cite{BressloffPRE2010}, up to stochastic differences in the simulations.}
\label{fig:Bressloffmodel}
\end{figure*}

Because both fixed points in this model are metastable, computation of the escape rates $r_-$ and $r_+$ allows for an explicit reduction of the dynamics of the model to a Markov chain: one can take the states of the Markov chain to be the two metastable states and the transition rates are simply given by the calculated escape rates $r_+$ and $r_-$:
\begin{equation}
\frac{d}{dt}\begin{bmatrix} P_+(t) \\ P_-(t) \end{bmatrix} = \begin{bmatrix} -r_- & r_+ \\ r_- & -r_+ \end{bmatrix}\begin{bmatrix} P_+(t) \\ P_-(t) \end{bmatrix},
\label{eqn:BressloffMarkov}
\end{equation}
where $P_{\pm}(t)$ is the probability of being near the $u^\ast_\pm$ state at time $t$. We plot the results of a simulation of the full master equation model for the single population compared to the reduced Markov chain model in Fig.~\ref{fig:Bressloffmodel}.

This procedure can in principle be extended to models with large populations or clusters of populations, such as the excitatory-inhibitory network considered in \cite{BressloffPRE2010}. In general the calculations become too complex to solve analytically, but can be useful for obtaining numerical estimates of the escape rates between different metastable states, which take a similar form to Eq.~(\ref{eqn:Bressloffescaperates}) under appropriate conditions. In principle, then, one could derive the effective Markov chain model for transitions between the metastable fixed point states, effectively providing a theoretical derivation of the reduced Markov model that one would seek to obtain by a Hidden Markov model analysis, as discussed in Sec.~\ref{sec:HMM}. Interestingly, this approach has also been used to study a reduced model of the interaction of metastable neural dynamics and synaptic modifications during learning \cite{Litwin-Kumar:2014fv}, a topic we review in Sec.~\ref{sec:learning_clusters}

\subsubsection{Non-equilibrium landscape and flux models of neural network dynamics}
\label{sec:fluxtheory}

Finally, we consider models based on extensions of the energy landscape picture to non-equilibrium stochastic dynamics. This attractor landscape metaphor is widely used in the top-down models suggested to understand and describe cognitive functions such as associative memory retrieval, classification and error correction, and more \cite{h82,amitBook89,YH18,YH20}.

One of the earliest such uses of the landscape picture in theoretical neuroscience was made by Hopfield \cite{h82}, who noticed that the dynamics of a spin-based model with symmetric connections (later generalized to a model with graded units \cite{cg83,h84}) could be understood as a gradient dynamics down an energy function Eq.~(\ref{eqn:spinHamiltonian}). In analogy to magnetic systems, Hopfield and other physicists suggested that the dynamics of a neural network for associative memory could be described by an energy function in the space of neural activity patterns. Moreover, under certain conditions the value of this function would always decrease as the system evolves in time, eventually achieving a stationary state \cite{h82,cg83,YH2,YH21,YH22}. Each of the minima of the energy function is a dynamical attractor of the system that involves the transformation of a given input stimulus to a specific output, namely, a memory. This energy is a global quantity, which is not felt by any individual neuron. Such a depiction is not merely schematic, but can be quantified in certain neural circuits with specific assumptions \cite{h82,amitBook89,YH2,YH21,YH22}.

The ability of a neural network to properly store and retrieve associative memories depends on whether the trajectories in the state space are strongly influenced by the initial states of the network or perturbations by external stimuli. More precisely, if the network is \emph{ergodic} it will eventually visit every state in its phase space, regardless of the initial preparation of the network or subsequent stimulus perturbations \cite{amitBook89,GoldenfeldBook1992}. In such a case associative memory retrieval is unreliable, if not impossible, as desired memory states (minima of the energy landscape) would only be accessible transiently before the network drifts towards a different memory. Moreover, even if deterministic network dynamics are not ergodic, neural networks must be able to function in the presence of stochastic noise. Strictly speaking, even small fluctuations will eventually cause the network to visit all possible states after a sufficiently long enough time. This ergodicity is not necessarily harmful for associative memories if the dwelling times in each metastable state are sufficiently long. Fortunately, as in Eqs.~(\ref{eqn:Isinglifetime}) and (\ref{eqn:Bressloffescaperates}), the cooperativity of many interacting neurons can lead to long dwelling times in metastable states, sufficient to break ergodicity for all timescales relevant to memory retrieval and other brain functions.

To understand the metastability of memory-like states in models like the Amari-Hopfield network, it is important to understand the properties of the energy landscape. For example, in the original Hopfield model \cite{h82} the energy landscape serves as a Lyapunov function, whose value always decreases monotonically \cite{h82,h84,YH21}. Thus the energy function provides a global measure and description of the dynamical system. However, this is true only if the neural interactions $J_{ij}$ are symmetric, which is unrealistic in real neural networks \cite{SongPLOSB2005}. 

In general, asymmetric networks can express a much richer repertoire of dynamics than symmetric networks, and therefore possess greater computational capabilities \cite{AsllanieaauSciAdv2018,Kerg2019,Orhan2020Improved}. For example, symmetric spin-networks with random $J_{ij} = J_{ji}$ exhibit glassy dynamics \cite{BinderRevModPhys1986,BOUCHAUDBook}, whereas networks with uncorrelated $J_{ij}$ and $J_{ji}$ may exhibit a transition to chaotic activity \cite{YH2}. Is it possible to construct something analogous to a Lyapunov function for general neural circuits? The answer is yes. In the following we define landscapes functions for general neural networks and the corresponding non-equilibrium dynamics associated with such landscapes.

To construct the Lyapunov function for general neural networks, it is convenient to start from a stochastic version of the dynamical system and then take the zero fluctuations limit to recover the original system. Therefore, consider the dynamics described by a set of ordinary differential equations, 
\begin{equation}
\frac{{d\bm{x}}}{{dt}} = \bm{F}(\bm{x}) + \bm{\xi}(t),
\label{eqn:xdynamics}
\end{equation}
where $\bm{x}=\{{{x}_{1}}, ..., {{x}_{n}}\}$ is the $n$-component state of the network, $\bm{F}(\bm{x})$ is the $n$-component `driving force' or interactions between components (neurons), and $\bm{\xi}(t)$ is a Gaussian stochastic noise of mean $\bm{0}$ and covariance $\langle \xi_i(t) \xi_j(t') \rangle = 2 \sigma^2 D_{ij}(\bm{x})\delta(t-t')$, for a noise strength $\sigma^2$ and diffusion matrix $\bm{D}(\bm{x})$, which may explicitly depend on the current state. Here, Eq.~(\ref{eqn:xdynamics}) is assumed to be given in the Stratonovich interpretation \footnote{The choice of interpretation of the stochastic differential equation (\ref{eqn:xdynamics}) is only important insofar as making sure the stochastic dynamics is consistent with the path integral formulation given in Eq.~(\ref{eqn:landscapeaction}). Both the Stratonovich interpretation and the Ito interpretation---the other common choice of interpretation---make equivalent physical predictions. It is sometimes erroneously claimed that the Ito interpretation may fail to predict metastability when the Stratonovich interpretation does, or vice versa \cite{SMYTHEPhysLettA1983}. However, such a discrepancy is merely a consequence of naively adding multiplicative noise to a deterministic model and assuming the driving force should be the same in the corresponding stochastic differential equation, which is only true for the Stratonovich interpretation. The appropriate generalization of a deterministic ODE to a stochastic DE in the Ito interpretation will introduce an additional term to the `deterministic' force $\bm{F}(\bm{x})$. When this additional term is properly accounted for, the two interpretations agree.}. This dynamics represents the temporal evolution of this neural network from one state to another. 

The stochastic differential equation (\ref{eqn:xdynamics}) can be mapped onto an equivalent Fokker-Planck equation for the probability density $P(\bm{x},t)$ of the network being in a state $\bm{x}$ at time $t$ (see, e.g., \cite{Kampen:2007cr} or \cite{Gardiner:2004pv}):
\begin{equation}
\frac{\partial P(\bm{x},t)}{\partial t} = - \nabla \cdot \bm{J},
\label{eqn:FPeqn}
\end{equation}
where $\nabla$ is a gradient with respect to the state variables $\bm{x}$ and the probability flux is given by
\begin{equation} \label{eqn:FPprobflux}
\bm{J} = \bm{F}(\bm{x}) P(\bm{x},t) - \nabla\cdot \left( \sigma^2 \bm{D}(\bm{x}) P(\bm{x},t)\right),
\end{equation}
where $(\nabla \cdot \bm{D})_i = \sum_j \partial_{x_j} \bm{D}_{ij}(\bm x)$. 
In a symmetric neural network, the driving force can be written as the gradient of a Lyapunov energy function \cite{YH21,Yan:2013rs}. For general networks, the driving force can be derived from Eq.~\ref{eqn:FPprobflux} to be
\begin{eqnarray} \nonumber
\bm{F} & = & \bm{J}_{ss}/P_{ss}+\sigma^2\bm{D} \cdot \nabla (P_{ss})/P_{ss}+ \nabla \cdot \sigma^2\bm{D} \\ 
& =& \bm{J}_{ss}/P_{ss}-\sigma^2\bm{D} \cdot \nabla U+ \nabla \cdot \sigma^2\bm{D}.
\end{eqnarray}
Here, the non-equilibrium potential is defined as $U =-\ln P_{ss}(\mathbf{x})$ in analogy to the Boltzmann law in equilibrium statistical mechanics, where  $P_{ss}$ is the steady state probability distribution. We see that for general (non-symmetric) neural networks, the dynamics of the non-equilibrium system is determined by the gradient of the potential landscape $U$ {\em and} by the curl flux.\footnote{It is notable that the Lypunov function in the original Hopfield model is {\em not} the potential landscape quantified by the stationary probability, $U =-\ln P_{ss}(\mathbf{x})$. As a consequence, the existence of  an energy-like function does not guarantee detailed balance in this case, i.e., the steady state flux measuring the degree of detailed balance breaking is not necessarily equal to zero. See \cite{Yan:2013rs} for details.} 

\begin{figure*}
\begin{center}
\includegraphics[width=0.75\textwidth]{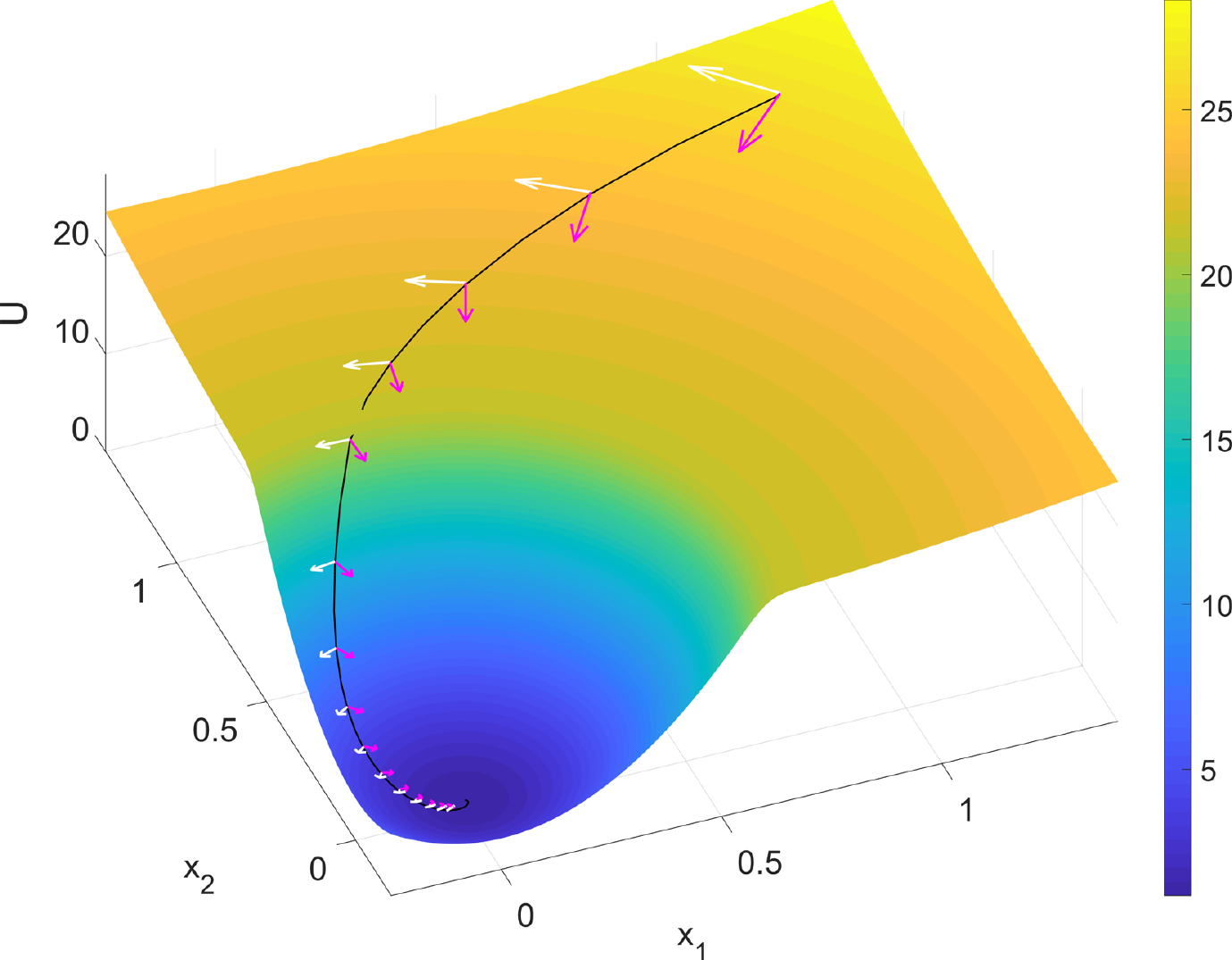}
\end{center}
\caption{\small An illustration of a driving force and flux on the underlying non-equilibrium landscape. $x_1$ and $x_2$ are state variables in arbitrary units. White arrows represent the flux, and pink arrows represent the force from the negative gradient of the potential landscape. See the text for details.}
\label{fig:driving-force}
\end{figure*}

The non-equilibrium potential $U$ can be used to quantify the global behavior of the non-equilibrium systems since $U$ is linked to the weight (or probability) of the state. However, $U$ is not a Lyapunov function. A Lyapunov function $\phi_0$ for general networks can be derived from the leading order expansion of the potential $U \doteq (1/\sigma^2)\sum_{k=0} (\sigma^2)^k \phi_k$ with respect to the scale of the fluctuations $\sigma^2$. At leading order $1/\sigma^2$ one obtains the equation \cite{Yan:2013rs,YH25,YH27,YH28,YH29}:
\begin{equation}
{\sum\limits_{i = 1}^{n}{F_{i}\left(\bm{x} \right)\frac{\partial\phi_{0}\left(\bm{x} \right)}{\partial x_{i}}}} + {\sum\limits_{i = 1}^{n}{\sum\limits_{j = 1}^{n}{D_{ij}\left(\bm{x} \right)\frac{\partial\phi_{0}\left(\bm{x} \right)}{\partial x_{i}}\frac{\partial\phi_{0}\left(\bm{x} \right)}{\partial x_{j}}}}} = 0
\end{equation}
This equation is called the Hamilton-Jacobi equation due to its resemblance to the Hamilton-Jacobi equation in classical mechanics. One can prove that the solution $\phi_0(\bm{x})$ is a monotonically decreasing function along the trajectory $\bm{x}(t)$ that is a solution of the dynamics Eq.~(\ref{eqn:xdynamics}). Thus, a solution $\phi_{0}(\bm{x})$ of the Hamilton-Jacobi equation is a Lyapunov function and can be used to quantify global stability. In the zero-fluctuation limit, the flux term of the driving force can also be expanded in terms of the fluctuation strength $\sigma^2$ to obtain its leading order as:
$\left. \bm{J}_{SS}(\bm{x})/P_{SS}(\bm{x}) \right|_{\sigma^2\rightarrow 0} =\bm{F}(\bm{x}) + \mathbf{D}(\bm{x})\cdot\nabla  \phi_{0}\left(\bm{x} \right)$. One can show that the intrinsic flux velocity $\bm{v} = \left. \bm{J}_{SS}/P_{SS} \right|_{\sigma^2\rightarrow 0}$ satisfies $\bm{v} \cdot \nabla\phi_{0} = 0$, which implies that the gradient of the non-equilibrium intrinsic potential  $\phi_{0}$  is orthogonal to the intrinsic flux (or intrinsic flux velocity) in the zero-fluctuation limit. The distinguishing feature of non-equilibrium systems is the presence of non-vanishing steady-state flux $\bm{J}_{SS}$, from which the appropriate generalization of the non-equilibrium driving force, in the  zero-fluctuation limit, can be determined to be \cite{YH23,Yan:2013rs,YH25,YH26}:
\begin{equation}
\bm{F}(\bm{x}) = - \bm{D}(\bm{x}) \cdot \nabla \phi_0(\bm{x}) + \bm{v}.
\label{eqn:noneqforce}
\end{equation}
Thus, we see that---unlike equilibrium systems in which the driving force is the gradient of an energy function---the non-equilibrium dynamics of a generic network can be globally determined by a driving force with three terms. The first term is the gradient of the potential $U$ related to the steady state probability landscape of the system, while the second term is associated to the curl steady state probability flux. The steady state probability landscape $P_{SS}$ quantifies the steady state probability of each state while the curl steady state probability flux  $\bm{J}_{SS}$ quantifies the flow around the states. The flux is a quantitative measure of the detailed balance breaking: a non-equilibrium signature of energy, material or information exchange between the environment and the system. If $\bm{J}_{SS}=\bm{0}$, then there is no net energy or particle flow into or out of the system. On the other hand, a nonzero $\bm{J}_{SS}$  leads to the net energy or particle flow into or out of the system. This is the cause of the detailed balance breaking that gives rise to intrinsically non-equilibrium dynamics. An illustration of the differential effects of the gradient and flux components of the driving forces is shown in Fig.~\ref{fig:driving-force}.

While the steady state probability landscape can identify higher probability states and correlate to biological functional states, the dynamical connections among those functional states are quantified by both the landscape \emph{and} the flux. The steady state probability landscape describes the probability of each state of the network, while steady state probability flux describes the net flow from or to each state. The steady state probability landscape will naturally generate a force moving towards higher probability while the steady state probability flux gives the contribution of the additional force along the direction of the flow. Intuitively, the stochastic dynamics of the entire network can be understood in analogy with a charged particle moving in an electric field generated by the electric potential (probability landscape) guiding motion along the electric field and a magnetic field (probability flux) giving a spiral or cyclic motion. 

\subsubsection{A combined approach: path integral analysis of the energy landscape of the clustered spiking network model}
\label{sec:combined}

The clustered spiking model of Sec.~\ref{sec:spiking_net} and the dynamics recovered from neural data analysis (Sec.~\ref{sec:latent-cont}) can be analyzed using the landscape and flux dynamics approach to unveil mechanisms underlying metastable sequence transitions. To apply this theoretical approach to the network model, a path integral formulation can be developed that takes into account additional non-equilibrium terms identified in Eq.~(\ref{eqn:noneqforce}). The path integral represents the probability of starting from an initial neural network state $\pmb{x}_i$ (basin of attraction) at time $0$ and ending up at a final state of $\pmb{x}_f$ at time $t$. This transition probability may be written as
\begin{equation}
P({\pmb{x}_f},t,{\pmb{x}_i},0)=\int \mathcal{D} {\pmb{x}}~\exp\left(-S[\pmb{x}(t)]\right)
\label{eqn:transitionprob}
\end{equation}
where $\mathcal D\bm{x}$ is the formal path integral measure and the weight of a path $\bm{x}(t)$ is determined by the `action'
\begin{equation}
S[\pmb{x}(t)]=\int dt~\Bigg\{\frac{1}{4} \frac{d\pmb{x}}{dt} \cdot \bm {D^{-1}}\cdot \frac{d\pmb{x}}{dt} -\frac{1}{2} \bm{F} \cdot \bm {D^{-1}} \cdot \frac{d\pmb{x}}{dt}+V_{\rm eff}(\bm{x})\Bigg\},
\label{eqn:landscapeaction}
\end{equation}
where $V_{\rm eff}(\bm{x})=\frac{1}{4}\bm{ F} \cdot \bm D^{-1} \cdot \bm{F} + \frac{1}{2}(\bm{D}\cdot \nabla)\cdot(\bm{D}^{-1} \cdot \bm{F})$; this final term comes from a Jacobian factor generated by choosing the Stratonovich interpretation of the Langevin dynamics Eq.~(\ref{eqn:xdynamics}). The path integral probability is equal to the sum of weights connecting all possible paths from the initial state $\bm{x}_i$ to the final state $\bm{x}_f$. Not every path gives the same weight, and there exists a dominant path that extremizes the action, and hence has the largest relative weight. Contributions from other sub-leading paths are exponentially smaller than the dominant paths, and one can therefore estimate the transition probability  Eq.~(\ref{eqn:transitionprob}) by $\exp(-S[\pmb{x}])$ evaluated at the dominant path $\bm{x}(t)$.

\begin{figure*}
\begin{center}
\includegraphics[width=0.8\textwidth,angle=0]{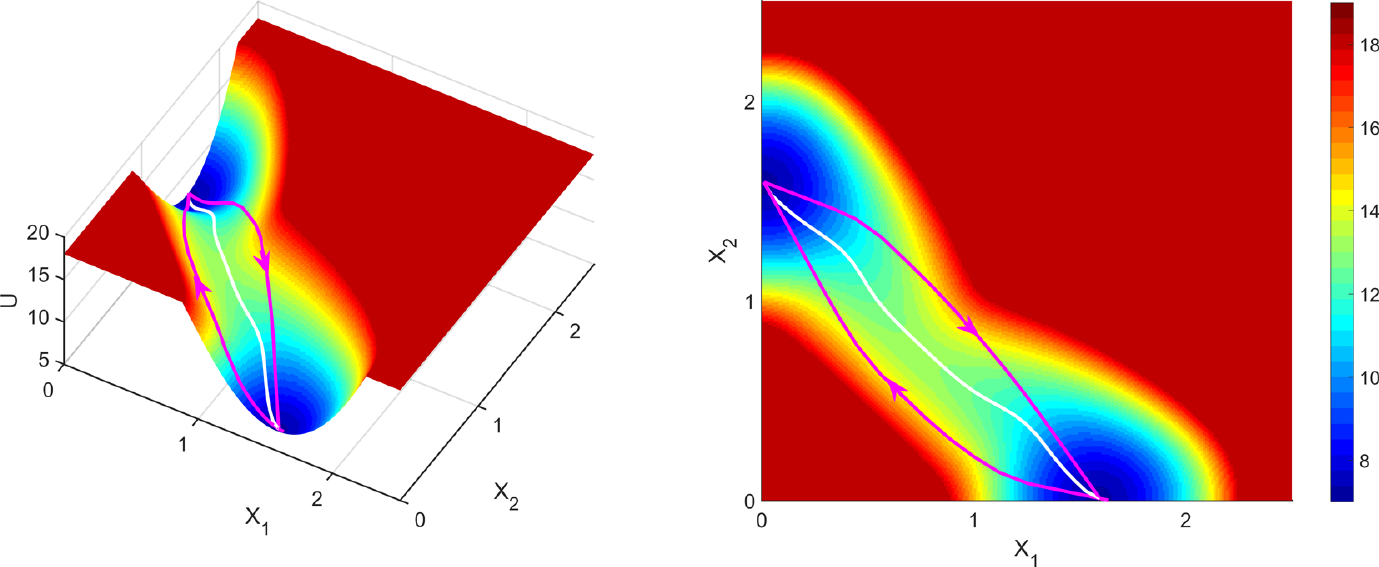}
\end{center}
\caption{\small 2D and 3D illustration of non-equilibrium landscape with the irreversible dominant transition paths between basins (purple lines with arrows) and the gradient path (white line). $x_1$ and $x_2$ are state variables (arbitrary units). $U$ is the underlying non-equilibrium potential. See the text for details. Reproduced with permission from H. Feng, K. Zhang, and J. Wang,  Chem. Sci. 5, 3761--3769 (2014). Copyright 2014 The Royal Society of Chemistry. \protect\cite{Feng_2014}}
\label{fig:actionpaths}
\end{figure*}

An interesting feature of the non-equilibrium dynamics is that the flux force is not invariant under time-reversal, and hence the forward path from $\pmb{x}_i$ to $\pmb{x}_f$ and backward path from $\pmb{x}_f$ to $\pmb{x}_i$ are not expected to follow the same route. This implies that network dynamics is in general irreversible as shown in Fig.~\ref{fig:actionpaths}. Although in generally these paths cannot be calculated analytically, one can numerically search for the dominant paths even in high dimensional space through the optimization of a line integral by Monte Carlo sampling. This greatly simplifies the computation and the method can be used for dealing with large neural networks. This enables identification of the path that is most likely to be taken to transition from one state to another during sequences of metastable transitions produced by the clustered network, and to extract details regarding the actual metastable state switching processes. Furthermore, to quantify the kinetics of metastable state switching, the transition states can be identified---they are shifted away from the saddle points on the underlying landscape due to the presence of the non-equilibrium rotational flux, as shown in Fig.~\ref{fig:actionpaths}. Results from this analysis can be further correlated with the underlying landscape topography as well as observations inferred from experimental data.

\subsubsection{Non-equilibrium thermodynamics, intrinsic energy, entropy, and free energy of general neural networks} \label{sec:noneq-thermo}

Through the landscape and flux approach, one can identify the functional states through the landscape minimum, quantify the stability by basin depths and barrier for the associated states and explore the switching speed between them. In addition to the nonequilibrium dynamics, nonequilibrium thermodynamics for general neural networks can be developed in a manner analogous to equilibrium thermodynamics \cite{Yan:2013rs,YH25,YH30,YH31,YH32,YH33,YH34}. We first relate the non-equilibrium intrinsic potential  $\phi_{0}(\bm{x})$ to the steady state probability distribution as 
\be
P_{SS}\left(\bm{x} \right) = \exp\left( - \phi_{0}(\bm{x})/\sigma^2 \right)/Z,
\ee 
where  $Z = \int d\bm{x}~ \exp\left( - \phi_{0}(\bm{x})/\sigma^2 \right)$ is defined as the time-independent (steady state) non-equilibrium partition function. The intrinsic energy and entropy of the nonequilibrium neural networks can be defined as 
\be
\mathcal{E} = \int d\bm{x}~\phi_{0}(\bm{x}) P\left(\bm{x},t \right) = - \sigma^2\int d\bm{x}~ \ln(ZP_{SS}(\bm{x})) P\left(\bm{x},t \right)
\ee
and 
\be
\mathcal{S} = - \int d\bm{x}~ P\left(\bm{x},t \right)\ln P\left(\bm{x},t \right),
\ee
respectively. Naturally, the intrinsic free energy can be defined as 
\be
\mathcal{F} = \mathcal{E} - \sigma^2 \mathcal{S} = \sigma^2 \left( \int d\bm{x}~ P \ln\left( P/P_{SS} \right) - \ln Z \right).
\ee
We can further investigate the derivative of the intrinsic free energy with respect to time and obtain
\be
\frac{d\mathcal{F}}{dt}=-\sigma^4\int d\bm{x}~ \nabla \ln (P/P_{ss}) \cdot \mathbf{D} \cdot  (\nabla \ln(P/P_{ss})) P \;   \leq 0.
\ee
This equation indicates that the intrinsic free energy of the non-equilibrium system always decreases in time until reaching the minimum value $\mathcal{F}=-\sigma^2 \ln Z $.  When the fluctuations $\sigma^2$ are finite, the non-equilibrium free energy defined as $\mathcal{F} = \mathcal{E} - \sigma^2 \mathcal{S} = \int d\bm{x}~ \sigma^2 U P - \sigma^2(-\int d\bm{x}~P \ln P)$ is also a Lyapunov function monotonically decreasing in time \cite{Yan:2013rs,YH25}.

The time derivative of the system entropy can be divided into two terms: $d{\mathcal{S}}/dt = d{\mathcal{S}}_{t}/dt - d{\mathcal{S}}_{e}/dt$.  The entropy production rate is  
\be
\frac{d{\mathcal{S}}_{t}}{dt} = \int d\bm{x}\left( \bm{J} \cdot \left( \sigma^2 \bm{D} \right)^{- 1} \cdot \bm{J} \right)/P,
\ee
which is either positive or zero \cite{Yan:2013rs,YH25,YH30,YH31,YH35}. The heat dissipation rate or entropy flow rate to the network from the environment is defined as 
\be
\frac{d{\mathcal{S}}_{e}}{dt} = \int d\bm{x}\left( \bm{J} \cdot \left( \sigma^2 \bm{D} \right)^{-1} \cdot \left(\bm{F} - \nabla \cdot \left( \sigma^2 \bm{D} \right)\right) \right)/P,
\ee
and can either be positive or negative. Although the total entropy change rate of the neural network (system plus environment) $d{\mathcal{S}}_{t}/dt$  is always non-negative, consistent with the second law of thermodynamics, the system entropy change rate $d{\mathcal{S}}/dt$ is not necessarily positive. This implies that the system entropy is not always maximized for general neural networks. Nevertheless, the system free energy does minimize itself for neural networks.
The thermodynamic cost for maintaining the function of the neural network can be quantified in terms of the entropy production rate, which is directly related to the indispensable part of the non-equilibrium driving force. The energy dissipation computed in this way has been successfully used to cost-performance trade-off in biological systems whose energy sources are ATP, GTP and SAM \cite{YH36,YH37}. Furthermore, since the landscape topography and flux in determining the functions and stability of the functional states can be globally quantified, one can explore which underlying neural network interactions or specific types of neurons these global dynamic and thermodynamic measures are sensitive to.  Through such global sensitivity analysis, the key types of neurons and neural interactions which are critical to the states' stability and switching dynamics could be identified, and hopefully this   global system perspective could be used to design strategies for perturbations of brain function that could mitigate pathologies caused by neurological disorders.

\section{Metastability, learning and neural network function} \label{sec:plasticity}

\subsection{Anatomical and functional underpinning of metastable dynamics}
As reviewed in Sec.~\ref{sec:theorymodels}, recent models point to the presence of neural clusters as originators of metastable activity. In this section, we will review evidence from experimental neuroscience suggesting that clustered architectures predicted by the model can indeed be found in cortical networks. 

\subsubsection{Preferential patterns of connectivity in cortex}

In cortical circuits, evidence for the presence of preferentially connected clusters of neurons has been reported using either paired recording electrophysiology, which allows for the direct measurement of the properties of a connection between two neurons \cite{Markram_1997}, or using a variety of approaches for circuit mapping including uncaging of neurotransmitters \cite{Yoshimura_2005} or optogenetic-assisted circuit mapping \cite{Petreanu_2007}. At the anatomical level, studies in sensory cortex have found that excitatory neurons in the superficial layer have a higher probability of being recurrently connected when they share a common input \cite{Yoshimura_2005}, and that the probability of connection is higher when the magnitudes of the incoming inputs are comparable \cite{Wang:2013pr}. While these results suggest the presence of preferential connectivity, the data were obtained \emph{ex vivo}, therefore there is no evidence that such groups of connected neurons respond to similar stimuli. At the functional level, Ref.~\cite{Ko:2011xi} reported that excitatory neurons responding to a specific feature of a visual stimulus were more likely to be connected, further supporting the presence of clusters of connected neurons. However, whether such connectivity represents clusters of neurons that participate in the functional aspects of neural processing remains controversial. Indeed, the connection probability reported by \cite{Ko:2011xi} is comparable to that reported by previous studies that simply measured the probability of finding any connected pairs of neurons in \emph{ex vivo} preparations \cite{Maffei_2004,Wang:2012xd}. Thus, the higher connection probability may rely on proximity, not so much on responsiveness to a common functional feature.  

\subsubsection{Metastability as a mechanism for multiplexing}

An additional level of complexity arises when considering that cortical neurons can respond to multiple features of a stimulus or encode multiple cognitive variables, a property sometimes called {\em multiplexing}. In the visual cortex, as well as auditory and somatosensory cortices, analysis of circuit organization and functional responsiveness has focused primarily on neurons responding to specific features; very little work investigated multiplexing. Coding for complex variables is often ascribed to high order cortical areas and areas of the cerebral cortex thought to be involved in bringing together sensory and cognitive information including attention, expectation, and reward. This hierarchical view of cortical circuits considers neurons as structures specialized to respond to specific stimulus features, process these features, and then transmit this information to neurons in the next hierarchical order, which in turn will bring together information they received from other circuits and encode a percept. 

Recent work in the rodent GC provides an alternative view of multiplexing: via metastability. First, neurons in GC are indeed involved in multiplexing: the same neuron can respond to multiple taste stimuli and to cues promoting expectations, and can participate in driving decisions \cite{Samuelsen:2012nr,Gardner:2014qf,Kusumoto-Yoshida:2015ip,Vincis:2016dz,Livneh:2017xu,Mukherjee:2019kn,Vincis:2020zv}. Second, taste-related metastable activity in GC is compatible with ongoing metastable dynamics (Sec.~\ref{sec:evidence}), a fact that the spiking network model of Sec.~\ref{sec:spiking_net} predicts to rely on the clustered architecture of the network, which provides a common mechanism for both taste-related and ongoing metastability. Can the latter also result in multiplexing capabilities? Metastable dynamics in GC may indeed facilitate the encoding of multiple variables by leveraging the temporal dynamics of the neurons' own activity: this could dynamically rearrange their participation in coding for one or the other variable by visiting different metastable states -- a form of dynamic population coding \cite{Meyers:2018uq}.

Further theoretical work is required to evaluate the plausibility of this proposal, but it is also crucial to find experimental evidence for its basic ingredient: recurrent neural clusters interacting so as to produce metastable activity. Although some indirect evidence of cortical clusters has been reported \cite{Kiani:2015gr}, ideally one would like to directly measure neural connections and their susceptibility to being modified by experience to form functional clusters. 

\subsubsection{Methods for mapping cortical circuits}

Paired recording approaches to local circuit mapping are highly effective in identifying specific connections \cite{Miles_1996,Markram_1997}, but suffer from a number of limitations. With paired recordings, a small number of neurons can be simultaneously recorded. The advantage of this approach is in allowing detailed identification of recorded neurons and full control over their membrane properties by using patch clamp electrophysiology. The resolution is on suprathreshold as well as subthreshold events underlying neuron-to-neuron communication, making it highly suitable for assessing connectivity. The primary limitation is that the analysis of connectivity works well for near-neighbor coupling, but is less effective in assessing less spatially restricted connectivity. Other approaches like channelrhodopsin-assisted circuit mapping \cite{Petreanu_2007} facilitate the identification of connectivity maps over a larger spatial scale, and can resolve subthreshold and suprathreshold events, but lose precise control over cell-to-cell connectivity as they can recruit both direct and indirect inputs to the recorded neurons. Both paired recordings and channelrhodopsin-assisted circuit mapping require the use of \emph{ex vivo} preparations, thus they are not ideal for determining whether or not connected neurons share common functional properties \emph{in vivo}. 

Multielectrode recordings in behaving animals allow for the simultaneous recording of spiking activity from multiple neurons and have been instrumental for the analyses that identified metastable dynamics in cortical circuits \cite{Seidemann1996-bf,Jones:2007ff,Kemere:2008uq,Ponce-Alvarez2012-zi,Mazzucato:2015jk,Engel2016-bb,Maboudi2018-ox}. This experimental approach provides spontaneous and stimulus-driven activity in animals that are engaged in a variety of tasks. Correlations of activity across groups of neurons facilitates the identification of functional clusters, although it cannot offer sufficient resolution to determine whether neurons that share common functional properties are recurrently connected or share a common input. If large scale recordings are needed to add a spatial component to the analysis of metastable dynamics, it is possible to increase the number of electrodes or recording sites, an approach that facilitates analysis of spiking activity along the vertical axis (depth) of the cortical mantle \cite{Jun:2017zd}, or to use calcium imaging for assessing neural activity along the horizontal axis \cite{Chen_2021}. While analysis techniques are well-established for identifying metastable dynamics in spiking activity, there are technical caveats that need to be resolved to extract these dynamics from the much slower calcium signals. More generally, there is a need to develop approaches to reliably link spiking activity to the activity detected with fluorescent signals emitted by calcium indicators \cite{Friedrich2017-qp,Pachitariu2017-sm,Giovannucci2019-rx,Barson:2020ef,Huang2021-gz}. Solving this set of technical problems will facilitate extraction of metastable states at the slower time scales typical of calcium signals and provide experimental evidence for the theorized link between metastability and cluster activation.

\subsection{Synaptic plasticity, learning and metastable dynamics} \label{sec:learning}

Cortical circuits subserving metastable dynamics may be genetically codified and partially formed at birth, but given their putative role in coding for sensory and cognitive processes (Sec.~\ref{sec:cognitive}), it is assumed that they would fully develop through experience via a process of learning and plasticity. As indirect evidence, changes in metastable dynamics due to learning has been reported in the rodent gustatory cortex \cite{Moran_2014}. Theoretical models suggest that metastability is generated and maintained by a clustered network architecture \cite{Deco2012-wg,Litwin-Kumar:2012ty,Mazzucato:2015jk,Cao:2016bf,Setareh:2017km,Mazzucato:2019wi,Rostami:2020wx}, and that neural plasticity may be involved in establishing such architecture and modulating it \cite{Litwin-Kumar:2014fv,Zenke:2015bv}. No study to date has directly assessed the biological correlates of such predictions. In the next section, we discuss experimental evidence from studies of experience-dependent circuit refinement and learning that may provide evidence for the involvement of plasticity in establishing, maintaining, or modulating clustered connectivity. While none of the experimental findings we present below is directly related to metastability, they help make inferences regarding neural circuit organization and its modulation by experience and learning that seems required to produce metastable dynamics in models. The hypotheses we discuss are speculative at this stage, but provide clues on whether or not a model of functional clustered connectivity is biologically plausible, on how clustered connectivity may originate during postnatal development, and plasticity and learning may affect it.

 Learning, memory formation and storage and other adaptations induced by experience are known to induce neural plasticity, which can manifest in changes in the efficacy of transmission of signals between neurons, alterations of neural connectivity or other modulation that affect a neuron's input/output function \cite{Hansel_2001,Malenka_2004,Nelson_2008}. Changes in the efficacy of neuron-to-neuron communication and circuit configurations that arise as a consequence of neural plasticity have the potential of destabilizing circuit dynamics. Indeed, many models of network function fail to reproduce network dynamics unless normalizing functions are built into the model, as we will discuss in Sec.~\ref{sec:learning_clusters}. However, healthy brain circuits clearly have the capability to preserve their activity throughout life, indicating that mechanisms are in place to maintain the system within a functional working range. How such dynamic patterns of activity may be maintained in the face of changes in neural activity driven by cognitive processes is an open experimental and theoretical question. 
\subsubsection{Circuit refinement by synaptic plasticity}

During postnatal development, neural plasticity is thought to help refine circuits that allow for the establishment of metastable dynamics. For example, visual experience is necessary for establishing preferential connectivity of neurons responding to a common property of an incoming visual input \cite{Ko:2013we}. These results, viewed in the context of the clustered network architecture that supports metastability (Sec.~\ref{sec:spiking_net}), suggest that an initial set of activity-dependent patterns may be needed to establish the circuit connectivity that facilitates the generation of metastable dynamics. Several studies have also reported evidence for the modulation of the efficacy of synaptic transmission in visual cortical circuits in response to the onset of visual experience \cite{Maffei_2004,Wang:2012xd,Tatti:2017sz}. Plastic changes like those reported in visual cortex have been observed in other sensory areas including somatosensory \cite{Foeller_2004} and auditory cortex \cite{Mowery_2016,Dorrn_2010}.

\subsubsection{Memory formation and Hebbian learning} \label{sec:hebb}

In addition to circuit refinement, synaptic plasticity is thought to be the neural underpinning of learning and memory. Changes in the efficacy of synaptic transmission between neurons can contribute to forming and storing new memories. A well established theory proposed by Donald O. Hebb in 1949 states that the formation of a memory requires the strengthening of synaptic transmission between neurons \cite{Morris_1999}. Extrapolated to the scale of entire circuits, this theory provided the basis for the idea that learning leads to the strengthening of synaptic transmission between groups of neurons that have all been activated by specific patterns of incoming activity \cite{Rao_Ruiz_2021}. Such groups of neurons could be easily reactivated if the learned pattern or stimulus is presented again after learning, possibly speeding up the neural coding of stimulus properties. These co-activated groups of neurons are thought to participate in the formation of what is sometimes defined as an `engram', a signature circuit for a memory \cite{Josselyn_2020} (see also Sec.~\ref{sec:assoc}). Storing of multiple memories, according to this idea, would occur by formation of multiple, distinct or partially overlapping engrams over time \cite{Abdou_2018}.

 This idea is the natural evolution of earlier proposals based on the concept of `Hebb assembly' \cite{amari72a,h82,a95,ab97,cmla04} which can be generalized in several ways. One such generalization is the inclusion of inhibitory  neurons. Theoretical work has proposed the possibility that inhibitory engrams may be recruited following the formation of a memory. The idea behind the inhibitory engram relies on the possibility that plasticity at a selected group of neurons that release GABA, the inhibitory neurotransmitter, may offer a counterbalance to the increased excitation and readjust circuit excitability within a sensitive and stable range \cite{Barron_2017}. No experimental evidence for the existence of inhibitory engrams is currently available, although there is extensive evidence that GABAergic neurons can change their efficacy in an activity-dependent fashion in response to patterned activity \cite{Maffei_2006,Nugent_2008,Maffei_2011}. The study of the functional significance of inhibitory synaptic plasticity has just began. Recent work demonstrated that a possible role for this form of plasticity is to determine the sign of the change in synaptic efficacy at converging excitatory inputs \cite{Wang:2014ya}, possibly providing constraints on how activity can modulate circuit excitability and connectivity. When considering these results in the context of the clustered theoretical model for metastable dynamics, one may speculate that functional connectivity of neurons recruited for the formation of a memory suggests the presence of a clustered architecture. Inhibitory circuits may contribute to separating the functional clusters, facilitating the formation and stabilization of memories. Studies on memory engrams may therefore support a role for plasticity in establishing and maintaining the cluster architecture that underlies metastable dynamics as predicted by the theoretical models (see Sec.~\ref{sec:learning}).
 
\subsubsection{The potential link with cluster formation}

The evidence discussed above provides grounds to speculate that experience-dependent potentiation of synapses among neurons belonging to the same engram may be part of the mechanism by which neural clusters are formed. Although it is natural to postulate such a relationship, much experimental work is required to clarify the link between cluster-based metastability and the formation of engrams via synaptic plasticity. Formation of engrams via Hebb-like synaptic plasticity is expected to partition cortical networks in subpopulations of neurons, which in turn could give rise to metastability via the mechanism reviewed in Sec.~\ref{sec:spiking_net}. However, the temporal and spatial constraints for co-activation of groups of neurons during the formation of memories are unknown. The identification of engrams so far is limited to the co-expression of genes known as `immediate early genes' that are used as a proxy for neural activity \cite{Tonegawa_2015}, therefore it does not provide any information for a role of these groups of neurons in circuit dynamics, nor on the activity patterns that may have led to their co-activation. 

While experiments can tell us a great deal about the learning rules and the induction of plasticity, the investigation of how changes in the efficacy of synaptic transmission can be integrated in a complex neural circuit without rapidly leading to instability can only be explored with theoretical models that can provide predictive hypotheses. In the case of metastable dynamics, if its maintenance is functional to the operation of a neural network, it is fundamental to determine how the implementation of learning rules observed experimentally may affect neural activity and circuit interactions, to determine how these learning rules modulate state transitions, and what are the neural and circuit requirements that preserve metastable dynamics in the face of experience and learning. In the next section, we review some of the theoretical efforts that have been made in this direction.

\subsection{Modeling cluster formation by experience-dependent plasticity} \label{sec:learning_clusters}

Clustered network models have been able to reproduce metastable dynamics observed experimentally and make useful predictions about possible circuit configurations that generate these dynamics. This is usually accomplished by tuning the synaptic weights and other parameters and requires some degree of ingenuity; mean field theory can aid this process, as reviewed in Sec.~\ref{sec:spiking_net}.  A goal of ongoing research is to understand how the clustered structure underlying metastable activity might develop and then be maintained in the brain. 

Synaptic plasticity is the most likely mechanism for generating metastable clustered neural circuits. Recent theoretical efforts have focused on variations of spike-timing dependent plasticity (STDP) combined with inhibitory plasticity and homeostatic mechanisms for the generation of neural clusters. The self-organization of neural circuits via STDP has been studied in a number of works (see e.g. the Discussion of \cite{Ocker:2015lq} and references therein); here we focus on a couple of recent works that are highly relevant to the development of metastable dynamics in neural circuits \cite{Litwin-Kumar:2014fv,Zenke:2015bv}. 

The basic STDP rule produces a synaptic change based on the relative timing of pre- and postsynaptic action potentials \cite{Markram:1997ee,Bi:1998kt}, though a dependence on the postsynaptic membrane potential can also be included \cite{Wang:2012xd}. However, it has long been recognized that this type of plasticity would quickly saturate the synaptic weights of co-activating neurons (e.g., \cite{von1973self,oja1982simplified,bcm82}), resulting in instabilities or, for finite synaptic weights, in the potential formation of a single, giant cluster of neurons. This problem can be countered by concurrent inhibitory plasticity \cite{Vogels:2013rp}, which acts to reduce the firing rates of excitatory neurons, reducing the interference among clusters and slowing down their formation. 

Let us illustrate these ideas by describing the model of \cite{Litwin-Kumar:2014fv}. The network comprises adaptive exponential integrate-and-fire neurons, a variation of the LIF model neuron described in Sec.~\ref{sec:spiking_net} (whose details are not relevant for the following). The synaptic weights between two excitatory neurons, $J_{ij}$, where $i$ is the postsynaptic neuron and $j$ is the presynaptic neuron, changes according to 
\begin{widetext}
\be \label{eq:ltp}
\frac{d}{dt} J_{ij}(t) = -A_d s_j(t) [u_i(t) - \theta_d]_+ + A_p x_j(t) [V_i(t) - \theta_p]_+ [v_i(t) - \theta_d]_+.
\ee
\end{widetext}
Here, $V_i$ is the membrane potential and $u_i, v_i$ are low-pass filters of $V_i$ with different time constants ($10$ and $7$~ms, respectively); $s_j(t)=\sum_k \delta(t-t_k^j)$ is the presynaptic spike train and $x_j(t)$ is a low-pass filter of $s_j(t)$ with a time constant of $15$~ms; $\theta_{d,p}$ is a threshold for inducing long term depression (LTD) or long term potentiation (LTP), respectively, with $\theta_d < \theta_p$; $[z]_+$ is the threshold linear function, equal to $z$ if $z \geq 0$ and zero otherwise; and $A_{d,p}$ are the strength of LTD and LTP, respectively, with $A_d < A_p$. 

The first term in the right hand side of Eq.~\ref{eq:ltp} causes LTD when the presynaptic  neuron fires an action potential and the (filtered) postsynaptic neuron activity $u_i$ is small but above $\theta_d=-70$~mV (equal to the neuron's resting potential). For large values of the membrane potential ($V_i>\theta_p=-49$~mV), the second term will dominate, leading to LTP. Synapses between neurons that co-activate in response to the same stimulus will undergo LTP, while synapses onto neurons that remain inactive will undergo LTD, in fairly Hebbian fashion (no change occurs if the presynaptic neuron is inactive).

This learning rule leads to synchronized oscillations \cite{Litwin-Kumar:2014fv}. To avoid this, inhibitory plasticity is implemented on the synapses {\em from inhibitory to excitatory} neurons:
\begin{eqnarray} \nonumber
J_{ij} & \leftarrow & J_{ij} + \eta (y_i(t) - 2 r_0 \tau_y) \quad \mbox{if the presynaptic neuron fired,} \\ \nonumber
J_{ij} & \leftarrow & J_{ij} + \eta y_j(t) \hspace{1.7cm} \mbox{if the postsynaptic neuron fired}.
\end{eqnarray}
Here, $J_{ij}$ is the absolute value of the synaptic weight, $y_{i,j}(t)$ is the low-pass filter of the postsynaptic or presynaptic spike train, respectively (with time constant $\tau_y=20$~ms), $\eta$ is the learning rate, and $r_0=3$~spikes/s is a target firing rate to which the inhibitory plasticity attempts to balance, homeostatically, the postsynaptic neuron. The inhibitory-to-excitatory weights are further kept within a bounded range. 

However, even this combination of excitatory and inhibitory plasticity seems insufficient to generate stable clusters, unless a process of `synaptic normalization' is also added \cite{von1973self,Renart:2003gt,Fiete:2010ub}. This is a procedure in which, at regular intervals, the excitatory synaptic weights onto the same postsynaptic neuron are reduced by a fixed amount (this will maintain a constant row sum in the synaptic matrix). This procedure depends on the current values of all the synaptic weights and it is not clear how it could occur in real synapses: as synapses do not have a way to communicate their strengths to one another,  learning rules should be {\em local}, i.e., they should make use of information coming only from the pre- and postsynaptic neuron \cite{Wang:2012xd} or from global neuromodulatory signals \cite{Reynolds:2002,Huang_2012} (see e.g. Ch.~10 of \cite{gk02} for theoretical implications of locality). The timing of synaptic normalization is also troublesome, since a mechanism that would inform all synapses in a circuit to normalize themselves based on an internal clock is not known to exist (although randomizing the times at which renormalization occurs would probably achieve the same goal \cite{luisa:2020uk}). However, there is evidence that neurons can sense their level of activity within the network and readjust their activity state to remain within a functional range \cite{Turrigiano_2017}, or sense network activity and engage distinct forms of plasticity to readjust it within a functional range \cite{Maffei_2009}. 

As experimental work investigating the functional implications of homeostatic regulation of circuit excitability is still ongoing, identifying biophysically plausible means of normalizing all synapses in a circuit to preserve circuit dynamics can at the moment be explored only in theoretical models. If well conceptualized, these models have predictive power and can provide strong working hypotheses for future experiments. The recent models reviewed above have a good degree of biological plausibility and have taught us many lessons, including: (i) once clusters are formed, metastable activity generates patterns of activity similar to those experienced by the neurons during training. By mimicking the conditions occurring during training, metastable activity can coexist with ongoing plasticity \cite{Litwin-Kumar:2014fv}. (ii) After training, the clustered structure of the network can be further modified by training with new stimuli while preserving metastability. (iii) Inhibitory plasticity is not strictly necessary for the formation of stable clusters. If the correct level of inhibitory activity is set prior to training, excitatory plasticity may be sufficient. However, inhibitory plasticity provides an adaptive mechanism to produce the necessary amount of inhibition to render learning stable \cite{Zenke:2015bv}.

\subsection{Landscape modifications induced by learning} \label{sec:learning_landscape}

In Sec.~\ref{sec:fluxtheory} we have introduced the non-equilibrium landscape and flux framework for general neural networks. As discussed there, local minima in neural network systems can significantly influence their dynamical behaviors by inducing sequences of metastable states. In fact, metastable dynamics induced by external stimuli, changes in synaptic properties and ubiquitous noise play a crucial role in different neural circuit functions. Below, we discuss the relationship between network dynamics and their corresponding neurobiological functions including associative memory, working memory, decision making, and fear response. The ability to produce the relevant dynamics can be predicted by the model landscape, which in turn can be modified by learning. We shall focus on two critical functional properties of neural systems that at first glance appear incompatible: robustness and sensitivity. As we shall see, these properties are closely associated with metastable dynamics.

\subsubsection{Associative memory} \label{sec:assoc}

\begin{figure*}
\begin{center}
\includegraphics[width=1\textwidth]{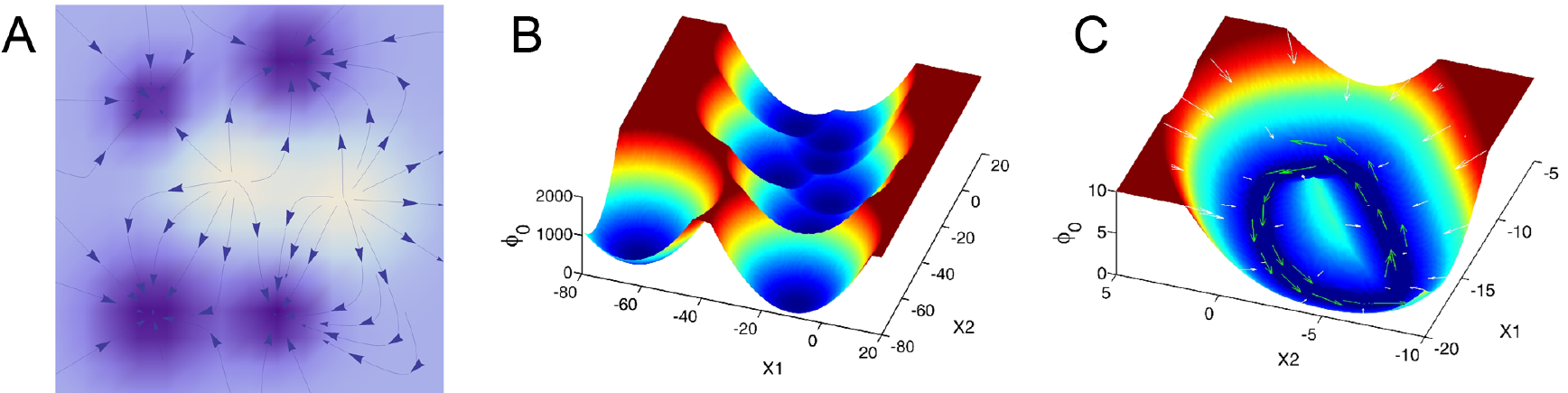}
\end{center}
\caption{\small (A) Schematic diagram of the energy function landscape of the Hopfield network. (B) The potential landscape $\phi_0(\bm{x})$ of a symmetric neural network. (C) The potential landscape $\phi_0(\bm{x})$ as well as the corresponding force for an asymmetric neural circuit: the green arrows represent the flux, and the white arrows represent the force from the negative gradient of the potential landscape. Reproduced from H. Yan, L. Zhao, L. Hu, X. Wang, E. Wang, and J. Wang, Proc Natl Acad Sci U S A 110, E4185--94 (2013). Copyright 2013 the authors. \protect \cite{Yan:2013rs}}
\label{fig3jin}
\end{figure*}

Associative memory is the ability to form memories that associate two stimuli, for example, a sound or light cue and a footshock in the case of fear conditioning. Once the association is learned, it can be retrieved with a partial cue close to that information. As discussed in Sec.~\ref{sec:hebb}, it is widely believed that associative memories are formed through the induction of synaptic plasticity and stored in synapses that have been modified during learning. Theoretically, the memories are represented by attractors of the network dynamics. Early modeling studies of associative memory have often focused on systems with symmetric couplings \cite{h82,YH21}, i.e., the connections between pairs of neurons satisfy $J_{ij}=J_{ji}$. In this case, memory states are local minima of a global energy function $E$. Once a partial cue about the desired memory is presented, the system is put in the initial state that is near the basin of attraction of the valley corresponding to that desired memory. Thus, memory retrieval can be performed by following the gradient of the computational energy $E$ (see Fig.~\ref{fig3jin}A). 

In real cortical networks synapses are not symmetric \cite{Buzsaki:2014pe}, a property that can be related to learning and recalling sequences rather than single items. Symmetric networks, whose dynamical behaviors are determined by purely potential energy, lack the ability to retrieve a temporally ordered sequence of memories using a single recalling input. Such a computation for temporal association can be achieved in asymmetric circuits through appropriate learning rules, which result in rapid transitions between quasi-stable states that represent individual memories \cite{YH2,amitBook89,YH38}. For general asymmetric circuits, the energy function $E$ defined in the symmetric cases is no longer a Lyapunov function. It cannot  describe collective computational properties such as temporal association. Fortunately, the intrinsic potential $\phi_0(\bm{x})$ we introduced in Sec.~\ref{sec:fluxtheory} is always a Lyapunov function regardless of the symmetry of the synaptic weights. The dynamics of general networks are dominated by both the nonequilibrium potential landscape related to the steady-state probability distribution and the steady-state rotational curl probability flux \cite{Yan:2013rs,YH25}. The flux breaking the detailed balance can provide the main driving force for transitions between states and oscillations.

Fig.~\ref{fig3jin}(B) and (C) shows landscapes of Lyapunov function $\phi_0(\bm{x})$ for symmetric and asymmetric circuits, respectively. Similar to the Hopfield symmetric network model, we can see multiple attractors in Fig.~\ref{fig3jin}(B). For asymmetric networks, sequential `oscillatory' motion can occur and the underlying potential landscape has a Mexican-hat topography (Fig.~\ref{fig3jin}(C)). After being attracted down to the oscillation ring (white arrows represent the driving force from the negative gradient of the potential landscape), the system is mainly dominated by the curl flux shown by the green arrows. Different memories can be associated by the flux, which may induce ordered sequences of memories, due to the asymmetry resulting from specific learning procedures. This produces associations among memories through the continuous attractor (oscillation ring) which is driven by the flux due to the asymmetry of the synaptic connections.

Oscillatory patterns of neural activity are widely distributed in our brain \cite{YH39,YH40,YH41,YH42,YH43}. Oscillations may play a role in various aspects of memory including spatial representation and memory maintenance \cite{YH44,YH45}. In addition to the context of storage and associative recall of information, the present approach can be useful in understanding mechanisms for generating rhythmic motor patterns (such as swimming and locomotion) by central pattern generators (CPG), the synchronization among different groups with coherent oscillations in cognitive functions such as and physiological rhythm regulations \cite{YH38,YH42,YH46,YH47,YH48,YH49,YH50}. An important example of this class of problems is the cycling of sleep phases. The rhythmic REM/non-REM cycle in human sleep is regulated by the activation-repression of two neural populations \cite{YH49,YH50}. A detailed description of the REM/non-REM sleep rhythm with the landscape and flux approach is shown in \cite{Yan:2013rs}. A global sensitivity analysis based on the global topography of the landscape shows the effects of key factors such as the release of the neurotransmitters acetylcholine and norepinephrine on the stability and function of the system.

\subsubsection{Working memory and decision making}
\begin{figure*}
\begin{center}
\includegraphics[width=1\textwidth]{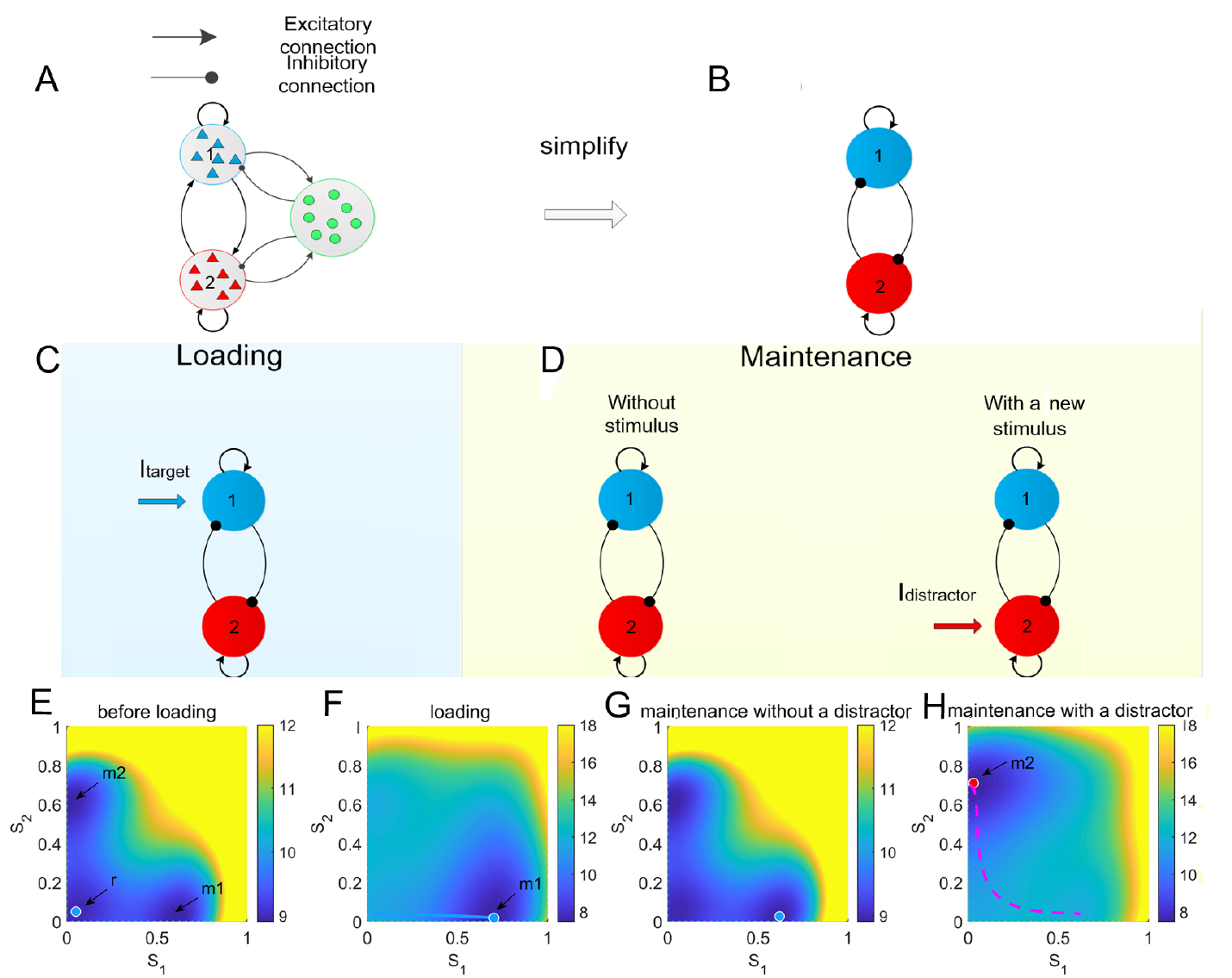}
\end{center}
\caption{\small (A-B) The schematic diagram of the circuit model for working memory (WM). The model comprises two selective, excitatory populations, labeled 1 and 2. Each excitatory population is recurrently connected and inhibits each other through a common pool of inhibitory interneurons. (C-D) The schematic diagrams for the WM during different phases in a WM task. (E-H) The corresponding potential landscapes in the (S1; S2) state space during different phases. The dimensionless quantities S1 and S2 are average synaptic gating variables of the two selective populations, which can represent the mean population activities. The label `r' indicates the attractor representing the resting state. The attractors representing the target-related and distractor-related memory state are labeled with `m1' and `m2', respectively. Reproduced from H. Yan and and J. Wang, PLoS. Comput. Biol.16(10): e1008209 (2020); licensed under a Creative Commons Attribution (CC BY) license. \protect \cite{YH26}}
\label{fig4jin}
\end{figure*}

Actively holding information online for a brief period of time (seconds) is an important ability of the brain. This capability is a part of working memory (WM), which is used for tasks such as planning, organizing, movement preparation and decision-making \cite{YH51,YH52,YH53,YH54}. In contrast to long-term memory, which requires structural changes in neural circuits and in the connections between neurons, the mechanisms underlying working memory are believed to depend on persistent neuronal activity \cite{YH55,YH56,YH57}. In general, positive reverberation driven by recurrent synaptic excitation in interconnected neural clusters can work as the basic principle for generating persistent activity. Triggered by incoming signals, working memory circuits can sustain an elevated firing even after the inputs are withdrawn. As discussed in Sec.~\ref{sec:spiking_net} (see e.g. Fig.~\ref{fig:mfrasters}A), dynamical models with local feedback excitation between principal neurons that are controlled by global feedback inhibition can exhibit multiple attractor states (each coding a particular memory item) that coexist with a background (resting) state. This is illustrated in Fig.~\ref{fig4jin} below for a circuit model of WM with two excitatory populations. 

\begin{figure*}
\begin{center}
\includegraphics[width=0.9\textwidth]{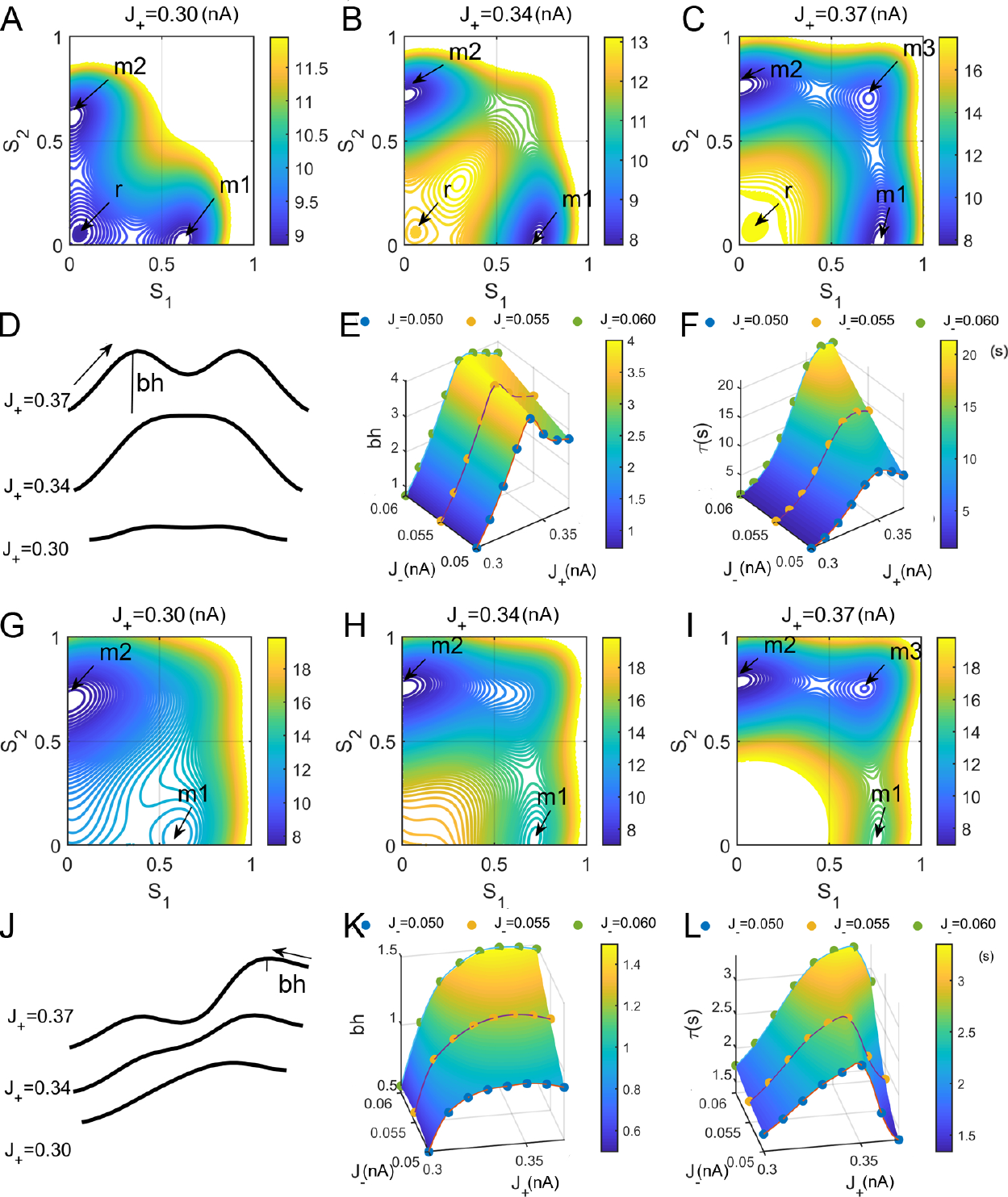}
\end{center}
\caption{\small (A-F) The robustness against random fluctuations during the maintenance phase. (A-C) The potential landscapes for different self-excitations $J_+$. (D) The schematic diagram of the barrier heights on the corresponding potential landscapes for increasing $J_+$. (E-F) Robustness of WM against random fluctuations as a function of self-excitations $J_+$ and mutual inhibition $J_-$ through quantifying the corresponding barrier height and the mean first passage time across the barrier. (G-L)  The robustness against distractors during the maintenance phase. Reproduced from H. Yan and J. Wang, PLoS. Comput. Biol.16(10): e1008209 (2020); licensed under a Creative Commons Attribution (CC BY) license. \protect \cite{YH26}}
\label{fig5jin}
\end{figure*}

Although robustness in the face of small perturbations or noise is an important requisite for a working memory system, the transient properties of the system are also important \cite{YH55,YH58,Miller_2001}. Sometimes, it is needed to tilt the balance in favor of increased flexibility rather than robustness depending on environmental conditions and/or behavioral task demands, e.g., a foraging task in which an animal uses visual input to catch a prey \cite{YH60}. In addition to robust maintenance of memory states, activity should be reset quickly when there is a novel sensory cue that needs to be stored. In other words, a working memory system should have the properties of robustness against fluctuations while being very sensitive to incoming stimuli. In recent theoretical works, this fundamental contradiction can be achieved by global inhibitory connections, where the system can exhibit structurally stable dynamics with fixed stimulus and qualitatively change its dynamics if the stimulus is changed\cite{YH26}. The non-equilibrium potential landscape and flux approach has been applied to a biophysically based working memory model composed of integrate-and-fire neurons through a mean-field approach, which can reproduce most of the psychophysical and physiological results in delayed response tasks \cite{YH26,YH54,wilson1972excitatory,Wong2006-ey}. This model comprises two selective, excitatory populations, labeled 1 and 2, whose dynamics are described by the following equations:
\begin{equation}
\frac{ds_{i}}{dt} = - \frac{s_{i}}{\tau_{s}} + \left( 1 - s_{i} \right)  \gamma f\left( I_{i,tot} \right),~i \in \{1,2\},
\end{equation}
where $s_{i}$ is the average gating variable of neural population $i$, which can represent the mean population activities, and $r_i = f(I_{i,tot})$ is the corresponding firing rate, a function of total input current $I_{i,tot}$, $\tau_S$ is the gating timescale, and $\gamma$ is the kinetic parameter that controls the rate of saturation of $s_i$. The robustness of working memory was quantified by the underlying landscape topography (barrier heights) and the corresponding mean transition time for varying recurrent excitations and mutual inhibitions, shown in Fig.~\ref{fig5jin} \cite{YH26}. A combination of both increased self-excitation and mutual inhibition can enhance flexibility to external signals without significantly reducing robustness to random fluctuations. The key element of the underlying mechanism for achieving good performance in working memory is the emergence of a new intermediate state with larger energy consumption.

\begin{figure*}
\begin{center}
\includegraphics[width=1\textwidth]{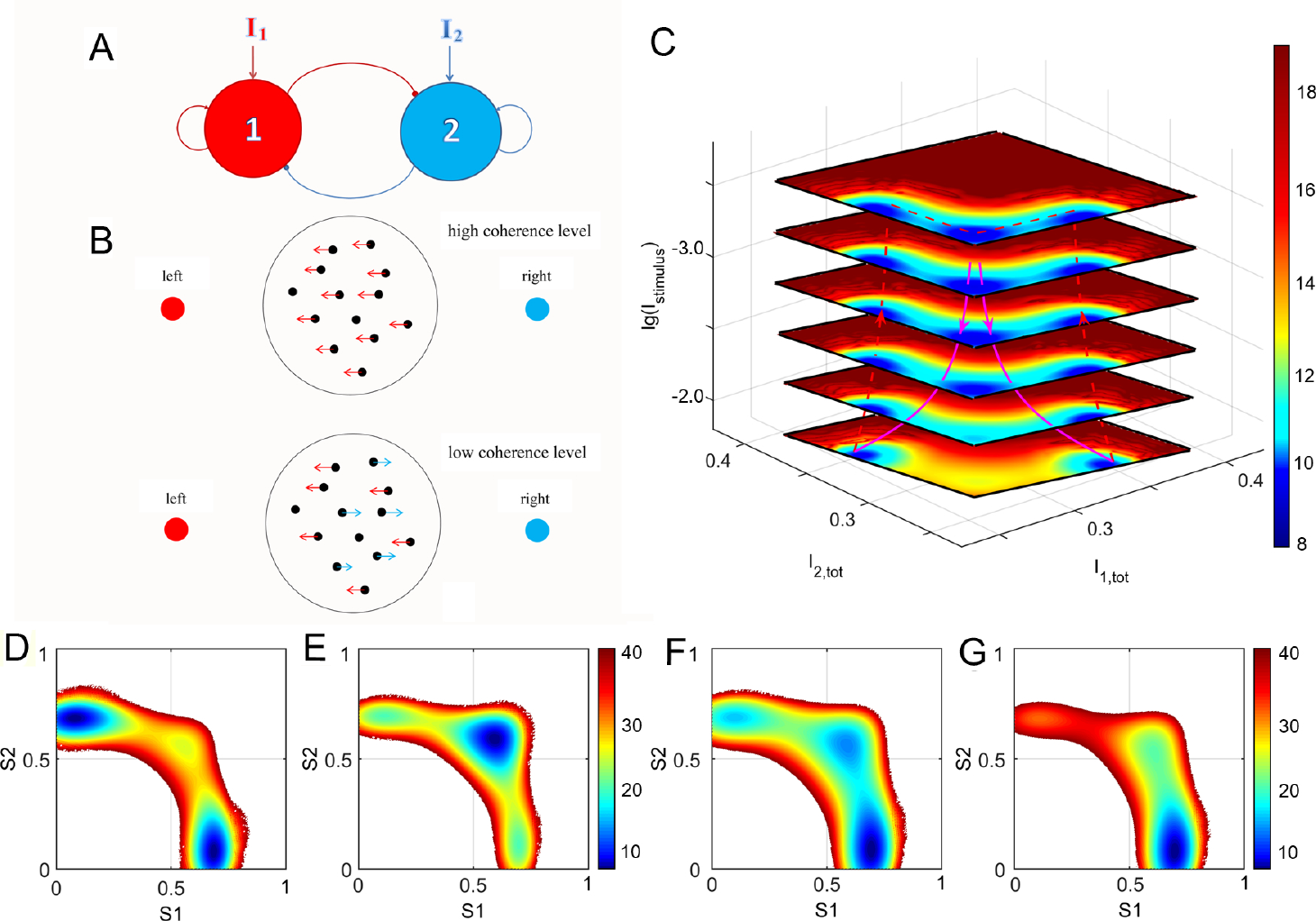}
\end{center}
\caption{\small (A) The schematic diagram of the reduced two population decision-making model. This reduced model consists of two competing neural populations that are selective for leftward or rightward directions, respectively. (B)The schematic representation of the random dots motion. For higher motion coherence, most dots move in one direction, whereas the dots move with no directional bias at a low motion coherence level. (C) The potential landscape of the decision-making network with varying inputs and pathways. The pink lines indicate the paths of decision making from undecided state to decided states. The red dotted lines represent the paths from the two decided states back to the undecided state. (D-G) The mechanism of changes of mind based on the emergence of the new intermediate state in the center. (D-E) The two-dimensional potential landscapes for different large inputs at zero coherence level. (F-G) The two-dimensional potential landscapes for large input  when the motion coherence $c'= 0.02$ and $0.06$, respectively. Reproduced with permission from Yan H, Zhang K and Wang J., Chin. Phys. B. 25, 078702 (2016). Copyright 2016 Chinese Physical Society and IOP Publishing. \protect \cite{YH68}.}
\label{fig6jin}
\end{figure*}

Cortical areas that are engaged in working memory, such as the prefrontal cortex, are also involved in other cognitive functions such as decision making, selective attention, and behavioral control \cite{YH52,YH54,Wong2006-ey,YH63,YH64}. In the two-choice visual motion discrimination task, trained monkeys were presented for a few seconds with a pattern of randomly moving dots and asked to make a decision regarding the direction of motion by saccadic eye movement \cite{YH64,YH65,YH66,YH67}. Models originally developed for working memory can account for decision-making processes. The only difference between a working memory simulation and a decision simulation is that in a delayed-response task only one stimulus is presented while for a perceptual discrimination task conflicting sensory inputs are fed into competing neural subpopulations in the circuit (Fig.~\ref{fig6jin}A) \cite{YH18,YH54}. The external sensory input to selective neural population in random dots motion tasks is determined by the motion coherence, which is used to indicate the degree of direction bias of moving dots (Fig.~\ref{fig6jin}B-C).

With the landscape approach, the decision-making processes can be quantified with the optimal paths from the undecided attractor states to the decided attractor states, which are identified as basins of attraction on the underlying landscape (Fig.~\ref{fig6jin}C) \cite{YH25,YH68}. In addition to the speed-accuracy tradeoff discussed in previous decision-making studies, actually there is a speed-accuracy-dissipation tradeoff. When additional signals are presented, reasonable accuracy performance can be reached with minimum dissipation cost and fast decision time \cite{YH68}. Moreover, making decisions is often accompanied by situations in which we change our minds. The mechanism of mind changes may be closely associated with new, intermediate states that emerge when large inputs are presented (Fig.~\ref{fig6jin}(D-G)). The initial incorrect choice is more likely to be changed due to the attraction by the new intermediate state, while the initial correct choice is still likely to be maintained. The speed-accuracy tradeoff always works. Although time pressure may lead to more initial errors, changes made to correct these errors are also more likely to happen due to the emergence of a new intermediate state. The mechanism for changes of mind guarantees a reasonable performance of the decision making process with emphasis on speed.

\subsubsection{Fear learning and expression}

The non-equilibrium landscape and flux approach can also be used to analyze the attractor landscape of cortical circuits involved in the phenomenon of fear conditioning. These circuits support the rapid selection of the appropriate response to a threat, a behavior that is crucial for animals' survival. Such ability involves associative learning from experience to predict, and then make appropriate responses, to danger. Pavlovian fear conditioning experiments have often been used to understand the neural circuitry underlying fear learning and expression \cite{YH69,YH70,YH71,YH72,YH73}. During conditioning, subjects are presented with a {\em conditioned stimulus} (CS; typically a sound or a light stimulus) paired with an aversive {\em unconditioned stimulus} (US; typically a mild electric shock) to elicit conditioned defensive responses, such as freezing behavior (a passive defensive response) \cite{YH73,YH74,YH75} or active responses such as startle, escape, flight and avoidance \cite{YH76,YH77}.

\begin{figure*}
\begin{center}
\includegraphics[width=1\textwidth]{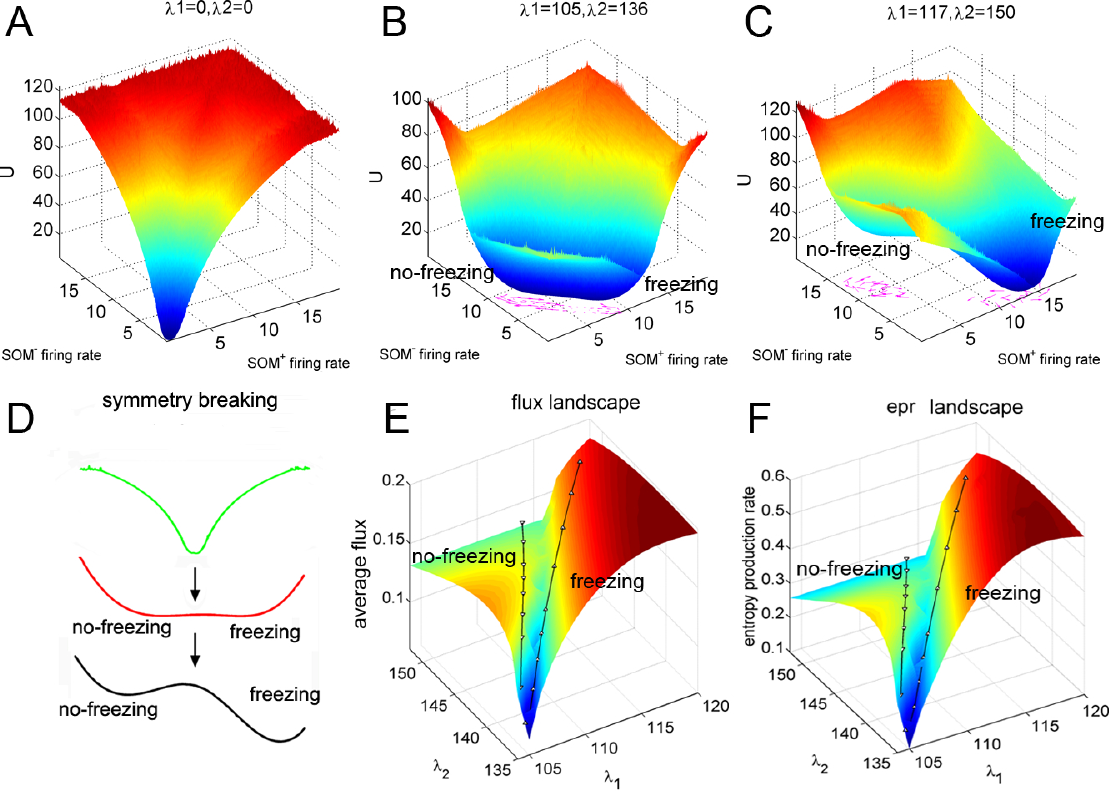}
\end{center}
\caption{\small (A-C) Potential landscapes for different inputs. The increase of the inputs induces symmetry breaking from the symmetric but featureless state to the biased state with biological functions. The fluxes are indicated by purple arrows. $\lambda_1$ and $\lambda_2$ represent the strength of the inputs to the SOM+ and SOM- neurons, respectively. (D) A diagram of how a one-dimensional potential landscape changes with stimulus inputs. (E) Average flux landscape in the space of different inputs. The average flux is significantly positively correlated with the external inputs, when the neural circuit is away from its equilibrium state. (F) The entropy production rate landscape in the space of different inputs. The neural circuit dissipates more energy with larger inputs. It costs more energy to maintain the dominant freezing responses than dominant no-freezing behaviors. $\lambda_1$ and $\lambda_2$: same as is panels A-C. Reproduced with permission from Yan H, Li B and Wang J., J. R. Soc. Interface. 16,20180756 (2019). Copyright 2019 Royal Society.\protect \cite{YH78}.}
\label{fig7jin}
\end{figure*}

The central amygdala plays a crucial role in both acquisition and expression of conditioned fear \cite{YH69,YH70,YH71,YH72,YH73}. Following fear conditioning, the excitatory synapses from neurons in the lateral amygdala onto neurons in central amygdala undergo different changes: those onto a category of inhibitory neurons known to express somatostatin (SOM+) are strengthened, while those onto non-somatostatin expressing neurons (SOM-) are weakened \cite{YH73,YH74,YH75}. In turn, this elicits freezing behavior. 

The mechanisms of active defensive response, as well as the rapid selection between passive and active responses, are less understood. It is possible that rapid selection is gated by the central amygdala. To understand the underlying dynamic mechanism of how the central amygdala gates passive and active defensive responses, \cite{YH78} used landscape and flux theory to study a model of the central amygdala dominated by local inhibitory connections between SOM+ and SOM- neurons. With this approach, the underlying attractor landscape of the circuit model with varying inputs can be quantified (Fig.~\ref{fig7jin}A-D). 

In the model, freezing states observed in experiment (due to activated SOM+ neurons and inhibited SOM- neurons) emerge in the presence of a CS with biased excitatory inputs to SOM+ neurons (due to fear conditioning-induced synaptic modifications) \cite{YH73,YH77}. However, the model also shows that, if excitatory synapses to SOM+ and SOM- neurons are {\em both} strengthened during fear conditioning, for a range of inputs there exists a bi-stable phase with both a freezing and a non-freezing state. In this bi-stable phase, the non-freezing (active defense) state is induced by the same inputs that can elicit freezing responses, but resulting instead in non-freezing active responses. Learning to selecting the type of response (passive vs. active) also requires the strengthened synaptic projections to both SOM+ and SOM-neurons. The underlying topology of non-equilibrium landscape shaped by such a set of inputs supports two distinct attractors with a clear barrier in between (Fig.~\ref{fig7jin}A-D). The switches between different defensive responses under threats can be physically characterized by the transitions between the two attractors. Furthermore, the maintenance of such bi-stable phase needs additional energy, which can be measured through the entropy production rate closely related to the non-equilibrium flux and discussed in Sec.~\ref{sec:noneq-thermo} (Fig.~\ref{fig7jin}E-F).

Based on this model, we predict that in situations where active responses are reinforced, such as in active avoidance learning, excitatory synaptic transmission onto the SOM- neurons would be more robustly potentiated than in situations where only passive responses are allowed, such as classical fear conditioning. If only the excitatory projections to SOM+ neurons were strengthened during fear conditioning, only the freezing state (passive fear response) would be possible in response to threats. 

\subsubsection{Lessons from the case studies}

The above case studies show how the non-equilibrium landscape and flux approach provides a general way to study neural circuit dynamics. As explained in Sec.~\ref{sec:fluxtheory}, this approach allows to analyze modern models of neural networks that have non-symmetric connections. This allows in particular to analyze models with synaptic connections that respect Dale's law (stating that neurons are either excitatory or inhibitory), and therefore obey important biological constraints. The relevant models reviewed in Sec.~\ref{sec:theorymodels} belong to this class and therefore require the landscape and flux theory approach for a proper analysis.

To summarize, the sequential functional states during metastable activity can be identified through the underlying potential landscapes. In particular, the robustness of these functional states can be quantified not only by the depth and the breadth of the corresponding basins of attraction, but also according to the distance between the basins and the transition times between states corresponding to different basins. Metastable neural dynamics can be induced by noise (whether from external sources or endogenously generated as in the deterministic spiking model of Sec.~\ref{sec:spiking_net}), or it can be induced by changes in the landscape topography. The latter may result from varying key ingredients such as relevant synaptic connections, inputs, or neurons. Regardless of its origin, metastable dynamics in cortical circuits can be described in terms of the optimal paths under the action of both the landscape and flux components of the driving forces, and has uncovered important potential mechanisms for various brain functions, including sleep cycle regulation \cite{Yan:2013rs}, stability-flexibility tradeoff in working memory \cite{YH26}, changes of mind in decision-making \cite{YH68}, the selection of passive and active fear responses \cite{YH78}, and network mechanisms of Parkinson disease \cite{YH79}. The theory also allows the quantification of the thermodynamic cost for maintaining neural network function and can help to facilitate the design of strategies achieving an optimal balance between performance and cost. These are not easily achieved through more conventional theoretical methods.

%
\section{Summary and conclusions} \label{sec:summary}

The field of neuroscience is moving towards an appreciation of the role of neural dynamics in coding and computation. In this article we have reviewed recent progress on the front of characterizing and modeling neural dynamics, with particular emphasis on a type of metastable dynamics that unfolds as a sequence of discrete states. This kind of metastable dynamics has been quantified in several cortical areas of rodents, monkeys and humans, and seems related to coding sensory stimuli as well as internal deliberations. Focusing on metastable dynamics signals a departure from earlier and more traditional views, which were centered on the notion of single neurons' input-output function and its modulation as predictors of stimulus features or behavioral outcomes. One of the most relevant implications of the fact that cortical activity evolves as a sequence of discrete, metastable states is that transitions in neural activity are not just triggered by external events, such as a stimulus or a reward, but are instead spontaneously generated and may occur at anytime, including when the subject is idling and not engaged in a task. This goes against the notion that neural activity is just a `reaction' to external events or a static representation of incoming stimuli, and is compatible with the presence of incessant `ongoing activity' observed in cortex. 

Among the most salient characteristics of the type of cortical metastable dynamics reviewed here are: (i) the hidden states are states of collective behavior in populations of neurons that are putative fixed points of the neural dynamics (Sec.~\ref{sec:models}); (ii) state transitions are typically one order of magnitude faster than the state durations, and are close to their theoretically observable lower bound (Sec.~\ref{sec:HMM}); (iii)  the ongoing dynamics of some cortical circuits is also highly structured and characterized by repeatable metastable transitions, and does not resemble metastability en route to a ground state configuration (Sec.~\ref{sec:ongoing}); (iv) neural dynamics evoked by a stimulus is often metastable, with some hidden states coding for specific stimulus features, internal decisions, or upcoming actions (Sec.~\ref{sec:evidence}); (v) metastable dynamics during a task can be modulated by task variables (such as trial difficulty or internal expectation) or behavioral outcome (correct vs. error) (Sec.~\ref{sec:modulations}).

We have also reviewed classical and contemporary approaches to infer hidden states from the neural dynamics (Sec.~\ref{sec:statmodels}), as well as mechanistic models of metastable dynamics based on clustered networks of spiking neurons (Sec.~\ref{sec:spiking_net}). A topological organization in potentiated clusters seem necessary to explain the observed metastable dynamics and is able to predict its most salient features. Network models are hard to study and are often analyzed with mean field techniques; we have reviewed basic mean field theory for networks of spiking neurons as well as the more general landscape and flux theory of network dynamics (Sec.~\ref{sec:theorymodels}). While the former allows to predict the fixed points of the dynamics in a situation of equilibrium, with the fixed points possibly becoming metastable in finite networks with random connectivity, landscape and flux theory also allows to study the metastable dynamics out of equilibrium.

Neural clusters have only been indirectly observed so far, and are presumably learned through experience. We have reviewed the possible links between plasticity and metastable dynamics in Sec.~\ref{sec:plasticity}. This included recent efforts to obtain metastable dynamics in spiking network models via experience-dependent synaptic plasticity (Sec.~\ref{sec:learning_clusters}), as well as theoretical investigations of the consequences of learning on the dynamics of cortical networks (Sec.~\ref{sec:learning_landscape}). 

With the improvement of recording techniques and the ability to perform a larger variety of behavioral tasks in the laboratory, we predict that evidence for neural clusters and metastability will continue to accrue, together with the refinement of theoretical tools for their analysis and modeling. In turn, these endeavors will help solidify a dynamics-centric view of cortical activity supporting sensory and cognitive processes. 

\acknowledgments

The authors wish to thank Tianshu Li for preparing Figure 1. This work was partially supported by the National Institute Of Neurological Disorders And Stroke of the National Institutes of Health under award number UF1NS115779 (BAWB, AM, IMP, AF, JW and GLC) and by the National Natural Science Foundation of China under award number NSFC 21721003 (HY). The content of this article is solely the responsibility of the authors and does not necessarily represent the official views of the National Institutes of Health or the National Natural Science Foundation of China. 

\section*{Author contributions}

All authors contributed to the design of the manuscript, literature research and manuscript writing.

\section*{Data availability statement} 

Data sharing is not applicable to this article as no new data were created or analyzed in this study.

\section*{References}
\input{APRreview.bbl}

\end{document}

%% file: APRreview.bbl
%